\documentclass[preprint,12pt]{elsarticle}

\usepackage{amssymb}
\usepackage{subfigure}
\usepackage{threeparttable}
\usepackage{array}
\usepackage{multirow}
\usepackage{amsmath}
\usepackage{bm}
\usepackage{graphicx}
\usepackage{float}
\usepackage{color}
\usepackage{multirow}
\usepackage{float}
\usepackage{hyperref}

\usepackage{amsmath}
\usepackage{gensymb}
\usepackage{verbatim}%to comment out a section
\usepackage{graphicx}
\usepackage{float}
\usepackage{subfigure}
\usepackage{hyperref}
\usepackage{threeparttable}
\usepackage{soul}
\usepackage{rotating, multirow}

\usepackage{color}
\usepackage[]{lineno}
\usepackage{bm}% bold math

\journal{Computers \& Mathematics with Applications}

\begin{document}

\begin{frontmatter}



\title{Investigation of nonlinear squeeze-film damping involving rarefied gas effect in micro-electro-mechanical-systems}

\author[]{Yong Wang$^a$}
\ead{wongyung@mail.nwpu.edu.cn}

\author[]{Sha Liu$^{a,b}$}
\ead{shaliu@nwpu.edu.cn}

\author[]{Congshan Zhuo$^{a,b}$}
\ead{zhuocs@nwpu.edu.cn}

\author[]{Chengwen Zhong$^{a,b}$\corref{cor1}}
\ead{zhongcw@nwpu.edu.cn}

\address{$^a$School of Aeronautics, Northwestern Polytechnical University, Xi'an, Shaanxi 710072, China\\
$^b$National Key Laboratory of Science and Technology on Aerodynamic Design and Research, Northwestern Polytechnical University, Xi'an, Shaanxi 710072, China}

\cortext[cor1]{Corresponding author}

\begin{abstract}
In this paper, the nonlinear squeeze-film damping (SFD) involving rarefied gas effect in the micro-electro-mechanical-systems (MEMS) is investigated. Considering the motion of structures (beam, cantilever, and membrane) in MEMS, the dynamic response of structure will be influenced largely by the squeeze-film damping. In the traditional model, a viscous damping assumption that damping force is linear with moving velocity is used. As the nonlinear damping phenomenon is observed for a micro-structure oscillating with a high-velocity, this assumption is invalid and will generates error result for predicting the response of micro-structure. In addition, due to the small size of device and the low pressure of encapsulation, the gas in MEMS usually is rarefied gas. Therefore, to correctly predict the damping force, the rarefied gas effect must be considered. To study the nonlinear SFD phenomenon involving the rarefied gas effect, a kinetic method, namely discrete unified gas kinetic scheme (DUGKS), is adopted. And based on DUGKS, two solving methods, a traditional decoupled method (Eulerian scheme) and a coupled framework (arbitrary Lagrangian-Eulerian scheme), are adoped. With these two methods, two basic motion forms, linear (perpendicular) and tilting motions of a rigid micro-beam, are studied with forced and free oscillations. For a forced oscillation, the nonlinear phenomenon of squeeze-film damping is investigated. And for a free oscillation, in the resonance regime, some numerical results at different maximum oscillating velocities are presented and discussed. Besides, the influence of oscillation frequency on the damping force or torque is also studied and the cause of the nonlinear damping phenomenon is investigated.
\end{abstract}

\begin{keyword}
Squeeze-Film Damping \sep Discrete Unified Gas Kinetic Scheme \sep Rarefied Gas Flow \sep Micro-Electro-Mechanical-Systems



\end{keyword}

\end{frontmatter}


\section{Introduction}\label{introduction}
\begin{figure}
	\centering
	\subfigure[]{
		\includegraphics[width=0.45 \textwidth]{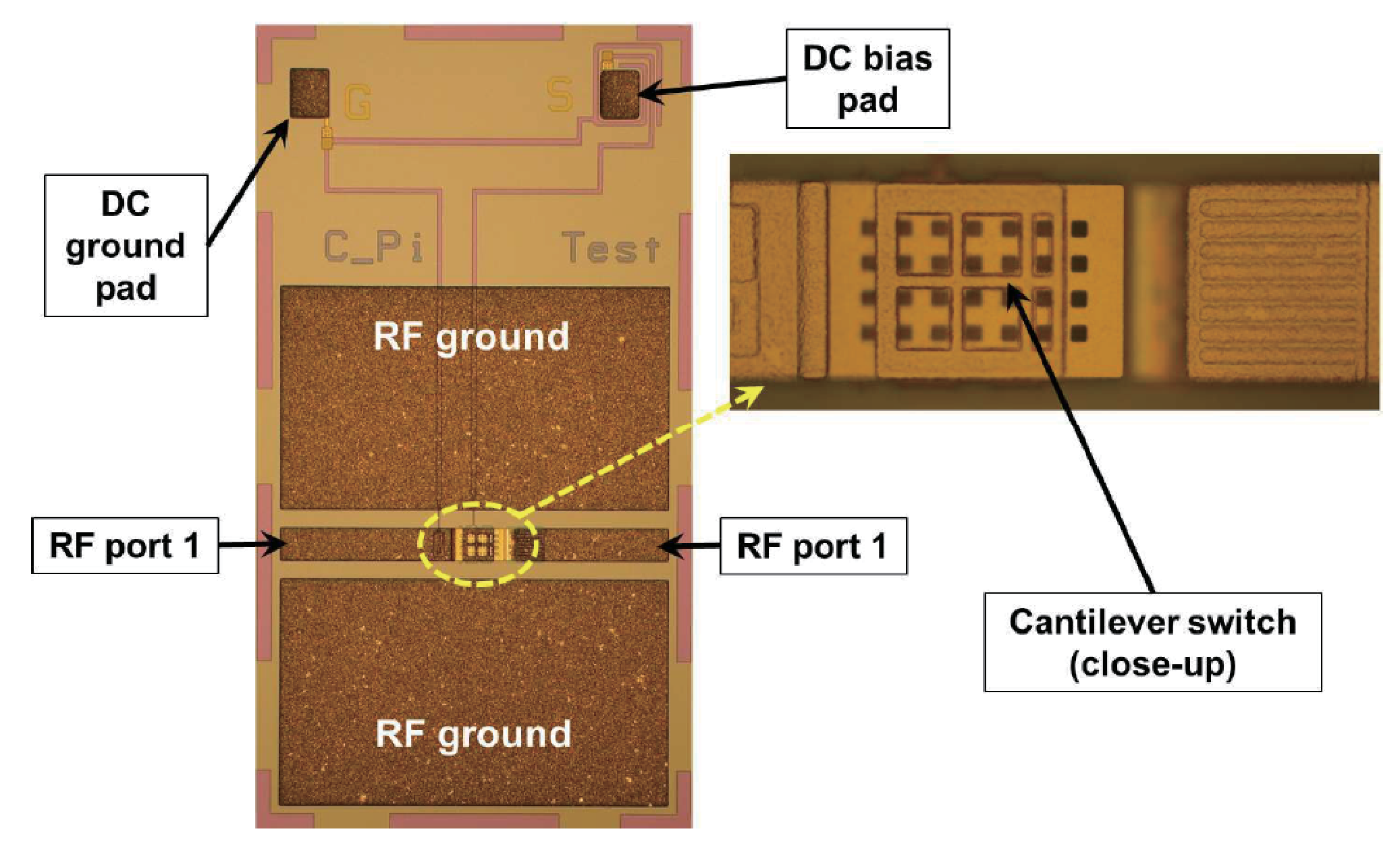}
		\label{rf_mems1}
	}
	\subfigure[]{
		\includegraphics[width=0.45 \textwidth]{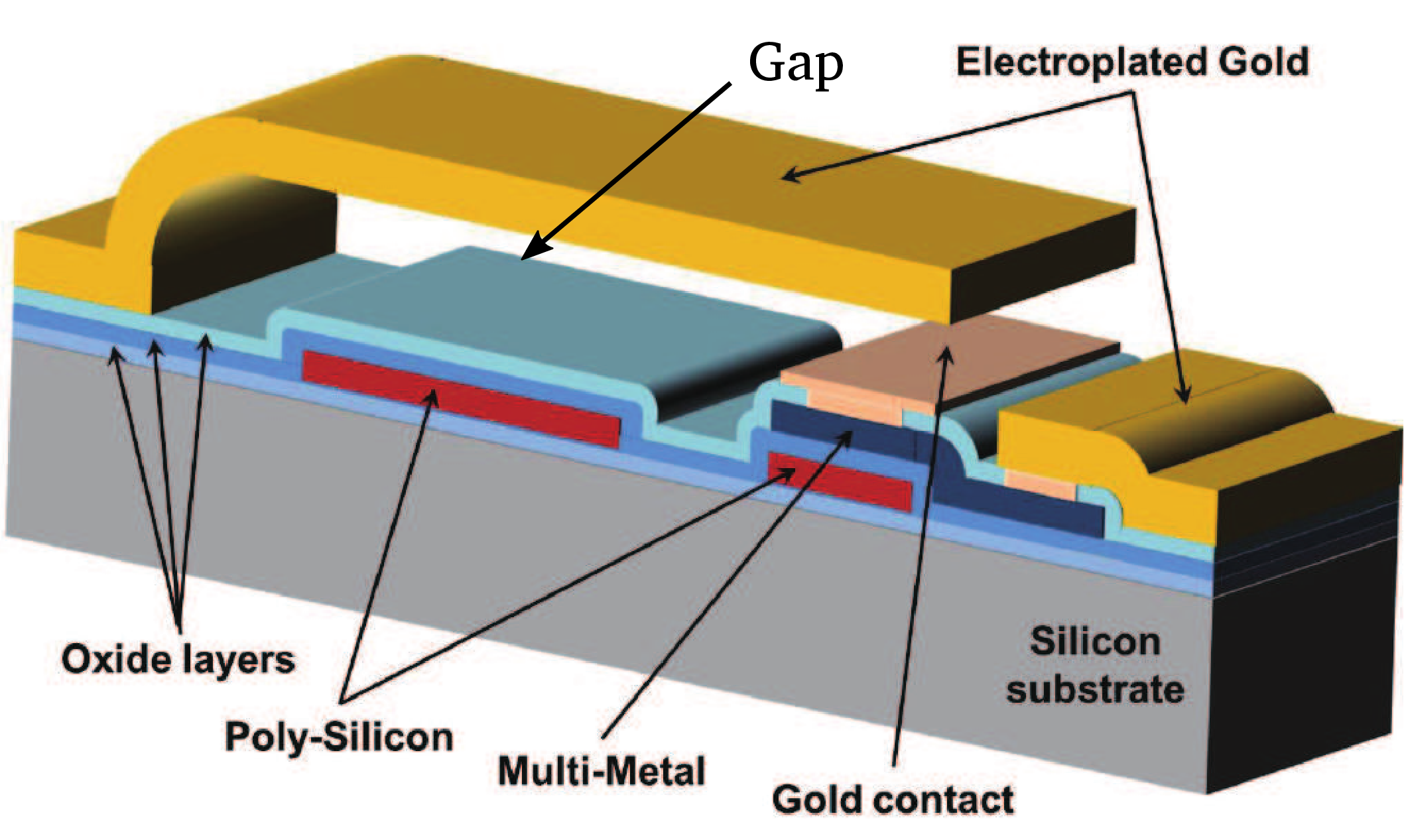}
		\label{rf_mems2}
	}	
	\caption{\label{rf_mems} \color{blue}(a) Microphotograph of a cantilever-type RF-MEMS series ohmic switch, and (b) 3D cross-section of a cantilever-type RF-MEMS switch (these two figures are presented by Iannacci et al. in Ref.~\cite{Iannacci2013RF}).}
\end{figure}

In recent years, with the rapid development of fabricating technologies, a variety of micro-electro-mechanical-systems (MEMS) devices have more applications in everyday life~\cite{Senturia1997Simulating}. Among these devices, a moving micro-structure is usually involved; for example, RF switches, high-$g$ acceleration sensor, and atomic force microscopy~\cite{Rebeiz2004RFMEMS,Lee2002Nonlinear}. Shown in Fig.~\ref{rf_mems}, for a RF switch, a micro-cantilever will oscillate perpendicularly above the substrate; the height of gap between the substrate and the micro-cantilever usually is about a few micrometers. With the motion of micro-structure, the gas will be pulled in or pushed out from the gap. Usually, as the variation of pressure in the gap is more dramatic than that at other areas of the device, the damping force or torque acting on the micro-structure  will be generated. Furthermore, due to the small size of structure and the low pressure of encapsulation, these forces or torques caused by the gas will increase significantly with a large surface-to-volume ratio of structure. Generally speaking, this type of damping is called the squeeze-film damping (SFD) and is the most important source of damping~\cite{Bao2007squeeze}.

To study the SFD problem, the traditional numerical methods usually are based on the Reynolds equation~\cite{Veijola2004Compact}, which is simplified from the Navier-Stokes equations. Following the definition of the Knudsen ($Kn$) number~\cite{Tsien1946superaerodynamics}, in theory, these methods are only available for flows in continuum and near-continuum regimes. Also due to the small size of micro-structure and the low pressure of encapsulation, the gas in MEMS devices usually is the rarefied gas. As the response of micro-structure will be influenced largely by rarefied gas~\cite{Guo2009Compact}, the influence of the rarefaction effect must be considered~\cite{Beskok2005microflows}. To consider the rarefied gas effect, one way is to improve the Reynolds equation with some techniques. For example, Veijola et al.~\cite{Veijola1995Equivalent} introduced an efficient viscosity based on the $Kn$ number to replace the real physical viscosity. Gallis et al.~\cite{Gallis2004improved} modified the coefficients used in the wall boundary condition of the Reynolds equation based on the Navier--Stokes slip--jump (NSSJ) simulations for flow at a small $Kn$ number and the direct simulation Monte Carlo (DSMC) molecular gas dynamics simulations for flow at a large $Kn$ number. Although these methods have lots of applications~\cite{Pandey2007effect, Lee2009Squeeze} and have satisfactory performance in some benchmark test cases, it can not make a conclusion that the methods based on the Reynolds equation can accurately describe the real physical flow at a high $Kn$ number~\cite{Guo2009Compact}. In addition, as several assumptions are introduced to derive the Reynolds equation, and very simple micro-structures are considered in these methods, further studies are needed to validate whether these methods can simulate the rarefied gas flows around the complex micro-structures or not. Another way to study the SFD problem is the molecular dynamics (MD) simulations~\cite{Li2010molecular}, but these methods are usually used in the free molecular regime. As a consequence, the computational cost for this type of method is higher when the $Kn$ number of the flow is moderate. Besides, the DSMC method also has been used to investigate the moving-boundary problem in MEMS devices~\cite{Rader2010DSMC}, but it still faces great challenges due to the ultra-low Mach number of the flow. The major disadvantages of the DSMC are slow convergence, large statistical noise, long time to reach steady state, and extensive number of molecules~\cite{Beskok2005microflows}. Usually, the typical speed of a micro-flow is about $1mm/s$ to $1m/s$, and the averaging molecular velocities are about $500m/s$, the five to two orders of magnitude difference between those two velocities results in large statistical noises, and requires $10^6$-$10^8$ samples to decrease the statistical noise~\cite{Bahukudumbi2003unified}. For example, Diab et al.~\cite{Diab2014Model} studied the SFD problem with the velocity of a moving micro-beam between $20m/s$ and $800m/s$, which is a much higher velocity for micro-flow. Although some improved DSMC algorithms~\cite{Fan2001statistical} have been developed, the restrictions on the time step and mesh size, and the statistical scatters of density and temperature are not relieved by most of those methods~\cite{Fei2013diffusive,Yao2011method}. In addition, for unsteady DSMC, the ensemble average at each time step replaces the time average used in a steady flow~\cite{Beskok2005microflows}, which put forward a higher requirement for parallel computation of complex flows. For example, when studying the oscillating Couette flow, over 5,000 realizations are implemented to ensemble average this stochastic process at every time step~\cite{Bahukudumbi2003unified}. Consequently, the DSMC method will not has a satisfactory performance for solving these problems, as the flow around a moving boundary is an inherently low-speed unsteady flow.

Recently, the discrete unified gas kinetic scheme (DUGKS) proposed by Guo et al.~\cite{Guo2013Discrete} has become a promising new method to simulate the flows in MEMS devices. Due to its kinetic nature, the DUGKS has many advantages over other numerical methods, and has ability to solve the flow problems in all flow regimes. For example, in the continuum flow regime, as it can adopt a larger Courant--Friedrichs--Lewy (CFL) number, and is less sensitive to mesh resolutions, so the DUGKS has a good performance than the traditional finite volume lattice Boltzmann method (FVLBM)~\cite{Zhu2017Performance, Wang1, Wang2}. And in the rarefied flow regime, the computational cost is declined compared to the unified gas kinetic scheme~\cite{Xu2010unified} (UGKS). Furthermore, as the DUGKS is an unsteady flow simulation method in nature, the ensemble averages are not required compared with the DSMC method. Currently, some improved and enhanced schemes based on the DUGKS also have been developed~\cite{Chen2019Conserved, Zhong2020simplified, Zhong2021simplified}. And, Wang et al.~\cite{Wang2019Arbitrary} proposed an arbitrary Lagrangian-Eulerian-type DUGKS for solving the moving boundary problems in continuum and rarefied gas flows. So, the original DUGKS and ALE-DUGKS are the promising numerical schemes for studying the micro-flows and the SFD phenomena in MEMS. In addition, several recently proposed schemes~\cite{Zhu2017Unified, Zhu2019Unified, Su2020Can, Yuan202novel, Yang2021direct, Liu2015unified, Wang2019generalized, Zhang2020Competition} also have advantages or can be further applied to study the rarefied gas flow and SFD problem in MEMS, which can be followed in the future.

For the SFD phenomenon in MEMS, in essence, it is a fluid-structure interaction (FSI) problem due to the motion of micro-structure. To investigate this problem, a decoupled method is usually used. Firstly, with an Eulerian-framework scheme (based on a stationary mesh), by imposing different velocities on a stationary micro-structure, the micro-flows are generated, then the squeeze-film damping forces can be calculated. And with these results, a damping force or torque coefficient can be obtained~\cite{Guo2009Compact}. Next, the structure dynamics equation is solved and the gas damping force or torque is considered as an equivalent internal structure damping~\cite{Lee2009Squeeze}. In the above procedure, an assumption that the squeeze-film damping force or torque is linear with moving velocity or angular velocity is used. Usually, this damping is referred to as a viscous damping~\cite{Geradin2015Mechanical}. As illustrated in Ref.~\cite{Chigullapalli2012Non}, when a micro-structure is moved at a high velocity, the downward and upward motions will generate different values of damping force. Therefore, the nonlinear phenomenon between the moving velocity and damping force is observed. Due to the assumption of viscous damping is not correct in some conditions, more accurate result will be obtained if the gas damping force or torque is treated as an external one. Consequently, introducing a new coupled framework that can simulate the moving micro-structure influenced by the squeeze-film damping for all flow regimes will has a great value in engineering applications. According to the above reasons, the main objective of this paper is to study the nonlinear SFD phenomenon with a coupled framework based on the ALE-DUGKS scheme. In the past several decades, a variety of numerical methods in the field of FSI have been proposed~\cite{Hou2012Numerical}. And in these methods, a loosely coupled method is usually adopted, that is the fluid and structure dynamic solvers are used alternately, and the data of force or torque are exchanged between those solvers in each iteration step. As this type of method is easy to implement, it will be adopted in this paper. To the best of the authors' knowledge, in corresponding studies, this is the first attempt to introduce a coupled FSI framework into the DUGKS or discrete velocity method (DVM) for solving the SFD phenomenon. Moreover, as the linear (perpendicular) and tilting motions of a rigid micro-beam are the basic two-dimensional motion forms~\cite{Hartono2007Squeeze} in MEMS, the FSI problem for these motions will be comprehensively studied in this work.

The rest of the paper is organized as follows. In Sec.~\ref{algorithm}, the ALE-DUGKS solution procedure, and the traditional decoupled method and the proposed new coupled framework for solving SFD problem are introduced. In Sec.~\ref{validation}, two test cases, the micro-Couette flow in rarefied gas and the free oscillation of a square cylinder in continuum flow, are conducted to validate the methods. In Sec.~\ref{cases}, the SFD coefficient is calculated based on the traditional decoupled method, and the nonlinear squeeze-film damping force and torque are studied based on the new coupled framework. Both the forced and free oscillations with the linear and tilting motions are considered. Finally, a brief conclusion is presented in Sec.~\ref{conclusion}.

\section{Numerical method}\label{algorithm}
In this section, to study the SFD phenomenon which involves the rarefied gas effect, the ALE-DUGKS is firstly introduced briefly, then the traditional decoupled method and the coupled framework for calculating the squeeze-film damping are presented.

\subsection{Boltzmann-BGK equation}
The original DUGKS proposed by Guo et al.~\cite{Guo2013Discrete} and the ALE-DUGKS proposed by Wang et al.~\cite{Wang2019Arbitrary} are the numerical schemes based on the Boltzmann model equation. In this work, the Boltzmann-BGK equation is used, which can be expressed as
\begin{equation}\label{Maxwell}
\frac{\partial{f}}{\partial{t}} + {\bm \xi} \cdot {\nabla{f}} = \Omega = -\frac{1}{\tau}[f-f^{eq}],
\end{equation}
where $f=f(\bm{x},\bm{\xi}, t)$ is the velocity distribution function for particles moving with velocity $\bm{\xi}$ at position $\bm{x}$ and time $t$, $\tau$ is the relaxation time depending on the fluid dynamic viscosity $\mu$ and pressure $p$ with $\tau={\mu}/p$. And, $f^{eq}$ is the Maxwell equilibrium function; for the two-dimensional flow, it is given by
\begin{equation}
f^{eq}=\frac{\rho}{2\pi{R}T}exp(-\frac{|\bm{\xi}-\bm{u}|^2}{2RT}),
\end{equation}
where $R$ is the gas constant, $\rho$ is the fluid density, $\bm{u}$ is the fluid velocity, and $T$ is the fluid temperature. Finally, the macro-physical quantities can be calculated as
\begin{equation}\label{dugks_marco}
\rho=\int{f}d\bm{\xi}, \rho\bm{u}=\int{\bm{\xi}f}d\bm{\xi}, p={\rho}RT,
\end{equation}
where the ideal gas law is used for the calculation of pressure.

\subsection{Arbitrary Lagrangian-Eulerian-type discrete unified gas kinetic scheme}
For the simulation of moving boundary problem, the original Boltzmann-BGK equation (Eq.~\eqref{Maxwell}) is extended to the ALE framework; then Eq.~\eqref{Maxwell} can be rewritten as~\cite{Wang2019Arbitrary,Chen2012a}
\begin{equation}\label{Maxwell_moving}
\frac{\partial{f}}{\partial{t}}+(\bm{\xi}-\bm{v})\cdot\nabla{f}=\Omega=-\frac{1}{\tau}[f-f^{eq}].
\end{equation}
where $\bm{v}$ is the mesh motion velocity. As explained in Ref.~\cite{Chen2012a}, the update rule of the equilibrium distribution function $f^{eq}$ does not depend on the mesh motion velocity $\bm{v}$, so only the calculation of the convection term is influenced in the ALE-DUGKS.

For the discretization of Eq.~\eqref{Maxwell_moving} in the particle velocity--space, a finite set of discretized particle-velocities is used; and $\bm{\xi}_i$ represents the $i$-th discretized velocity. As shown in Eq.~\eqref{dugks_marco}, to integrate the macro-quantities, the micro-velocities of particle can be set to coincide with the abscissas of the quadrature rule. In this study, for a low-speed continuum flow, the D2Q9 discretized velocity model, weights and corresponding equilibrium function developed in the LBM~\cite{qian1992lattice} are used. And for the rarefied gas flows in MEMS, the Gauss-Hermit and Newton-Cotes quadrature rules are used.

\begin{figure}
	\centering
	\subfigure[]{
		\includegraphics[width=0.32 \textwidth]{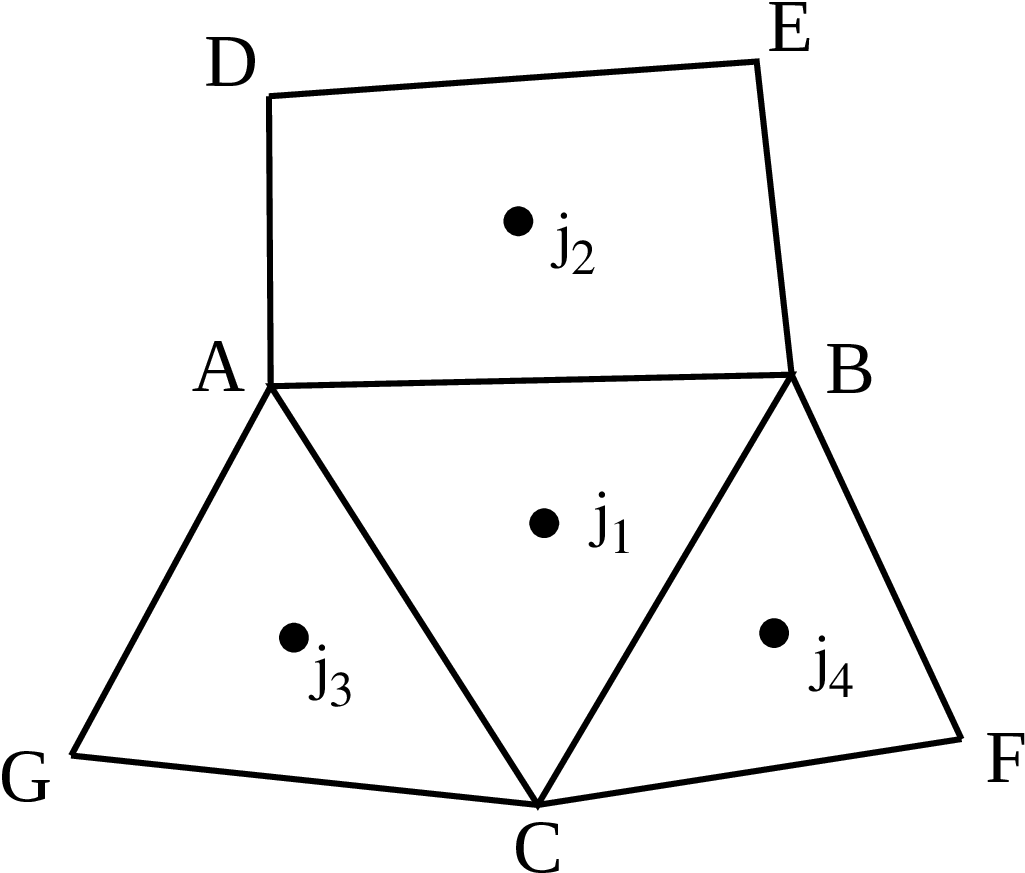}
		\label{structure1}
	}
	\subfigure[]{
		\includegraphics[width=0.32 \textwidth]{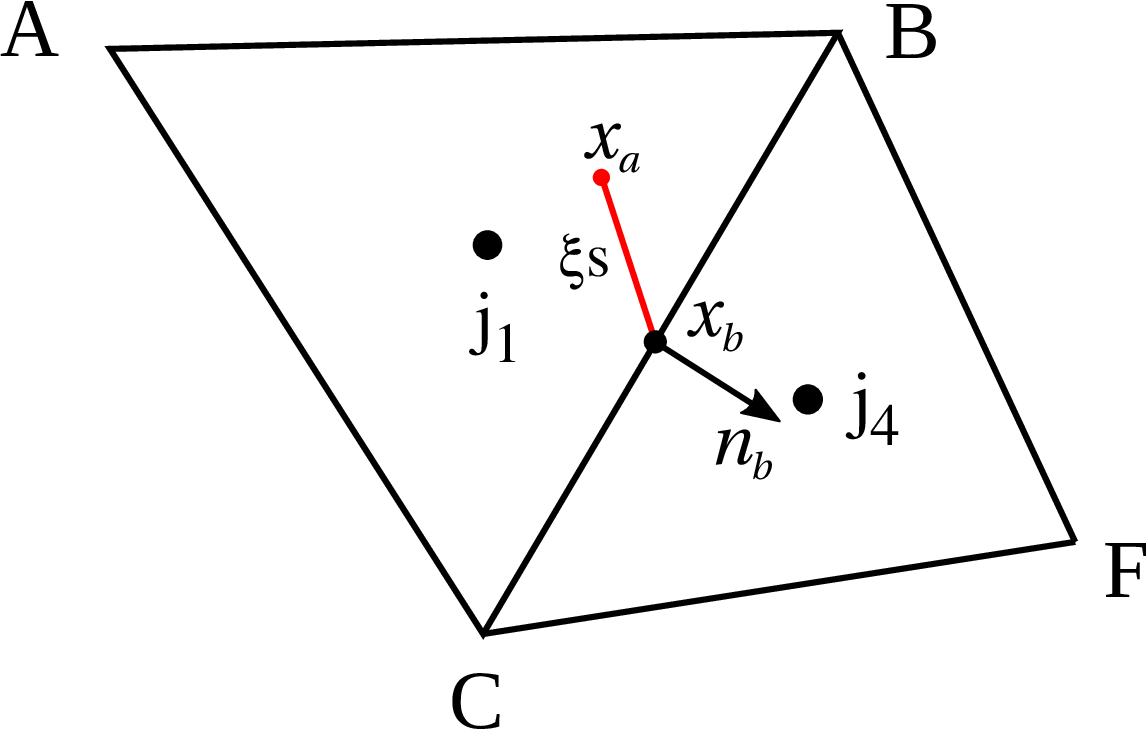}
		\label{structure2}
	}	
	\caption{\label{unstructure-mesh} Sketches of (a) an unstructured mesh used in the ALE-DUGKS and (b) micro-flux calculation for a cell interface.}
\end{figure}

For the discretization of Eq.~\eqref{Maxwell_moving} in the macro physical-space, an unstructured mesh finite volume scheme is used. Fig.~\ref{structure1} shows the sketch of an unstructured mesh, where $j$ is the center of triangular cell $ABC$ and subscript represents the
index number of a cell. If the mid-point rule is used for the integration of the convection term, and the trapezoidal rule is used for the calculation of the collision term, Eq.~\eqref{Maxwell_moving} can be discretized as
\begin{equation}\label{semi_equ_ale}
f_j^{n+1}(\bm{\xi})\left|V_j^{n+1,*}\right|-f_j^n(\bm{\xi})\left|V_j^{n,*}\right|+\Delta{t}\bm{F}_{ALE}^{n+1/2}(\bm{\xi})=\frac{\Delta{t}}{2}[\Omega_j^{n+1}(\bm{\xi})\left|V_j^{n+1,*}\right|+\Omega_j^n(\bm{\xi})\left|V_j^{n,*}\right|],
\end{equation}
and the micro-flux of a cell surface $\bm{F}_{ALE}^{n+1/2}(\bm{\xi})$ is given as
\begin{equation}\label{flux_ale}
\bm{F}_{ALE}^{n+1/2}(\bm{\xi})=\int_{\partial{V_j}}(\bm{\xi}-\bm{v})\cdot{\bm{n}}f(\bm{x},\bm{\xi},t_{n+1/2})dS=\sum_{k}(\bm{\xi}-\bm{v}_{b,k}^{n+1/2})\cdot\bm{n}_{b,k}^*f^{n+1/2}(\bm{x}_{b,k},\bm{\xi})S_k^*,
\end{equation}
where $n$ is the time level, $\Delta{t}=t^{n+1}-t^n$ is the time step, $\bm{x}_b$ is the center of cell interface, and $k$ is the total number of cell interfaces. Under the ALE framework, as the geometrical information of a grid cell changes temporally during the simulation, the cell volumes $V^{n+1,*}$ and $V^{n,*}$ at $n$ and $n+1$ time levels, the moving velocity of cell interface $\bm{v}_b^{n+1/2}$ at $n+1/2$ time, the outward unit normal vector $\bm{n}_b^*$, and the area of cell interface $S_b^*$ must be calculated by the discretized geometric conservation law (DGCL)~\cite{Wang2019Arbitrary,thomas1978gcl}, where superscript $^*$ means that the values of variables at corresponding times maybe not equal to the real values of variables at those times. In this study, the DGCL scheme-2 presented in Ref.~\cite{Wang2019Arbitrary} is used, where
\begin{equation}
    \bm{v}_b^{n+1/2}=\frac{\bm{x}_b^{n+1}-\bm{x}_b^n}{\Delta{t}},
\end{equation}
and
\begin{equation}\label{dgcl2_s}
    V_j^{n,*}=V_j^n, \bm{S}_b^*=\bm{S}_b^n, V_j^{n+1,*}=V_j^n+\Delta{t}\sum_k{\bm{v}_{b,k}^{n+1/2}\bm{S}_{b,k}^{n}}.
\end{equation}

To remove the implicit collision term, two new distribution functions are introduced:
\begin{equation}\label{tilde_phi}
    \tilde{f}=f-\frac{\Delta{t}}{2}\Omega=\frac{2\tau+\Delta{t}}{2\tau}f-\frac{\Delta{t}}{2\tau}f^{eq},
\end{equation}
\begin{equation}\label{tilde_phi_plus}
    \tilde{f}^+=f+\frac{\Delta{t}}{2}\Omega=\frac{2\tau-\Delta{t}}{2\tau+\Delta{t}}\tilde{f}+\frac{2\Delta{t}}{2\tau+\Delta{t}}f^{eq},
\end{equation}
then Eq.~\eqref{semi_equ_ale} can be rewritten as
\begin{equation}\label{equ_phi_ale}
    \tilde{f}_j^{n+1}=\left|\frac{V_j^{n,*}}{V_j^{n+1,*}}\right|\tilde{f}_j^{+,n}-\frac{\Delta{t}}{\left|V_j^{n+1,*}\right|}\bm{F}_{ALE}^{n+1/2}(\bm{\xi}).
\end{equation}
And with the conservative property of the collision term:
\begin{equation}
	\int{\Omega}d\bm{\xi}=0, \int{\bm{\xi}\Omega}d\bm{\xi}=0,
\end{equation}
the calculation of macro-quantities in Eq.~\eqref{dugks_marco} can be replaced by
\begin{equation}\label{macro_phi}
    \rho=\int{\tilde{f}}d\bm{\xi}, \rho\bm{u}=\int{\bm{\xi}\tilde{f}}d\bm{\xi},
\end{equation}
and $\tilde{f}$ will be solved instead of $f$ in the practical computation.

For the calculation of the interface micro-flux shown in Fig.~\ref{structure2}, if integrate Eq.~\eqref{Maxwell} along the characteristic line within a
half time step $s=\Delta{t}/2$, the original distribution function $f(\bm{x}_b,\bm{\xi},t_n+s)$ in Eq.~\eqref{flux_ale} can be updated as
\begin{equation}\label{equ_phi_plus}
    f(\bm{x}_b,\bm{\xi},t_n+s)-f(\bm{x}_b-\bm{\xi}s,\bm{\xi},t_n)=\frac{s}{2}[\Omega(\bm{x}_b,\bm{\xi},t_n+s)+\Omega(\bm{x}_b-\bm{\xi}s,\bm{\xi},t_n)].
\end{equation}
Here, two additional distribution functions are introduced:
\begin{equation}\label{bar_phi}
    \bar{f}=f-\frac{s}{2}\Omega=\frac{2\tau+s}{2\tau}f-\frac{s}{2\tau}f^{eq},
\end{equation}
\begin{equation}\label{bar_phi_plus}
    \bar{f}^+=f+\frac{s}{2}\Omega=\frac{2\tau-s}{2\tau+s}\bar{f}+\frac{2s}{2\tau+s}f^{eq}.
\end{equation}
Then, Eq.~\eqref{equ_phi_plus} can be rewritten as
\begin{equation}\label{bar_phi_stream}
    \bar{f}(\bm{x}_b,\bm{\xi},t_{n}+s)=\bar{f}^+(\bm{x}_b-\bm{\xi}s,\bm{\xi},t_n),
\end{equation}
and the original distribution function at $(\bm{x}_b,t_{n+1/2})$ is given by
\begin{equation}
    f(\bm{x}_b,\bm{\xi},t_{n}+s)=\frac{2\tau}{2\tau+s}\bar{f}(\bm{x}_b,\bm{\xi},t_n+s)+\frac{s}{2\tau+s}f^{eq}(\bm{x}_b,\bm{\xi},t_n+s),
\end{equation}
where the macro-quantities at a cell interface can be calculated if $\tilde{f}$ in Eq.~\eqref{macro_phi} is replaced by $\bar{f}$. Besides, the time step $\Delta{t}$ used in this paper is given by
\begin{equation}\label{CFL}
	\Delta{t}=\alpha\frac{\Delta{x}}{|\bm{\xi}_{max}|},
\end{equation}
where $0<\alpha<1$ is the $CFL$ number, and $\Delta{x}$ is the minimum size of grid cells. Finally, two relations are used in the practical computation:
\begin{equation}\label{bar_phi_plus_a}
    \bar{f}^+=\frac{2\tau-s}{2\tau+\Delta{t}}\tilde{f}+\frac{3s}{2\tau+\Delta{t}}f^{eq}, \tilde{f}^+=\frac{4}{3}\bar{f}^+-\frac{1}{3}\tilde{f},
\end{equation}
and the Taylor expansion and the least-squares method are used to reconstruct the $\bar{f}^+$ at location $\bm{x}_b-\bm{\xi}s$. For the details of the implementation of the ALE-DUGKS, it can be found in Ref.~\cite{Wang2019Arbitrary}.

\subsection{Decoupled method and coupled framework for calculating the squeeze-film damping force and torque}
\begin{figure}
	\centering
	\subfigure[]{
		\includegraphics[width=0.45 \textwidth]{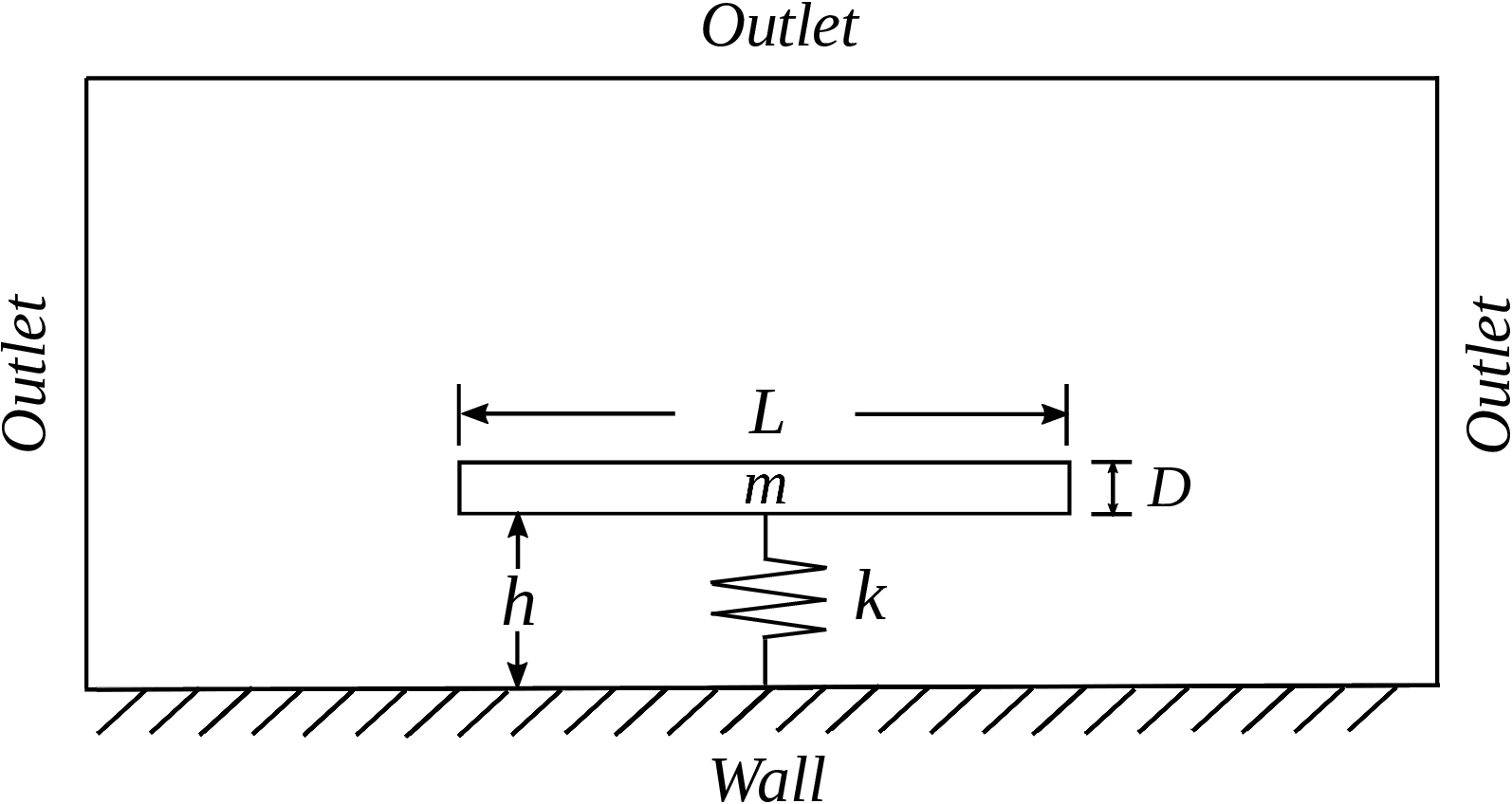}
	}
	\subfigure[]{
		\includegraphics[width=0.45 \textwidth]{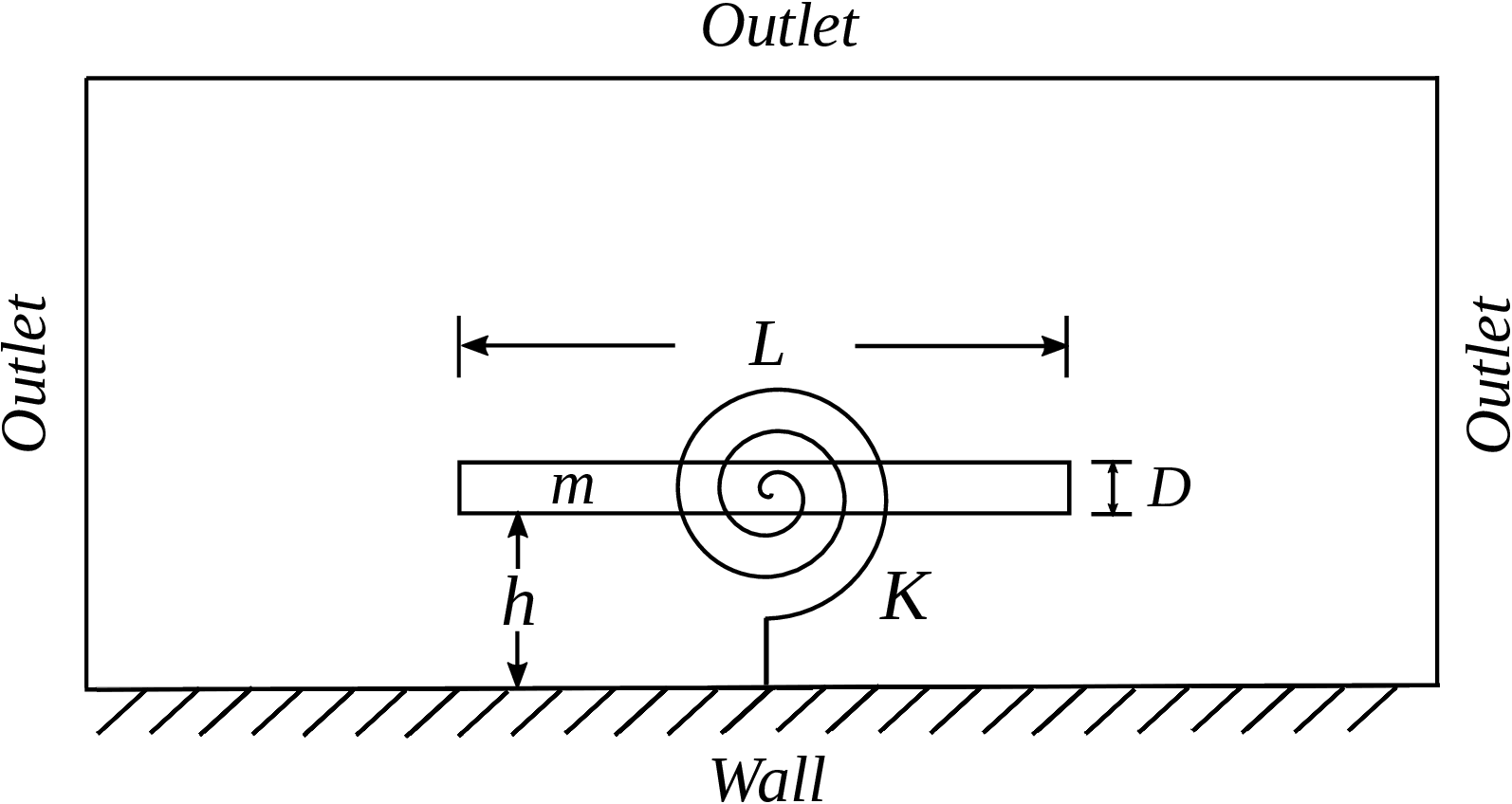}
	}
	\caption{\label{beam-config} Configuration for an elastically mounted micro-beam oscillating in rarefied gas (not drawn to scale) with (a) linear (perpendicular) motion and (b) tilting motion.}
\end{figure}

Shown in Fig.~\ref{rf_mems2}, for the oscillation of a micro-cantilever, a two-dimensional flow simulation can be adopted with the micro-flow far away from the anchor. Fig.~\ref{beam-config} shows two states of a micro-beam: linear (perpendicular) motion and tilting motion; $L$ and $D$ are the width and thickness of a micro-beam, respectively, and $h$ is the height of gap. For linear motion, the structure dynamic equation is given by
\begin{equation}\label{linear-struct}
 m\ddot{y}(t)+c\dot{y}(t)+ky=F_{ext},
\end{equation}
where $y$ is the perpendicular displacement of structure, $m$ is the mass of structure, $c$ is the damping coefficient of structure, $k$ is the stiffness coefficient of structure, and $F_{ext}$ is the external excitation force. And for tilting motion, the corresponding equation is given by~\cite{Pan1999Squeeze}
\begin{equation}\label{tilt-struct}
 I\ddot{\theta}(t)+\eta\dot{\theta}(t)+K\theta=T_{ext},
\end{equation}
where $\theta$ is the rotation angle of structure, $I$ is the polar moment of inertia, $\eta$ is the torsional damping coefficient, $K$ is the torsional stiffness coefficient, and $T_{ext}$ is the external excitation torque. To predict the response of micro-beam with the given external force or torque, both the structure intrinsic damping and the gas squeeze-film damping must be determined. In this paper, only the gas SFD is considered, and the structure intrinsic damping is ignored to encourage a larger amplitude oscillation.

\subsubsection{Decoupled method}
\begin{figure}
	\centering
	\subfigure[]{
		\label{plate-linear-decoupled}
		\includegraphics[width=0.45 \textwidth]{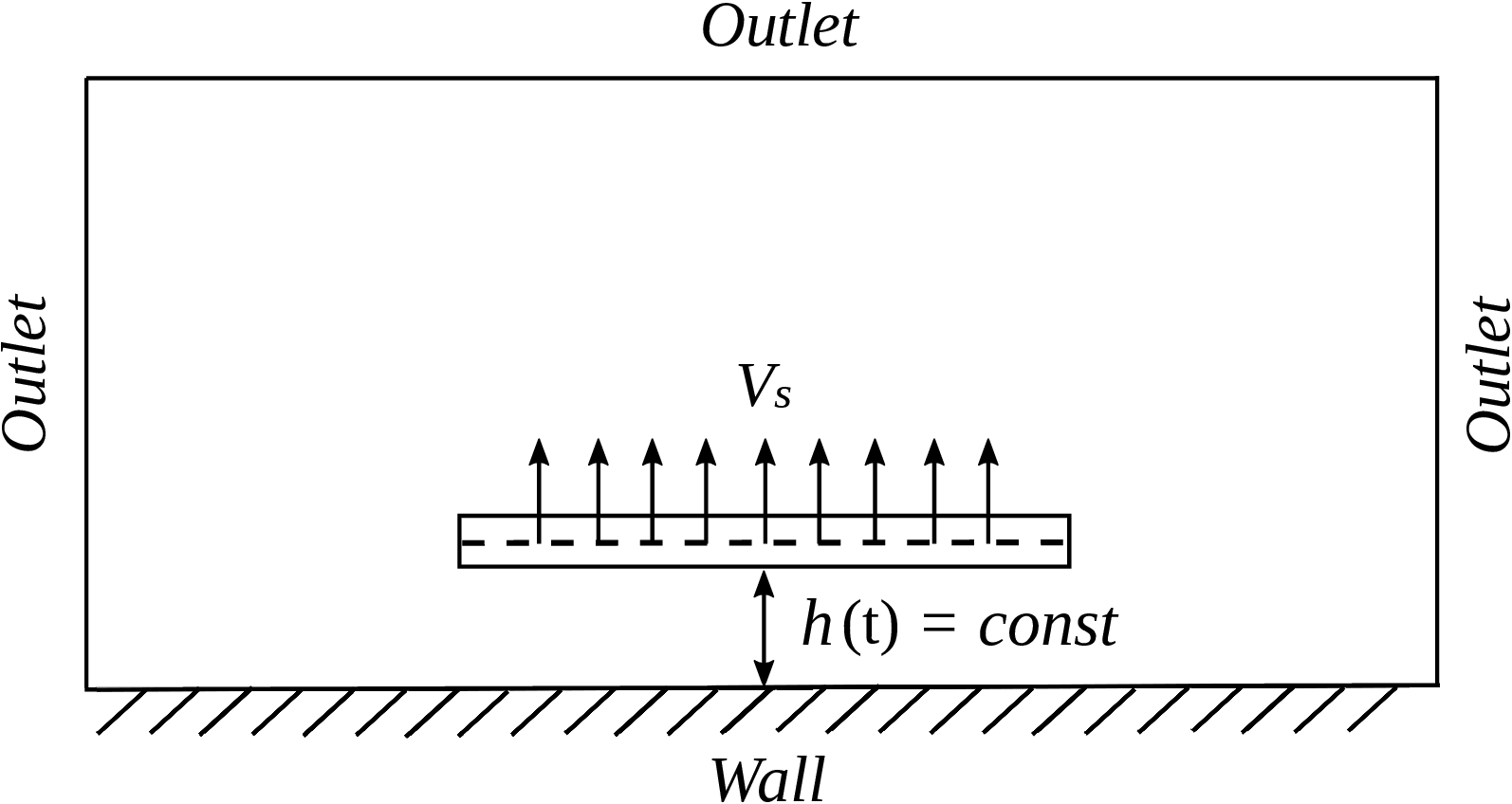}
	}
	\subfigure[]{
		\label{plate-tilt-decoupled}
		\includegraphics[width=0.45 \textwidth]{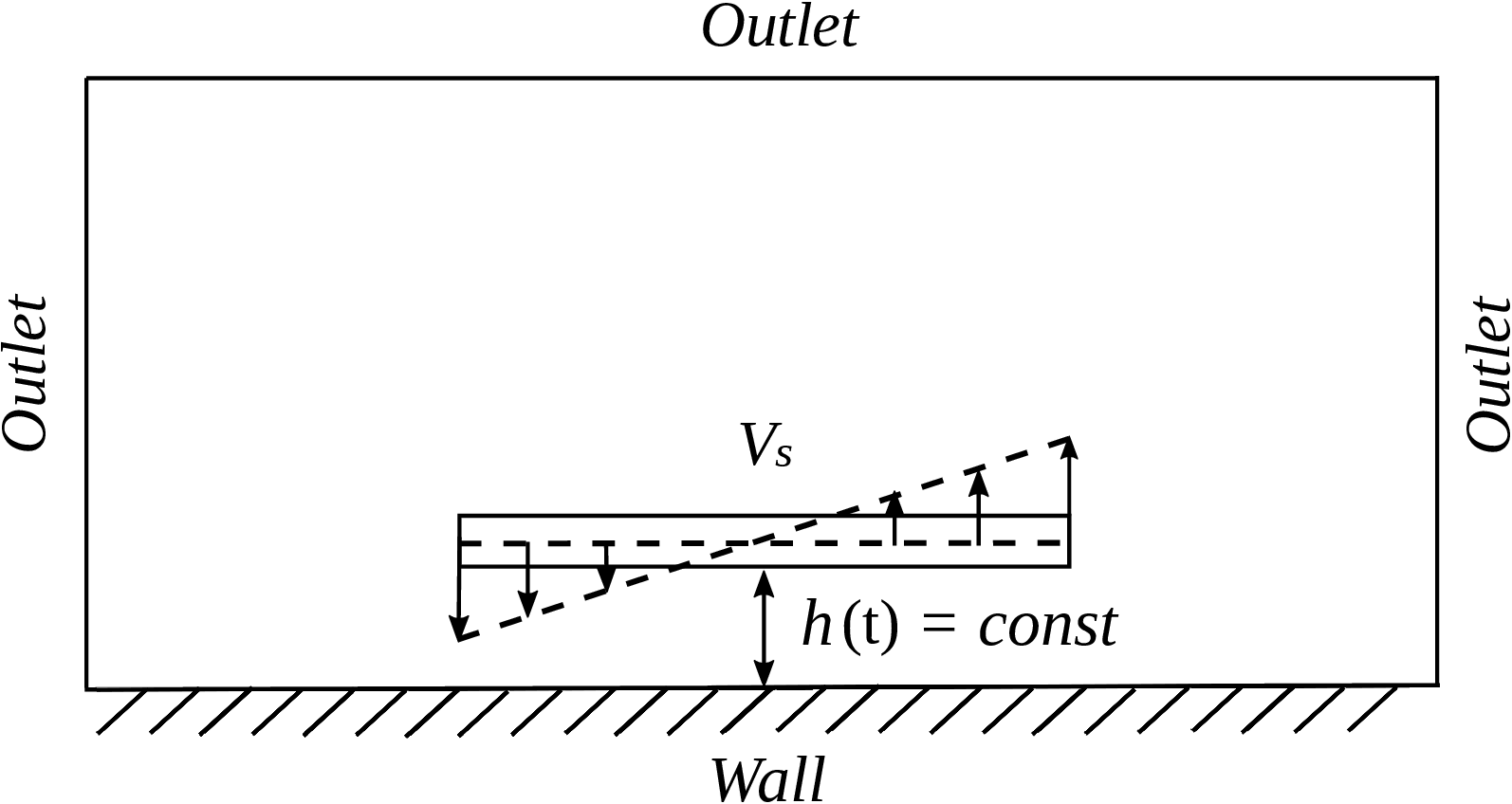}
	}
	\caption{\label{plate-decoupled} \color{blue} Sketches of decoupled method for solving squeeze-film damping at (a) linear (perpendicular) motion and (b) tilting motion.}
\end{figure}

Similar to the traditional method, the DUGKS also can be adopted to calculate the SFD coefficient ($c$ and $\eta$ in Eqs.~\eqref{linear-struct} and \eqref{tilt-struct}). Fig.~\ref{plate-decoupled} shows the sketches of decoupled method based on an Eulerian-framework scheme, where the height of gap $h$ is constant, and the profile of velocity on the surface of a stationary beam is given by the physical condition. Then, Eq.~\eqref{Maxwell} or Eq.~\eqref{Maxwell_moving} with $\bm{v}=0$ can be used to describe the micro-flow. By calculating the force and torque acting on the micro-beam, damping coefficients, $c_f$ and $c_{\eta}$, are given as
\begin{equation}\label{cf}
  c_f=\frac{F}{V_sL}, c_{\eta}=\frac{T}{L},
\end{equation}
respectively, where $F$ is the rarefied gas damping force, and $T$ is the damping torque. As discussed in Sec.~\ref{introduction}, for the decoupled method, when the damping coefficient is determined, it will be treated as the structure intrinsic damping; then, Eq.~\eqref{linear-struct} or \eqref{tilt-struct} will be solved to predict the response the micro-beam. As the assumption of viscous damping~\cite{Geradin2015Mechanical} is used, the damping force is always linear with the velocity, no matter what the value of motion velocity. Consequently, nonlinear SFD force~\cite{Chigullapalli2012Non} can not be predicted by the decoupled method.

\subsubsection{Coupled framework}
In this paper, based on a loosely-coupled FSI algorithm~\cite{Farhat2006Provably}, a new framework for solving SFD is used. As the damping force or torque are treated as an external one, Eqs.~\eqref{linear-struct} and \eqref{tilt-struct} are rewritten as
\begin{equation}\label{linear-struct-coulped}
m\ddot{y}(t)+ky=F_{ext}+F,
\end{equation}
and
\begin{equation}\label{tilt-struct-coupled}
I\ddot{\theta}(t)+K\theta=T_{ext}+T,
\end{equation}
respectively, where the structure intrinsic damping is ignored. For the discretization of structure dynamic equation, the implicit Newmark scheme~\cite{Newmark1959a} is introduced. And a second-order extrapolation scheme is used to predict the force or torque at $n$+1 time level:
\begin{equation}\label{force-predict}
   F^{n+1,*}=2F^{n}-F^{n-1}, T^{n+1,*}=2T^{n}-T^{n-1}.
\end{equation}
The advantage of above coupled framework is that the nonlinear damping force or torque can be calculated dynamically. Finally, the detailed implementation procedure of this FSI framework is as follows:
\begin{description}
  \item[(1)] predict the damping force $F$ or torque $T$ at new time level with Eq.~\eqref{force-predict};
  \item[(2)] update the structure displacement $y$ or rotation angle $\theta$ with the implicit Newmark scheme~\cite{Newmark1959a};
  \item[(3)] deform the mesh with a new structure location by the Laplace smoothing equation~\cite{Rainald1996Improved};
  \item[(4)] update the distribution function $\tilde{f}$ from $n$ to $n+1$ time level according to Eq.~\eqref{equ_phi_ale};
  \item[(5)] calculate the force or torque acting on a micro-structure based on the distribution function (Eq. (2.18) in Ref.~\cite{Xu2015Direct}).
\end{description}

From our numerical tests, one inner iteration of the above procedure in one time step is enough to obtain a convergent displacement of the structure ($|y^{n+1,*}-y^{n+1}|<10^{-6}$), so the inner iterative cycle used in the traditional FSI framework~\cite{Farhat2006Provably} is not required. The reason is that the present ALE-DUGKS is an explicit numerical scheme, and the coupled time step is set to $\Delta{t}_{CFD}=\Delta{t}_{CSD}$, where $\Delta{t}_{CFD}$ is time step used for computational fluid dynamics (CFD) simulation and $\Delta{t}_{CSD}$ is that used for computational structural dynamics (CSD) simulation, so the coupled numerical error is small for the implicit Newmark scheme at a small time step. The present coupled FSI framework has been coded with the help of Code\underline{ }Saturne~\cite{archambeau2004code}, an open-source computational fluid dynamics software of Electricite De France (EDF), France (\url{http://www.code-saturne.org/cms/}). We appreciate the development team of Code\underline{ }Saturne for their great works.

\section{\color{blue}Validation of the numerical framework}\label{validation}
In this section, to validate the decoupled method and coupled framework based on the DUGKS, two test cases, namely micro-Couette flow in rarefied gas and free oscillation of a square cylinder in continuum flow, are conducted.
\subsection{Micro-Couette flow in rarefied gas}
\begin{figure}
	\centering
	\subfigure[]{
		\includegraphics[width=0.45 \textwidth]{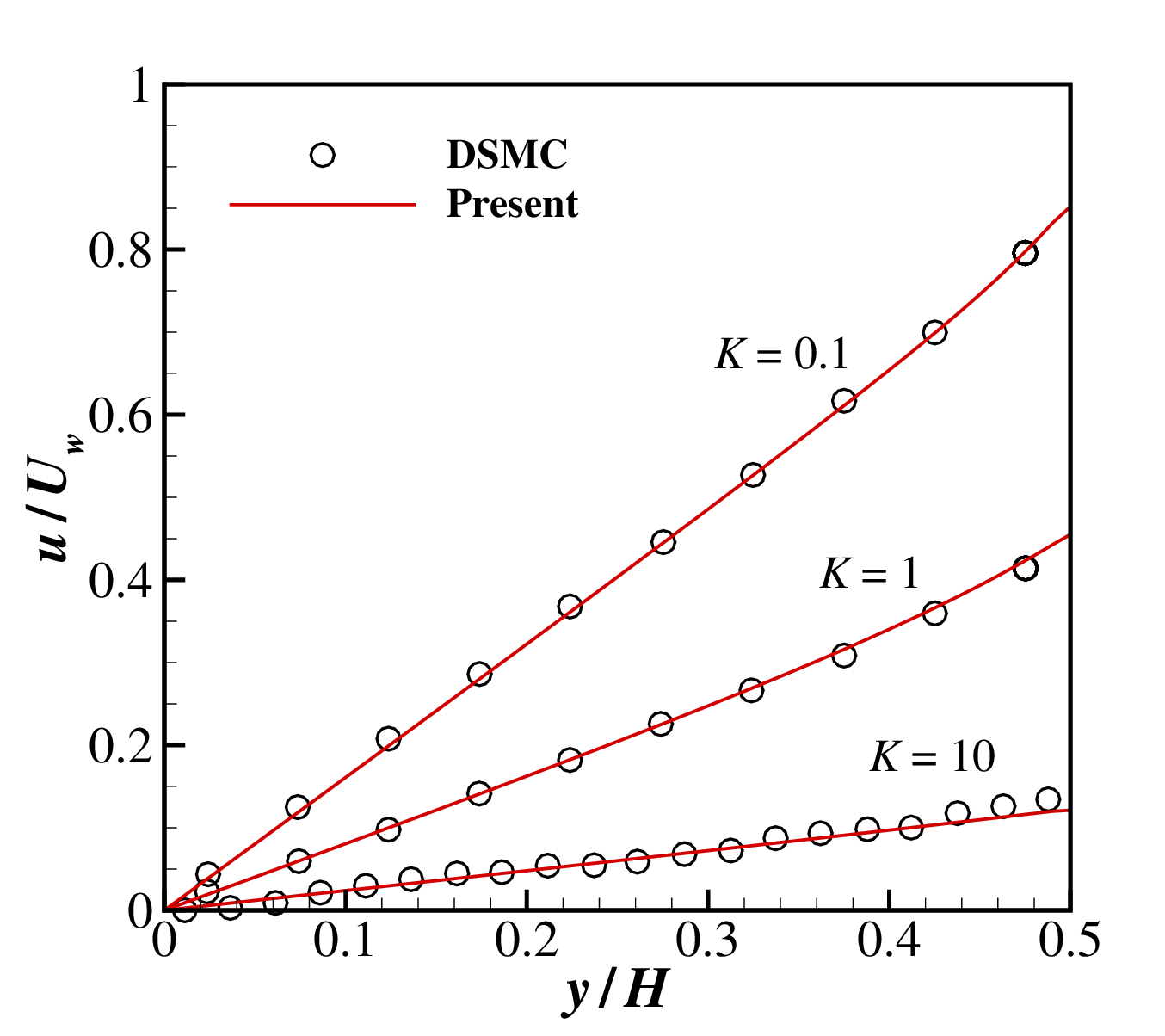}
	}
	\subfigure[]{
		\includegraphics[width=0.45 \textwidth]{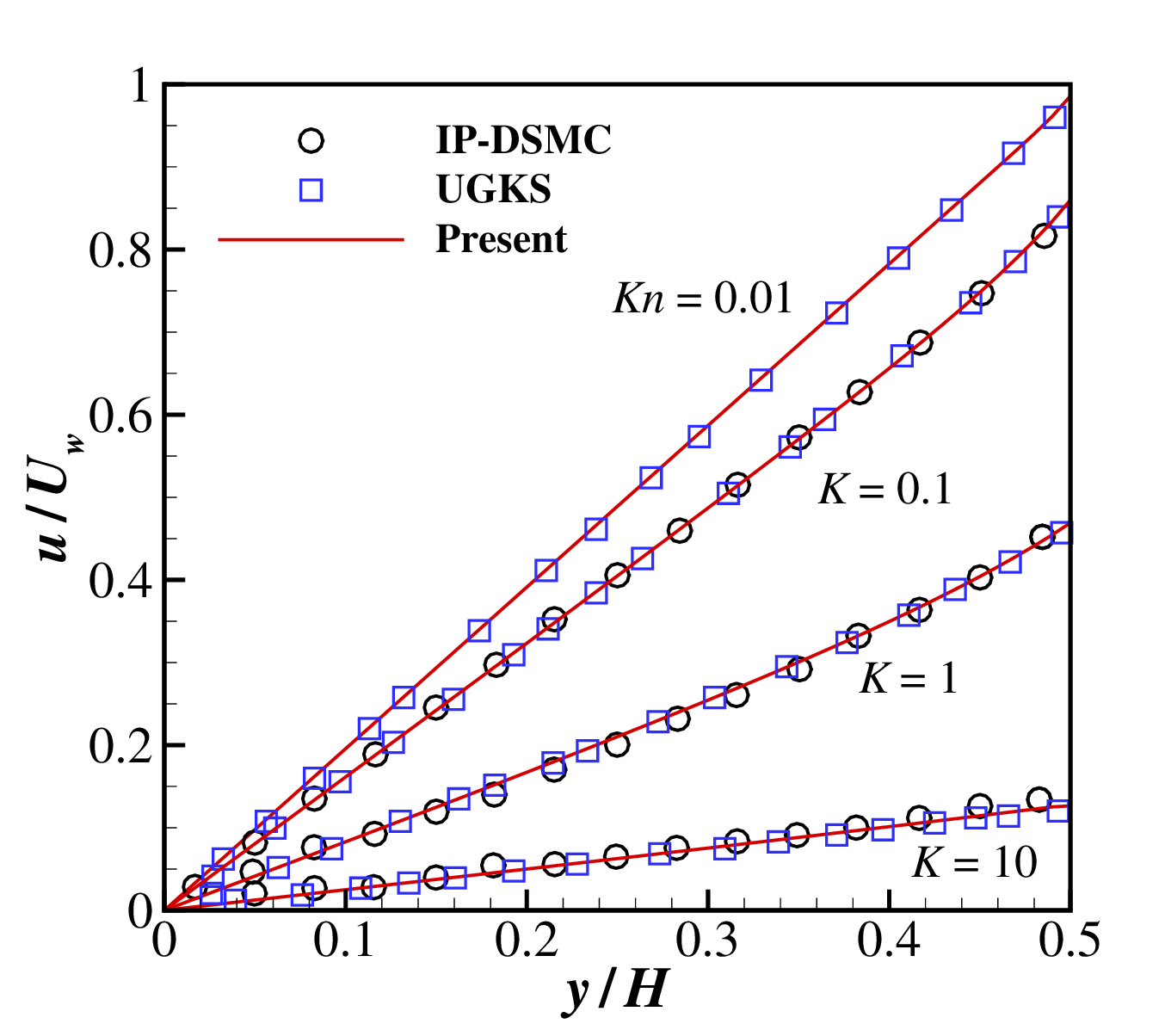}
	}
	\caption{\label{couette}\color{blue} $u$-velocity profile of micro-Couette flow at (a) $U_w=16.85m/s$ (DSMC:~\cite{Bahukudumbi2003unified}) and (b) $U_w=119.15m/s$ (IP-DSMC~\cite{Fan2001statistical}, UGKS~\cite{Zhu2016implicit}), where $Kn$ is set to $Kn=2K/{\sqrt{\pi}}$.}
\end{figure}

\begin{figure}
	\centering
	\subfigure[]{
		\includegraphics[width=0.15 \textwidth]{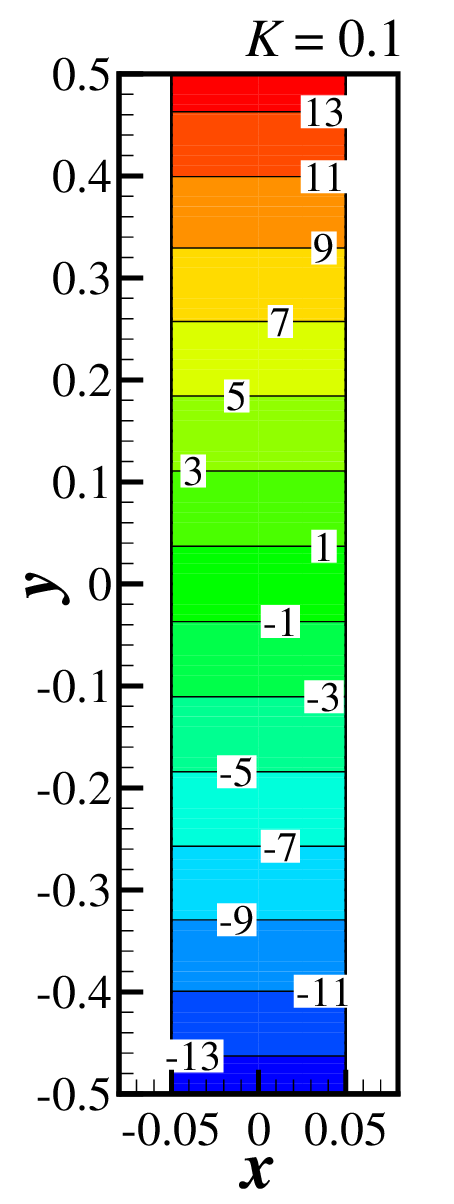}
		\includegraphics[width=0.15 \textwidth]{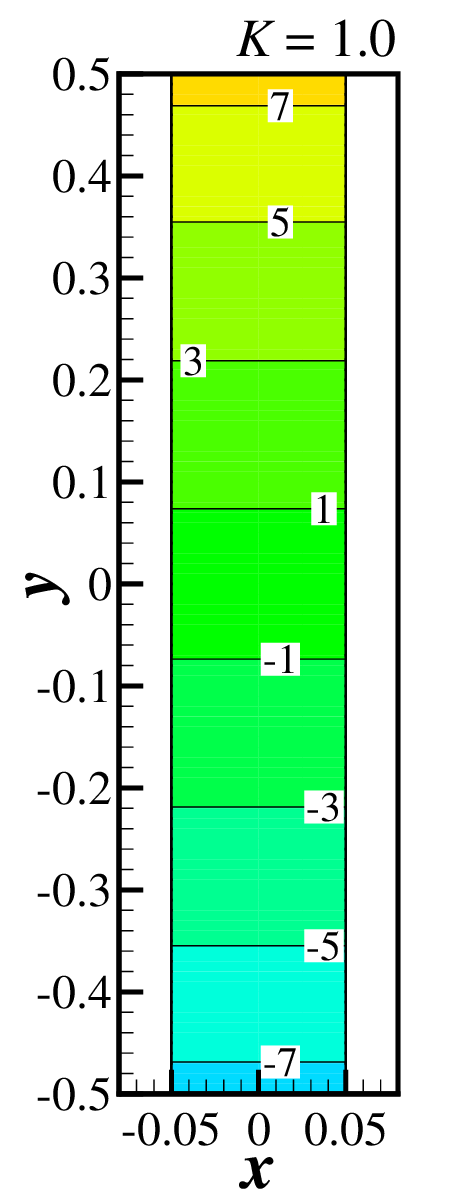}
	}
	\subfigure[]{
		\includegraphics[width=0.45 \textwidth]{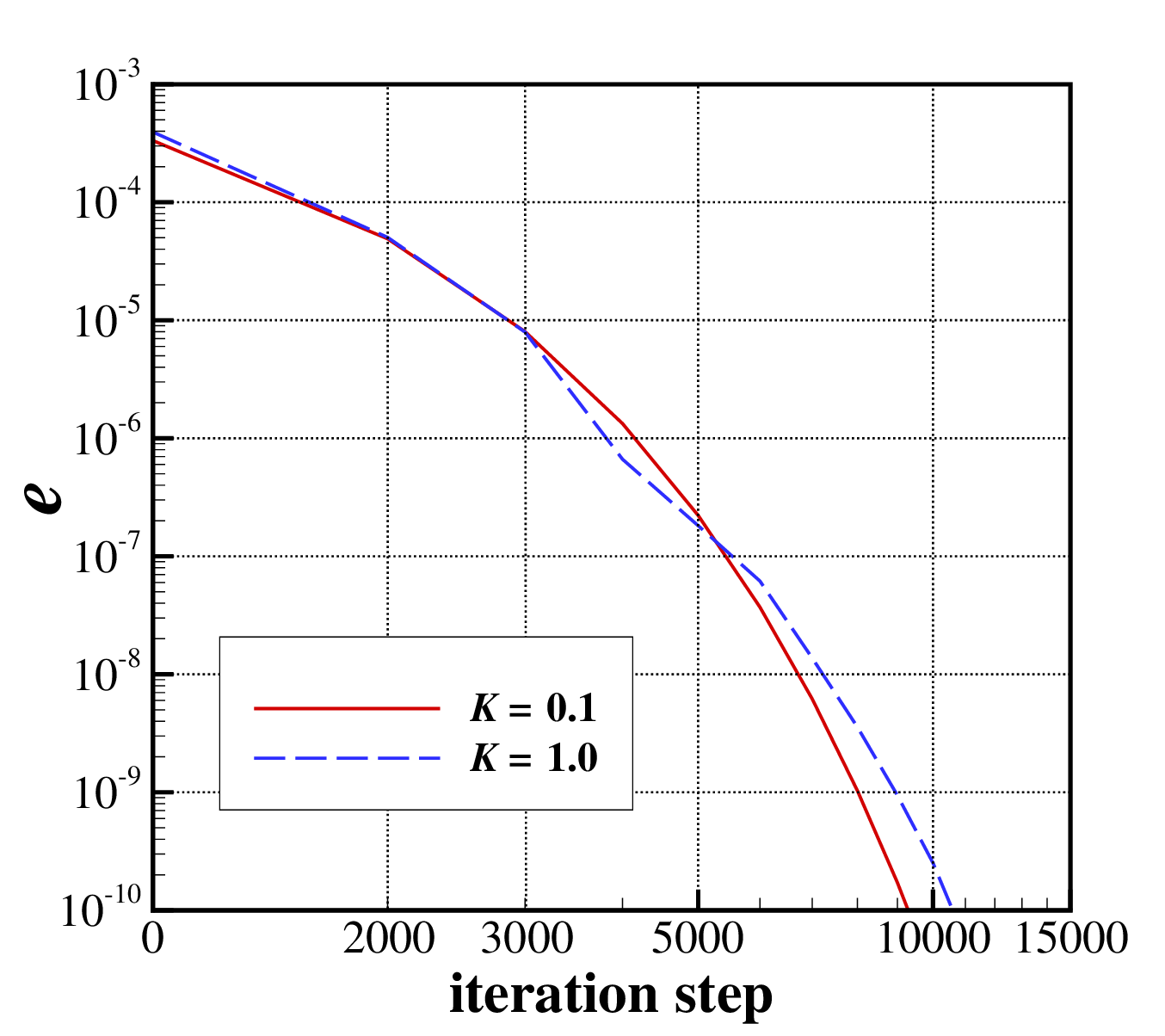}
	}
	\caption{\label{couette-contor}\color{blue} (a) $u$ velocity contours and (b) convergence history of the micro-Couette flow at two $Kn$ numbers, where $Kn$ is set to $Kn=2K/{\sqrt{\pi}}$, and wall moving velocity is $U_w=16.85m/s$.}
\end{figure}

The micro-Couette flow is driven by two parallel moving plates with a distant $H$. It can be treated as a benchmark test case for the decoupled method, as the displacement of moving wall is set to zero. In the simulation, 400 quadrangular cells are used, with 101 grid nodes are placed in $y$-direction and 5 in $x$-direction. For the boundary conditions, the top and bottom plates are set to wall boundaries with moving velocities $\pm{U_w}$, and left and right sides of the channel are set to periodic boundaries. The working gas is argon (the specific gas constant $R=208J/kg/K$), and four $Kn$ numbers, $0.01$, $0.2/\sqrt{\pi}$, $2.0/\sqrt{\pi}$ and $20/\sqrt{\pi}$, are carried out. The initial temperature $T_0$ in the channel and $T_w$ at the walls are set to $273K$. For the moving velocities of walls $U_w$, two values, $\pm{16.85}m/s$ and $\pm{119.15}m/s$, are used. Besides, the reference temperature and velocity are $T_{ref}=273K$ and $U_{ref}=\sqrt{2RT_{ref}}=337m/s$, respectively. The $CFL$ number used in Eq.~\eqref{CFL} is 0.8 for $Kn=0.01$ and $0.2/\sqrt{\pi}$, and 0.7 for other $Kn$ numbers. Finally, the 28-points Gauss-Hermite quadrature rule is used for flow at $Kn=0.01$, and $80\times{80}$ points of Newton-Cotes quadrature rules with a range of $[-4,4]\times{[-4,4]}$ is used for flows at other $Kn$ numbers (for flow at $119.15m/s$, the compressible DUGKS is used~\cite{Wang2019Arbitrary}). Fig.~\ref{couette} shows the comparisons of velocity profiles with those of the DSMC~\cite{Bahukudumbi2003unified,Fan2001statistical} and the UGKS~\cite{Zhu2016implicit}. Fig.~\ref{couette-contor} shows the contours of $x$-direction velocity $u$ and the convergence history $e$ of velocity at two $Kn$ numbers, respectively, where $e$ is given by
\begin{equation}
e=\frac{\sqrt{\sum_{i}\left[({u_i^{n+1000}-u_i^n})^2+({v_i^{n+1000}-v_i^n})^2\right]}}{\sqrt{\sum_{i}[(u_i^n)^2+(v_i^n)^2]}},
\end{equation}
and $i$ is the index number of grid cells. As shown in the figures, the DUGKS obtains satisfactory results in all flow regimes compared with other numerical methods, and also shows good convergence property without statistics noise. Consequently, the DUGKS demonstrates great potentials in simulating the low-speed micro-flows in MEMS.

\subsection{Free vibration of a square cylinder in continuum flow}
\begin{figure}
	\centering
	\includegraphics[width=0.4 \textwidth]{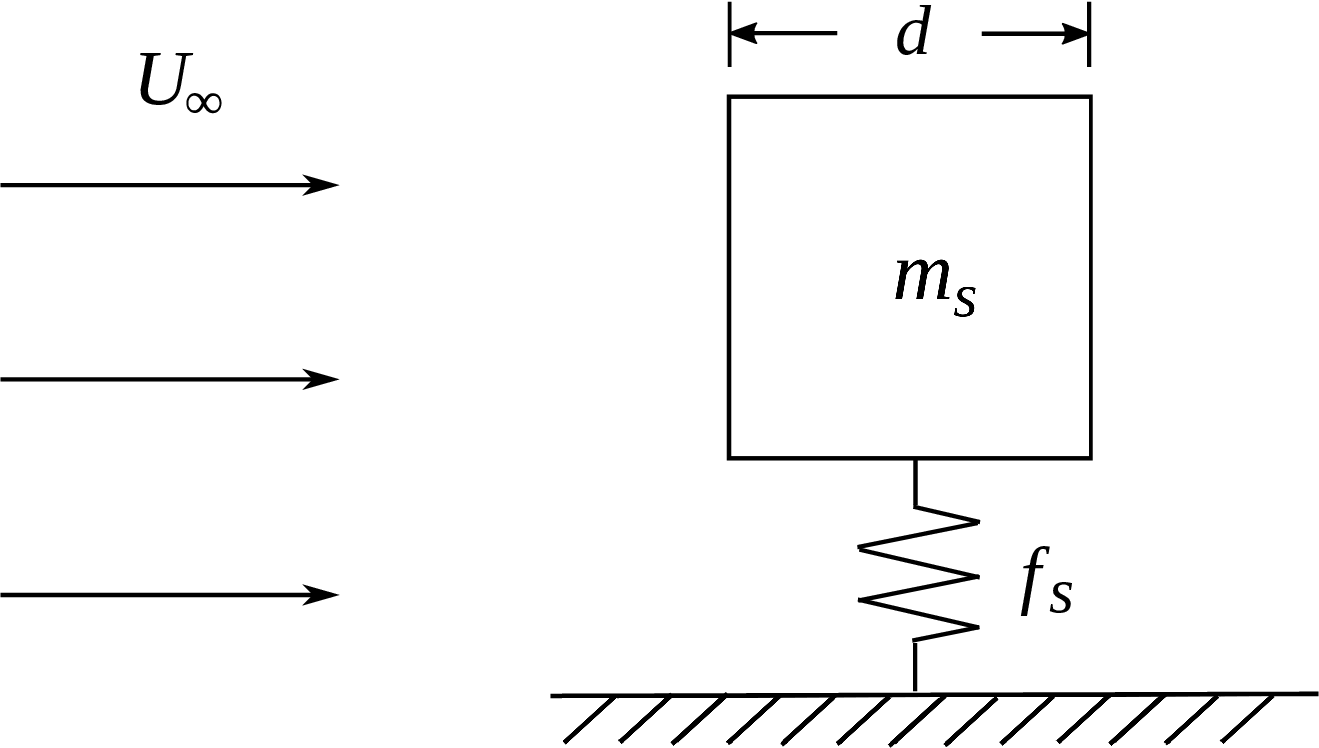}
	\caption{\label{square-config} Schematic diagram of flow around an elastically mounted square cylinder.}
\end{figure}

\begin{figure}
	\centering
	\subfigure[]{
		\includegraphics[width=0.5 \textwidth]{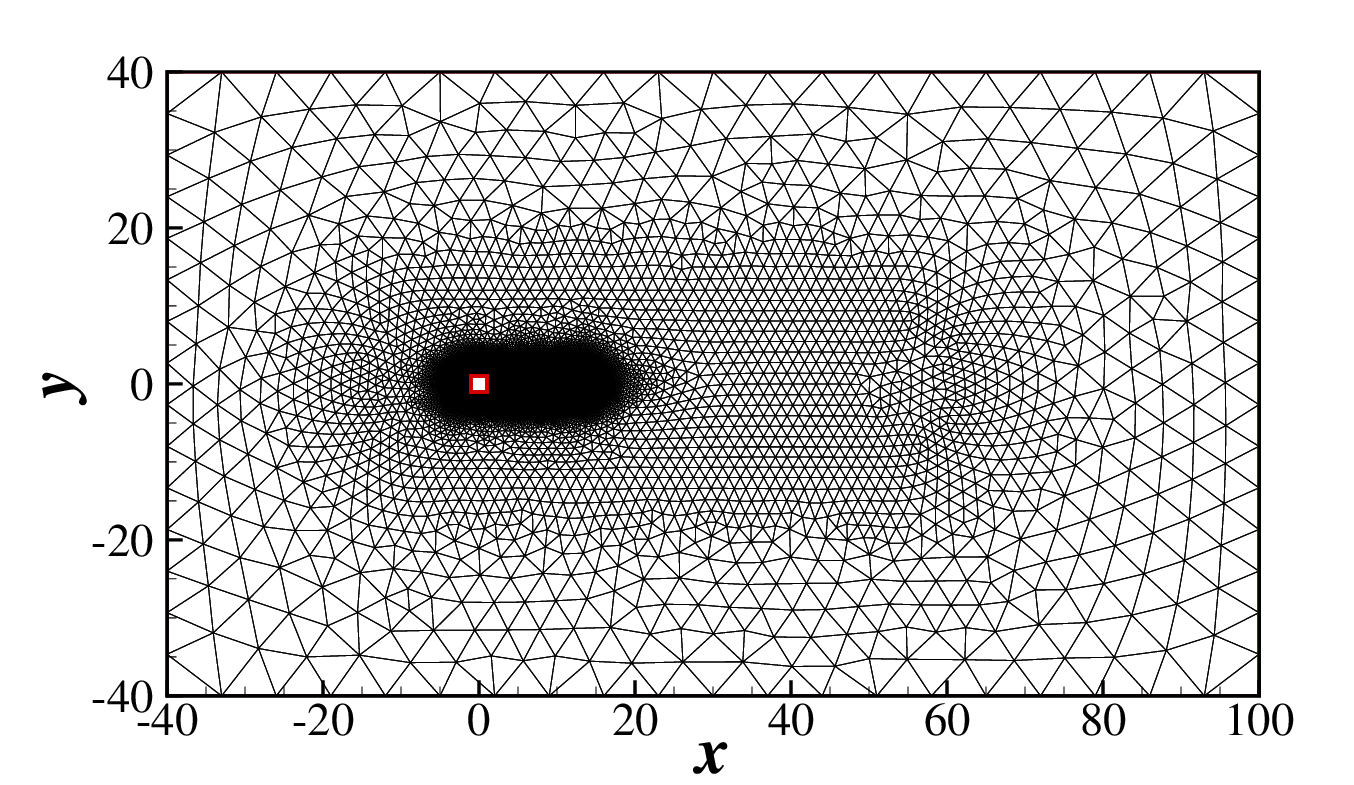}
	}
	\subfigure[]{
		\includegraphics[width=0.405 \textwidth]{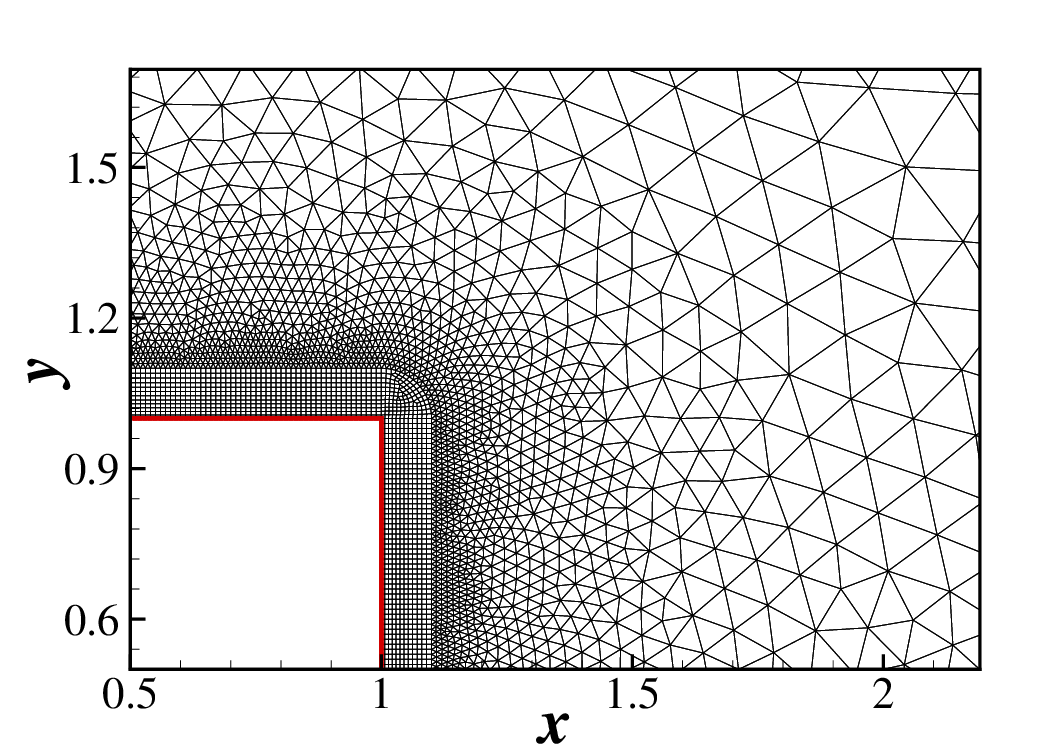}
	}	
	\caption{\label{square-mesh} Mesh for flow around an elastically mounted square cylinder: (a) full domain and (b) near the cylinder surface.}
\end{figure}

\begin{figure}
	\centering
	\includegraphics[width=0.6 \textwidth]{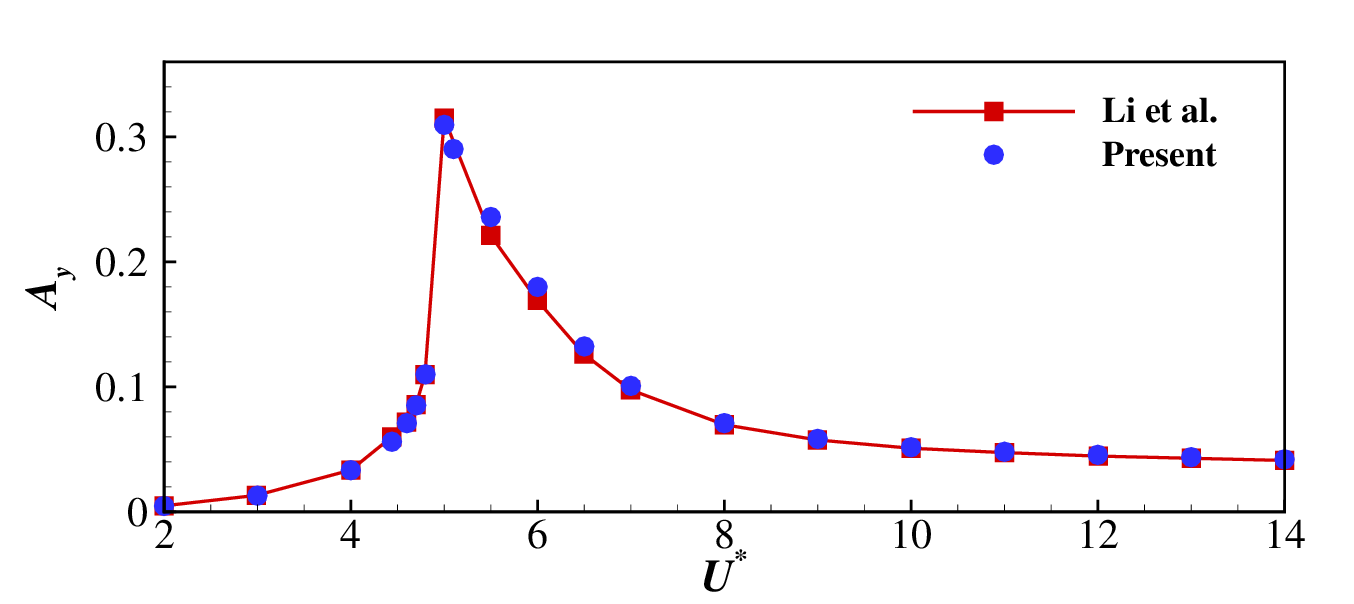}
	\caption{\label{square-fsi} Transverse vibration amplitude of flow around an elastically mounted square cylinder at $Re=100$ and $m^*=3$.}
\end{figure}

For laminar flow around a cylinder at Reynolds number $Re>50$ ($Re\sim47$ for circular cylinder~\cite{Williamson1996}), the flow is unsteady and vortex shedding can be observed. Then if a cylinder is elastically mounted in a uniform flow, the unsteady aerodynamic force will leads to the vortex-induced vibrations (VIV)~\cite{Singh2005Vortexs}. In this work, flow around an elastically mounted square cylinder shown in Fig.~\ref{square-config} is studied, and $Re$ is set to $100$. It can be treated as a benchmark test case for the coupled framework in continuum flow region. Fig.~\ref{square-mesh} shows the hybrid unstructured mesh used in this case. The total number of grid cells is 44992, with 800 points at the surface of cylinder. The region near the surface of cylinder is discretized into quadrangular cells with the minimum size of grid cells being $d/150$, where $d$ is the side length of the square cylinder. The computational domain is set to $[80d\times140d]$, which is large enough to eliminate the influence of the far-field boundary condition. In this test case, two-degree-of-freedom structure dynamics equations is used~\cite{Li2019Mode}:
\begin{equation}
   \ddot{x}+(2\pi{F_s})^2x=\frac{C_d}{2m^*}, \ddot{y}+(2\pi{F_s})^2y=\frac{C_l}{2m^*},
\end{equation}
where $F_s=f_sd/U_\infty$ is the reduced natural frequency relating to the natural frequency of a mass--spring system $f_s$ ($U_\infty$ is velocity of free stream), $m^*$ is the non-dimensional mass of square with $m^*=m_s/\rho{d}^2$ ($m_s$ is the mass of square per unit length and $\rho$ is the density of free stream) and is set to $m^*=3$ in this case, and $C_d$ and $C_l$ are the drag and lift coefficients of a square, respectively. Fig.~\ref{square-fsi} shows the maximum vibration amplitude $A_y$ of square in transverse direction, where $A_y$ is given by $A_y=(y_{max}-y_{min})/2$, and $U^*$ is the reduced velocity defined as $U^*=1/F_s=U_\infty/f_sd$. In general, our result agrees well with the numerical result of Li et al.~\cite{Li2019Mode}, and demonstrates the capability of the present coupled framework to further simulate the SFD problem in MEMS. Furthermore, although the focus of this paper is rarefied gas flow, this continuum flow test also shows the performance of the DUGKS for simulating the unsteady flow; on the contrary, higher computational cost is required for the DSMC method to obtain a smooth result of the flow field.

\section{Results and discussion}\label{cases}
\begin{figure}
	\centering
	\includegraphics[width=0.6 \textwidth]{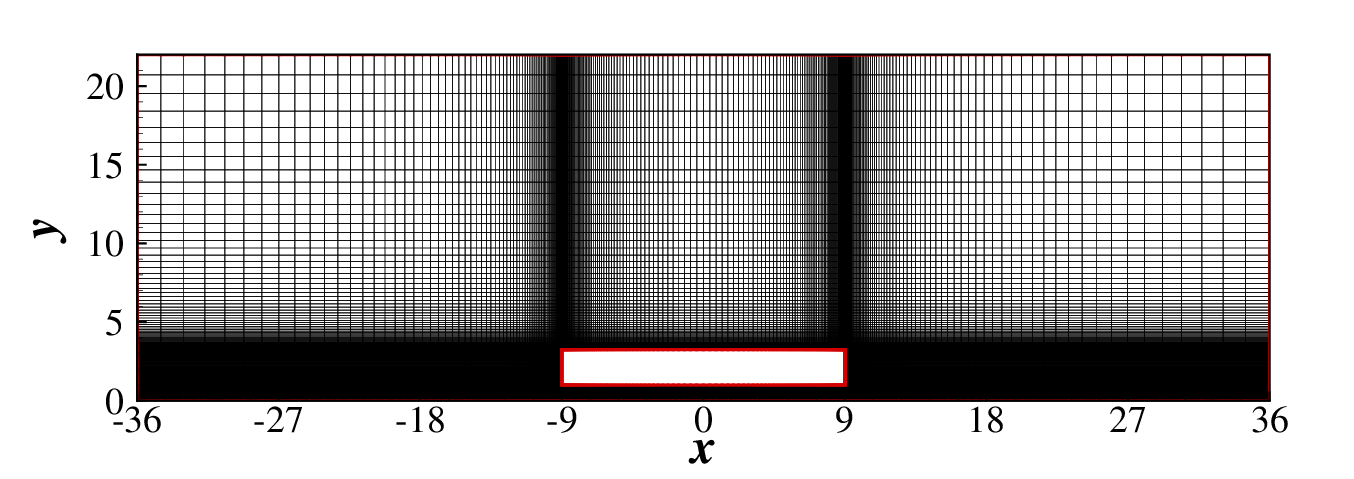}
	\caption{\label{beam-mesh} Mesh for an elastically mounted micro-beam oscillating in rarefied gas.}
\end{figure}

In this section, two-dimensional forced or free oscillation of a micro-beam with linear (perpendicular) or tilting motion in rarefied gas is fully studied. Fig.~\ref{beam-config} shows the configuration of computational domain and boundary conditions. The substrate and the surface of micro-beam are set to diffuse-scattering wall boundary conditions, and the left, right and top boundaries of the computational domain are set to outlet boundary conditions. Following the setup described in Ref.~\cite{Guo2009Compact}, the width $L$ and thickness $D$ of micro-beam are set to $18.0\times10^{-6}m$ and $2.25\times10^{-6}m$, respectively, and the gap height $h$ is $1.0\times10^{-6}m$. Fig.~\ref{beam-mesh} shows the mesh used in this study, the total number of grid cells is 37200, and the minimum size of grid cells near the wall is $0.025h$, with 40 grid cells are located in the gap. The working gas also is argon, and the reference temperature $T_{ref}$ and velocity $U_{ref}$ are $273K$ and $307.6m/s$ ($U_{ref}=\sqrt{\gamma{RT_{ref}}}$, and $\gamma=5/3$ is the specific heat ratio), respectively. The relation between mean-free-path $\lambda$ and viscosity $\mu$~\cite{Guo2015Discrete} is given by
\begin{equation}
	\lambda=\frac{2\mu(7-2\omega)(5-2\omega)}{15\rho\sqrt{2\pi{RT}}},
\end{equation}
where $\omega=0.5$ is the index related to the HS model. The Knudsen number is defined as $Kn^{(h)}=\lambda/h$ in the following section, where $h$ is the reference length of flow. For the initial conditions of rarefied gas flow, the reference density $\rho_{ref}$ and viscosity $\mu_{ref}$ are set to the corresponding values at $Kn^{(h)} = 3.61\times10^{-4}$, Mach number $Ma=2.19\times{10}^{-4}$ and Reynolds number $Re=1.0$. And by keeping a constant value of gas viscosity, the density of rarefied gas at other Knudsen numbers can be obtained. Besides, the Gauss-Hermite quadrature rule is used for all the considered Knudsen numbers.

\subsection{\color{blue}Decoupled method: squeeze-film damping at different Knudsen numbers}\label{SecIII-I}
\subsubsection{Squeeze-film damping coefficient and nonlinear damping phenomenon}
\begin{figure}
	\centering
	\includegraphics[width=0.4 \textwidth]{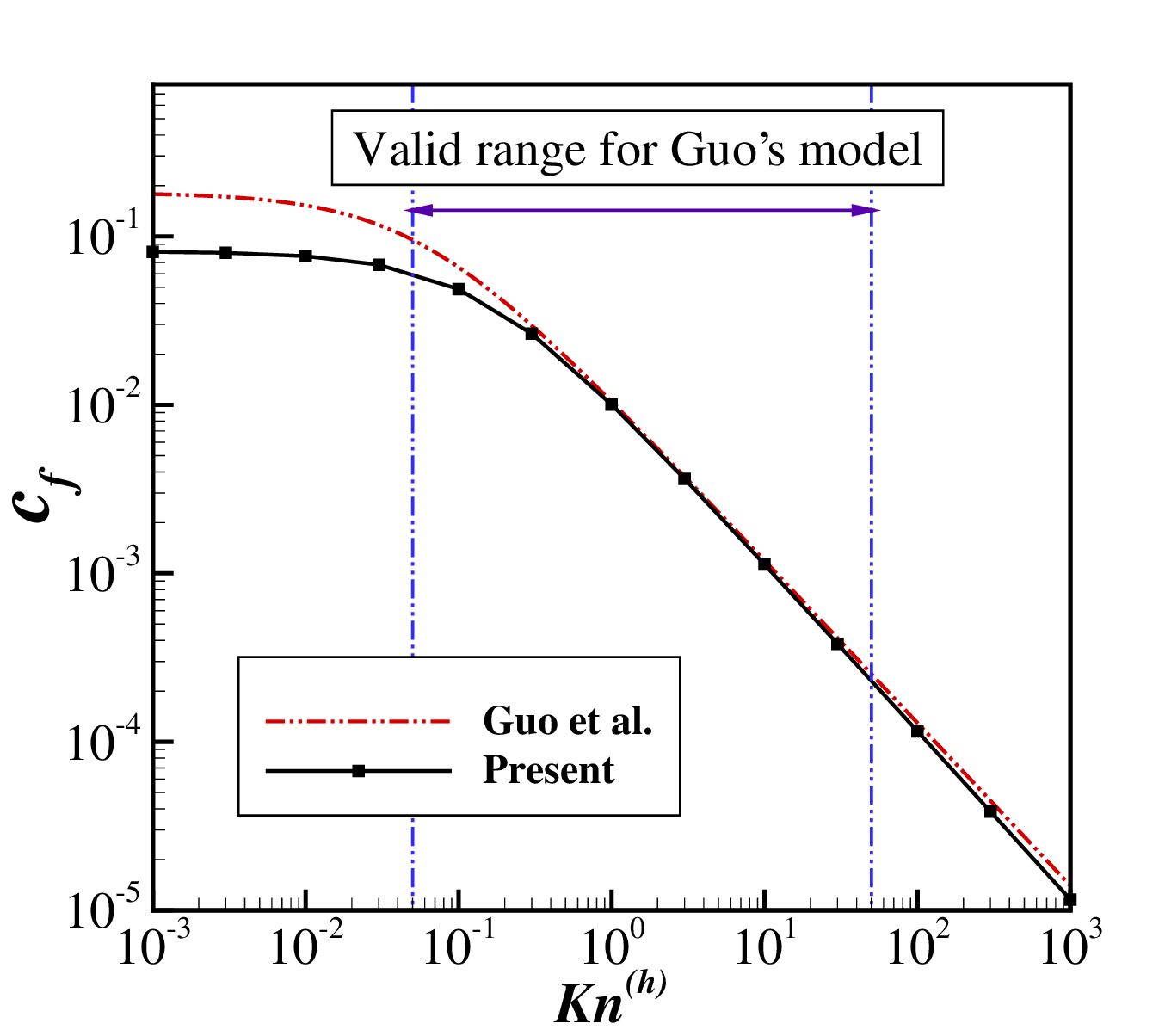}
	\caption{\label{Cf-dkn} Comparison of squeeze-film damping coefficient $c_f$ with Guo et al.'s compact model~\cite{Guo2009Compact}.}
\end{figure}

Firstly, the calculation of the SFD coefficient~\cite{Guo2009Compact} is considered, which is based on the traditional decoupled method. Shown in Fig.~\ref{plate-linear-decoupled}, during the simulation, by imposing a constant moving velocity on the surface of micro-beam, the damping force $F$ acting on the micro-beam from continuum to free-molecule flow regimes is calculated. The moving velocity $V_s$ is set to $-0.0674m/s$ in this case (according to the setup described in Ref.~\cite{Guo2009Compact}, this value is about $-0.075m/s$ based on the sonic speed of air). Fig.~\ref{Cf-dkn} shows a comparison of present result with Guo et al.'s compact model~\cite{Guo2009Compact} (based on the Boltzmann ellipsoidal statistical BGK equation). And this compact model is given as
\begin{equation}\label{Guo_model}
    c_f(x_1,x_2)=\frac{F}{V_sL}=\frac{ax_1^c}{1+bx_1^ex_2^f}{D},
\end{equation}
where $x_1=L/h$, $x_2={Kn^{(h)}}/x_1$ (or equals to the Knudsen number $Kn^{(L)}$ based on the width of micro-beam), and the constant parameters are set to $a=10.39$, $b=1.374$, $c=3.100$, $e=1.825$ and $f=0.9660$, respectively. Clearly, for $Kn^{(h)}>0.3$, our result agrees well with Guo et al.'s compact model. The differences between these two results at $Kn^{(h)}>100$ maybe is the different collision models used in the schemes. Besides, as the Guo et al.'s model is constructed based on the rarefied gas flow simulations ($0.05<Kn^{(h)}<50$), it is difficult to identify which result is better at near-continuum and continuum flow regimes. {\color{blue}Consequently, same as other decoupled method, a similar compact model also can be constructed based on the Eulerian-framework DUGKS}.

\begin{figure}
	\centering
	\subfigure[]{
		\includegraphics[width=0.5 \textwidth]{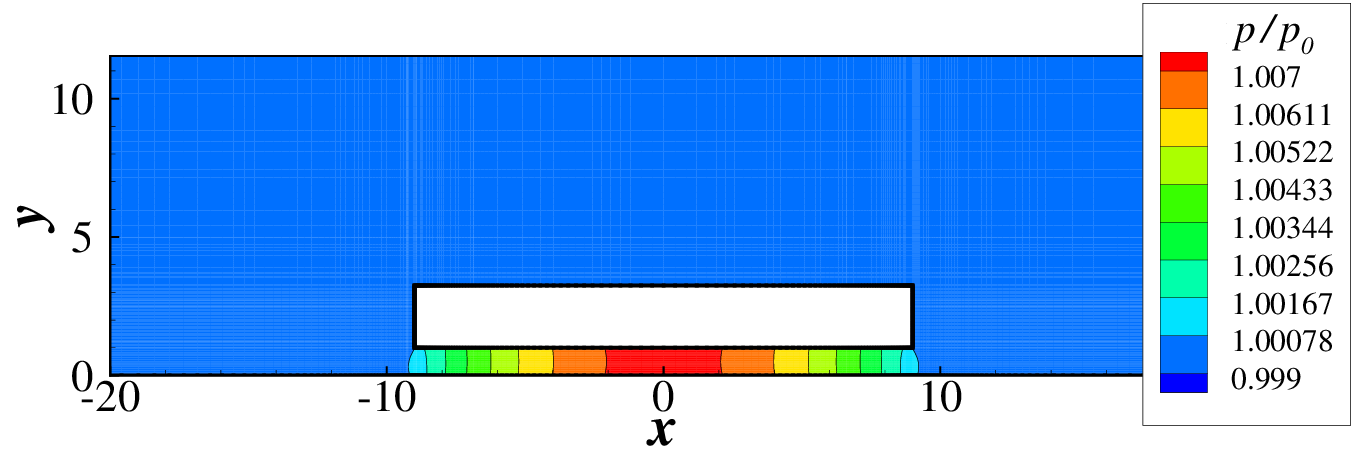}
		\includegraphics[width=0.5 \textwidth]{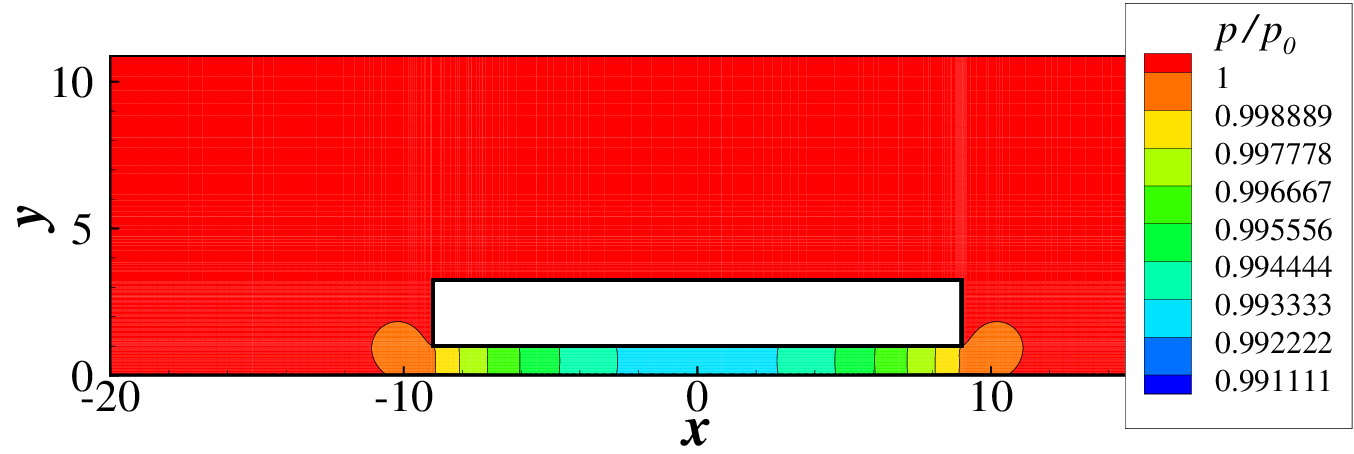}
	}
	\subfigure[]{
		\includegraphics[width=0.5 \textwidth]{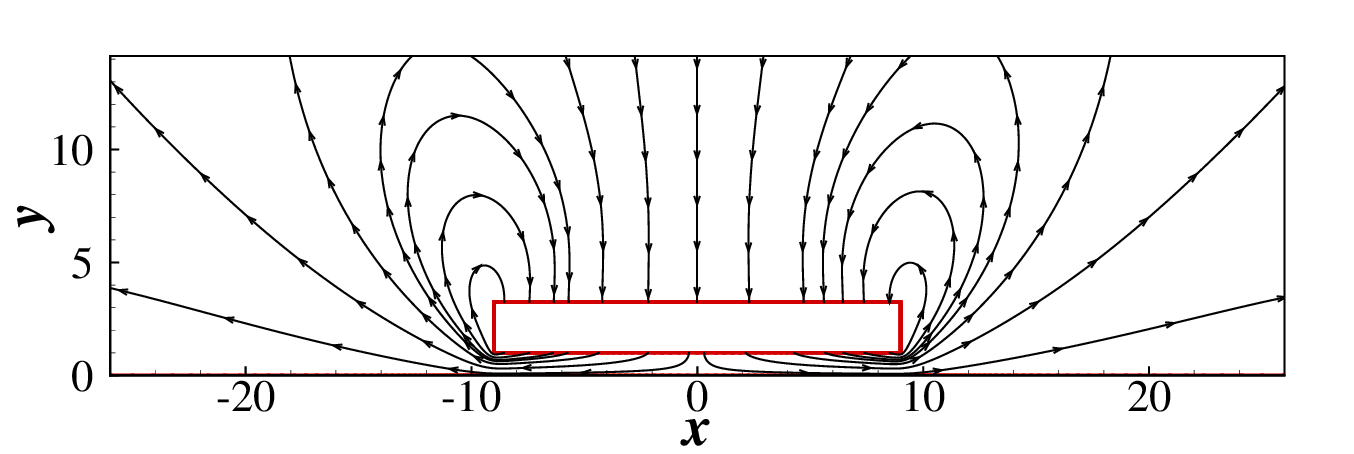}
		\includegraphics[width=0.5 \textwidth]{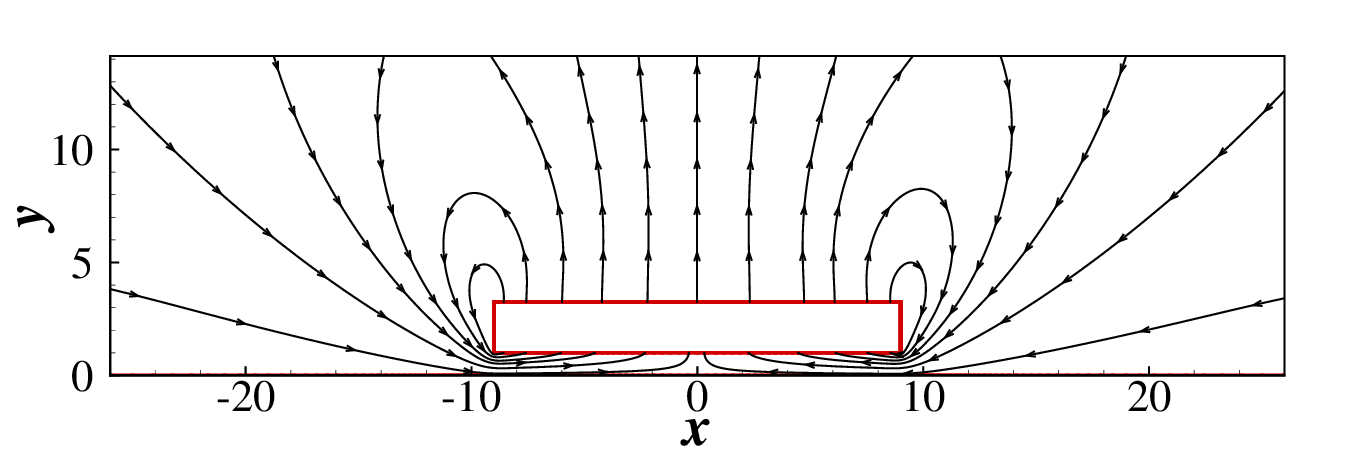}
	}	
	\caption{\label{Cf-downup} (a) Pressure contours and (b) streamlines for flow around a micro-beam in rarefied gas with downward (left) and upward (right) moving velocities at $Kn^{(h)}=0.1$ and $|V_s|=0.0674m/s$.}
\end{figure}

\begin{figure}
	\centering
	\subfigure[]{
		\includegraphics[width=0.4 \textwidth]{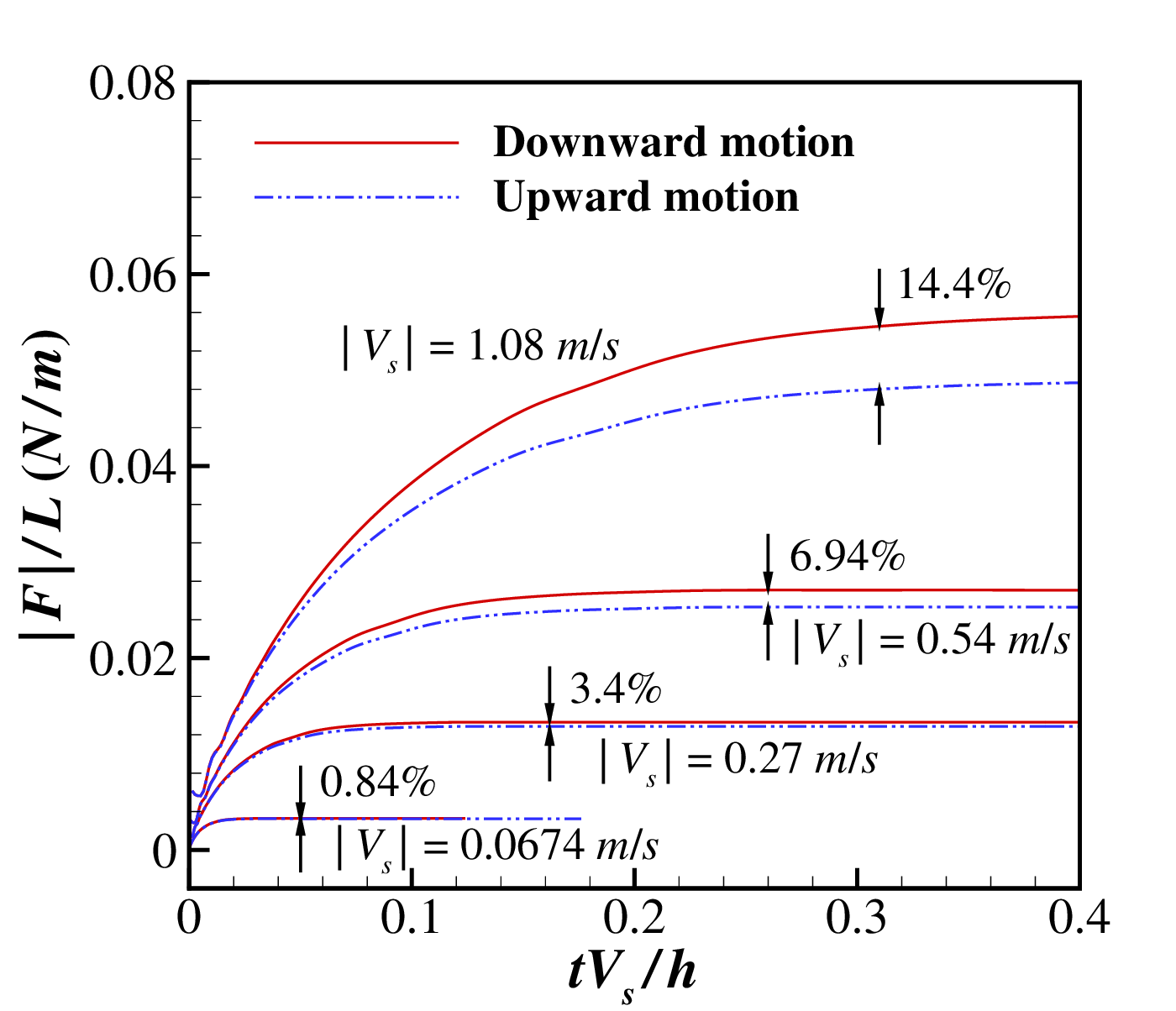}
		\includegraphics[width=0.4 \textwidth]{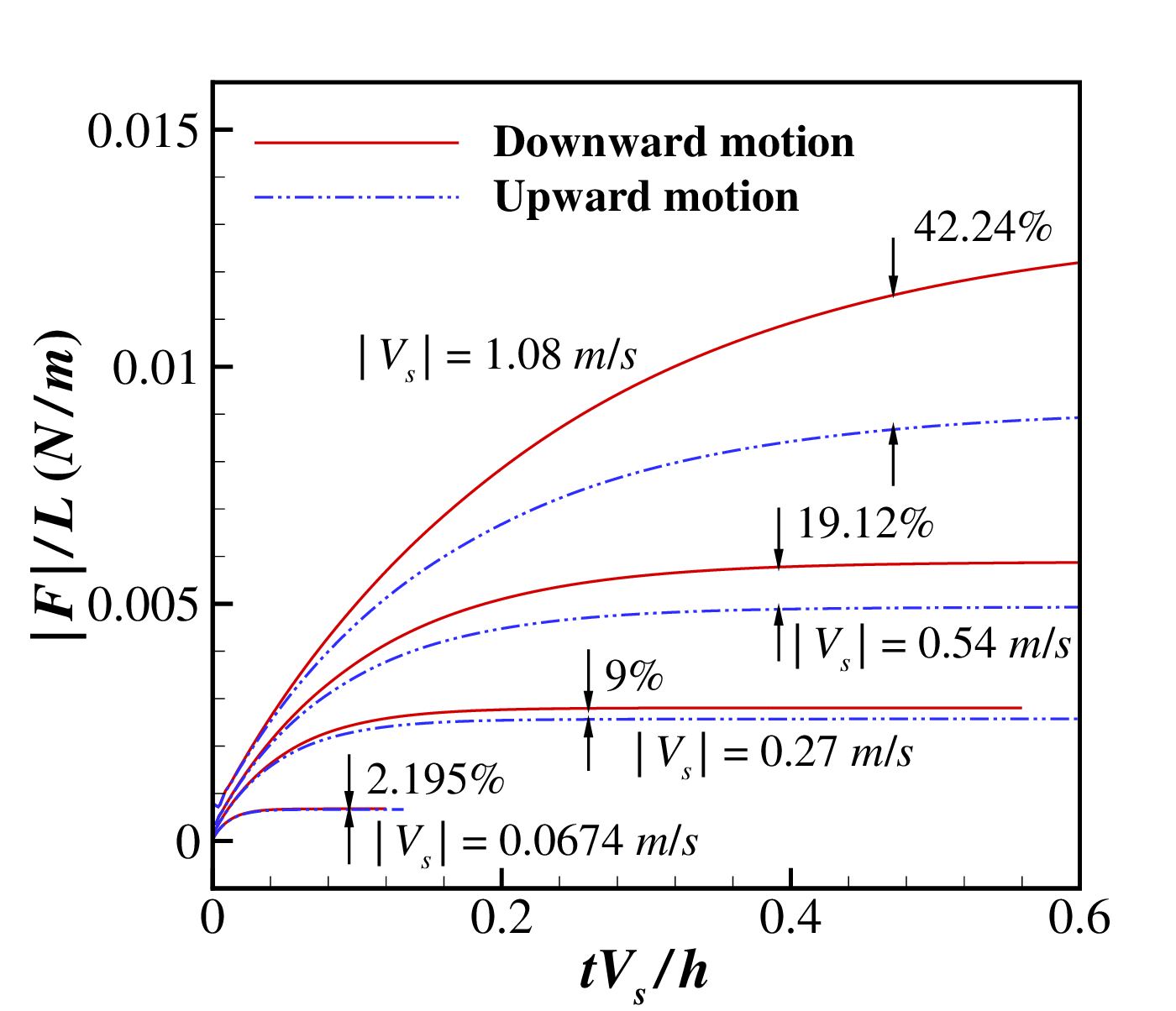}		
		\label{kn-press-convergent}
	}
	\subfigure[]{
		\includegraphics[width=0.4 \textwidth]{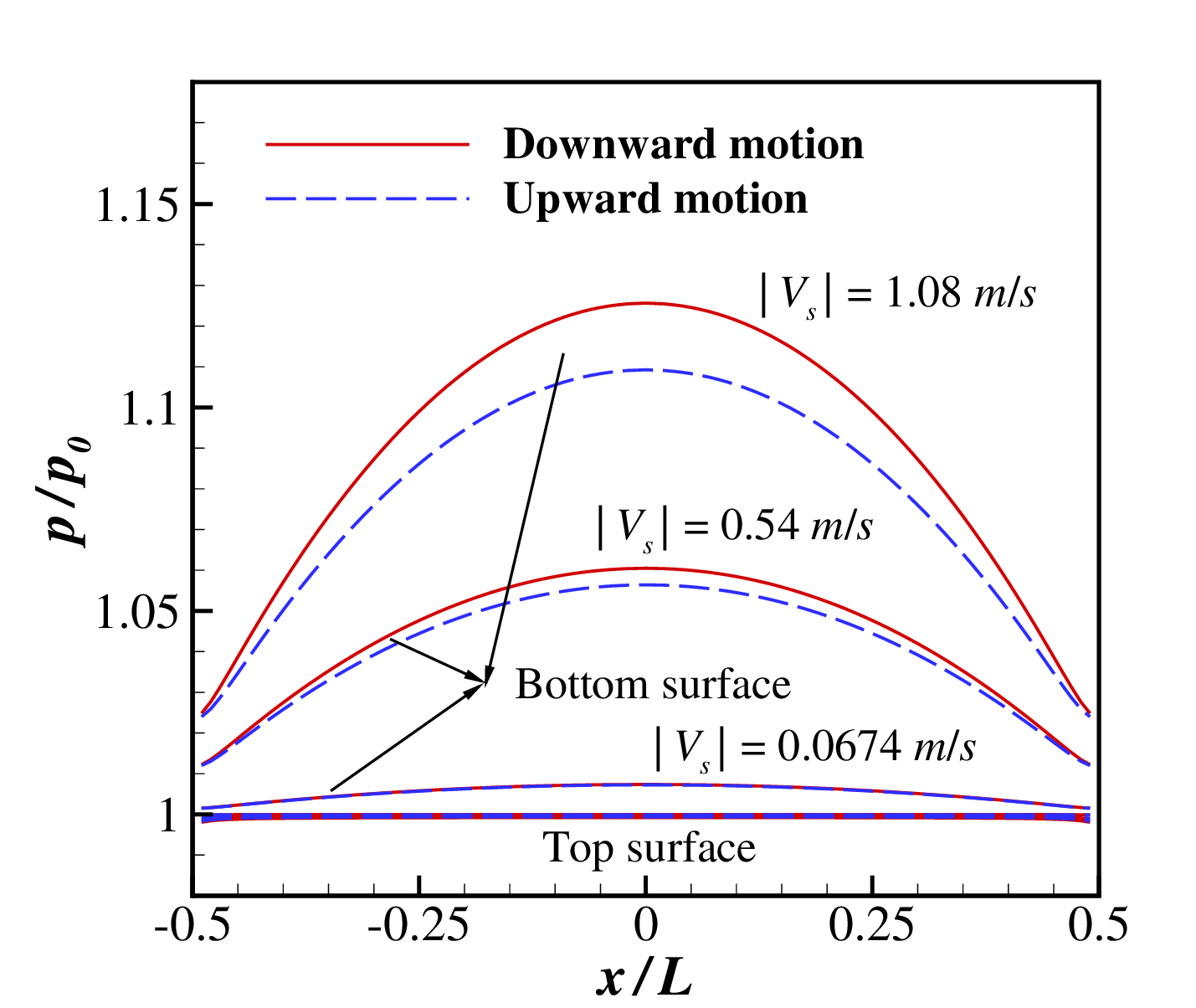}
		\includegraphics[width=0.4 \textwidth]{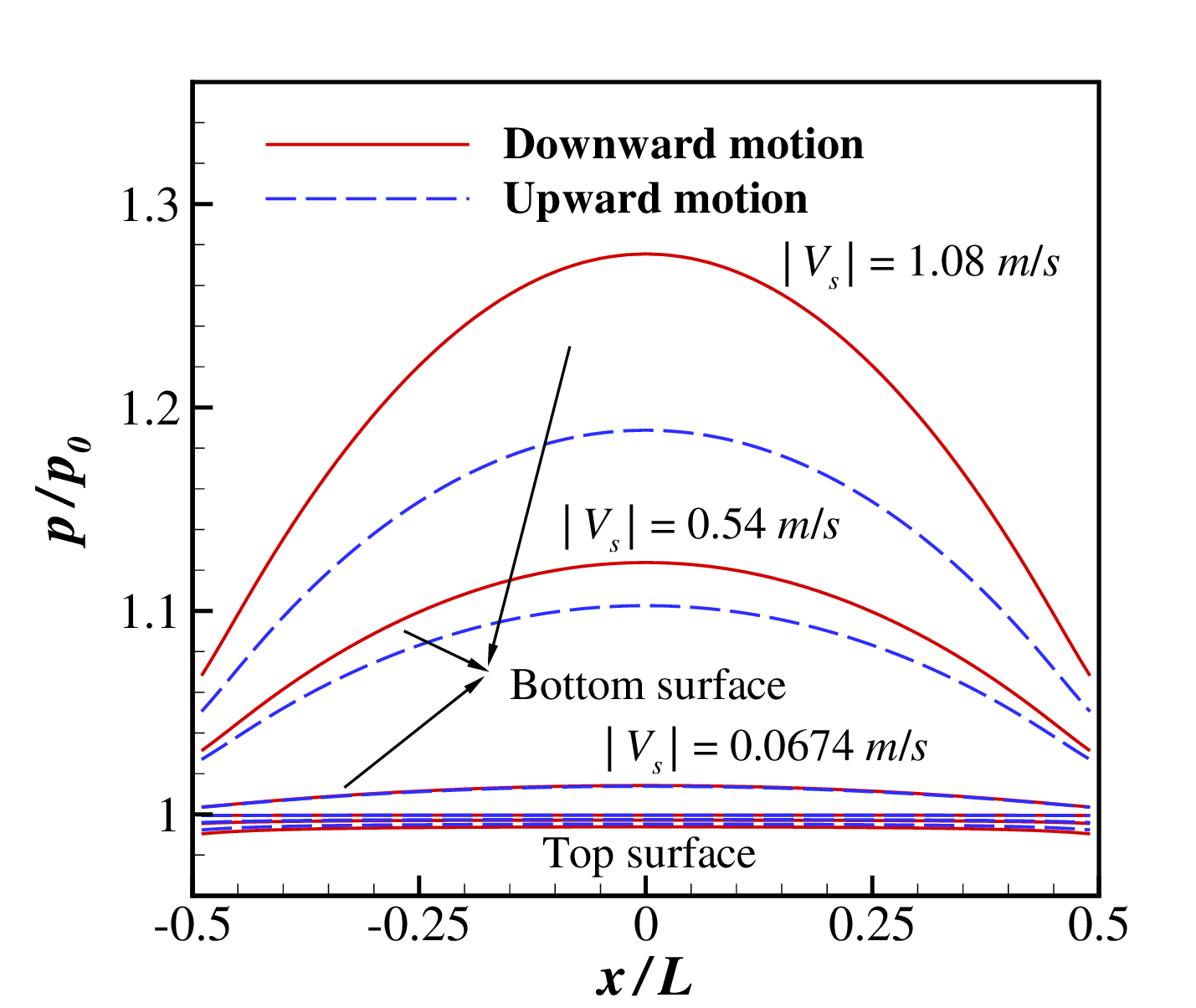}
		\label{kn-press-distribution}
	}	
	\subfigure[]{
		\includegraphics[width=0.4 \textwidth]{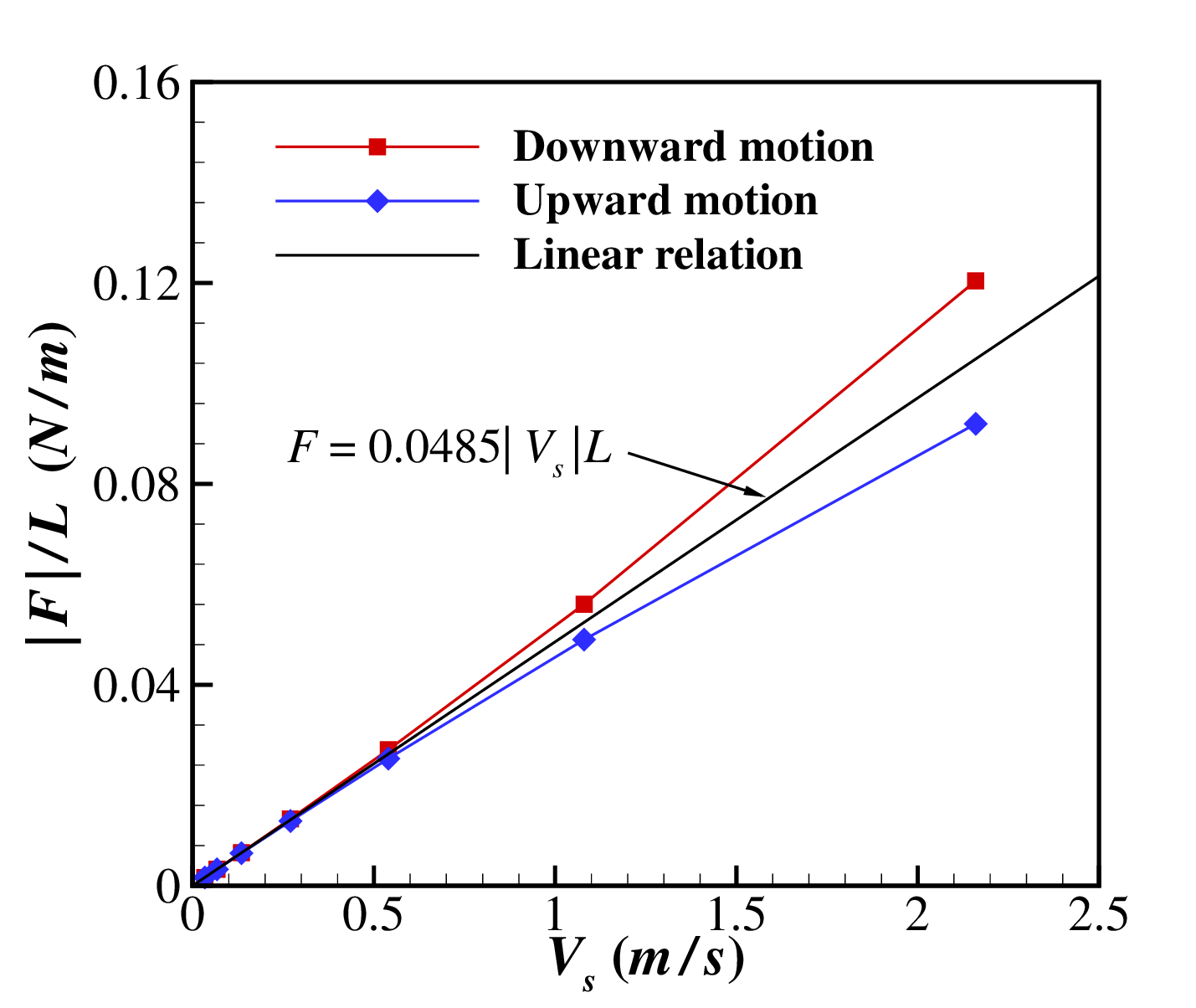}
		\includegraphics[width=0.4 \textwidth]{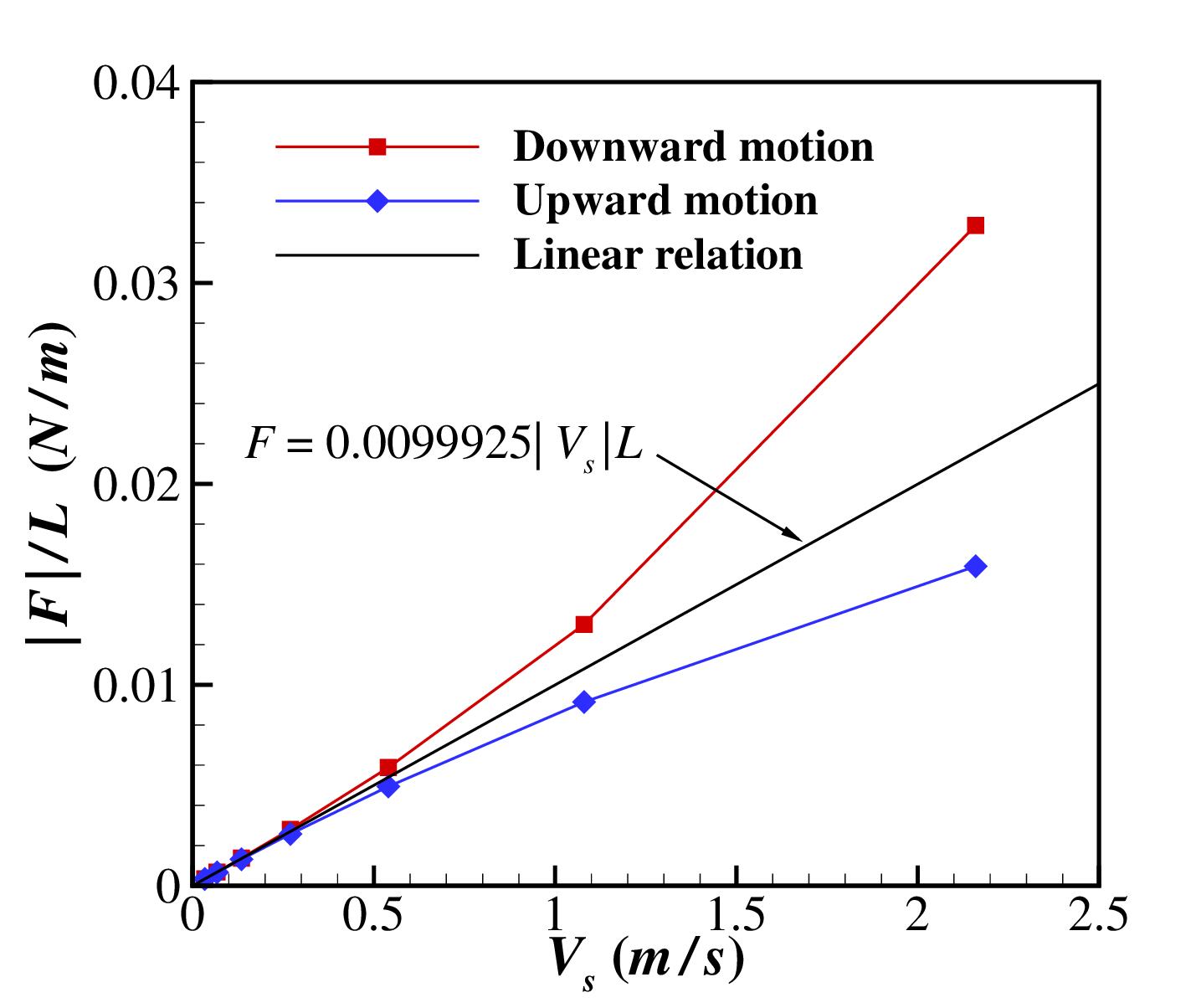}
		\label{kn-press-velocity}
	}
	\caption{\label{kn-press} Comparisons of flow around a micro-beam in rarefied gas at $Kn^{(h)}=0.1$ (left) and $Kn^{(h)}=1.0$ (right): (a) convergent history of damping force acting on a micro-beam at different velocities $V_s$, (b) pressure distribution along the top and bottom surfaces of micro-beam at different velocities (to make a comparison, symmetrical results are presented for upward motion) and (c) the variation of damping force at different velocities.}
\end{figure}

Next, the nonlinear damping phenomenon discussed in Ref.~\cite{Chigullapalli2012Non} is also studied. As illustrated above, the squeeze-film damping can be treated as an equivalent structure damping in the decoupled method. To verify the correction of the model, a set of simulations with velocities $V_s$ at different directions (downward or upward motion) and magnitudes is conducted. In the following sections, the damping force is positive with a downward moving velocity, and is negative with an upward moving one. Fig.~\ref{Cf-downup} shows the pressure contours and streamlines near a micro-beam with the downward and upward moving velocities at $Kn^{(h)}=0.1$ and $|V_s|=0.0674m/s$. Clearly, compared with other regions, the variation of pressure in the gap is obvious. And the gas in the gap is driven out by the micro-beam with a downward moving velocity and vice versa. Fig.~\ref{kn-press} shows the comparison results at two $Kn^{(h)}$ numbers. As shown in Fig.~\ref{kn-press-convergent}, at a small value of $|V_s|$ (about $0.0674m/s$), the difference of damping force between downward and upward motions is little. When the magnitude of $V_s$ is increased, the differences of those are obvious, and these tendencies are more significant at a large $Kn$ number. Fig.~\ref{kn-press-distribution} can be used to explain the reason, that is due to the influence of the substrate, the variation of pressure on the bottom surface is much more significant than that on the top surface. Furthermore, a conclusion can be made from Fig.~\ref{kn-press-velocity}, that is the traditional equivalent damping model is only available at a low-speed motion ($|V_s|<0.1m/s$), as the assumption of the linear relation cannot be maintained at a high-speed motion.

\subsubsection{Nonlinear damping phenomenon at different oscillation frequencies}
\begin{figure}
	\centering
	\subfigure[]{
		\includegraphics[width=0.4 \textwidth]{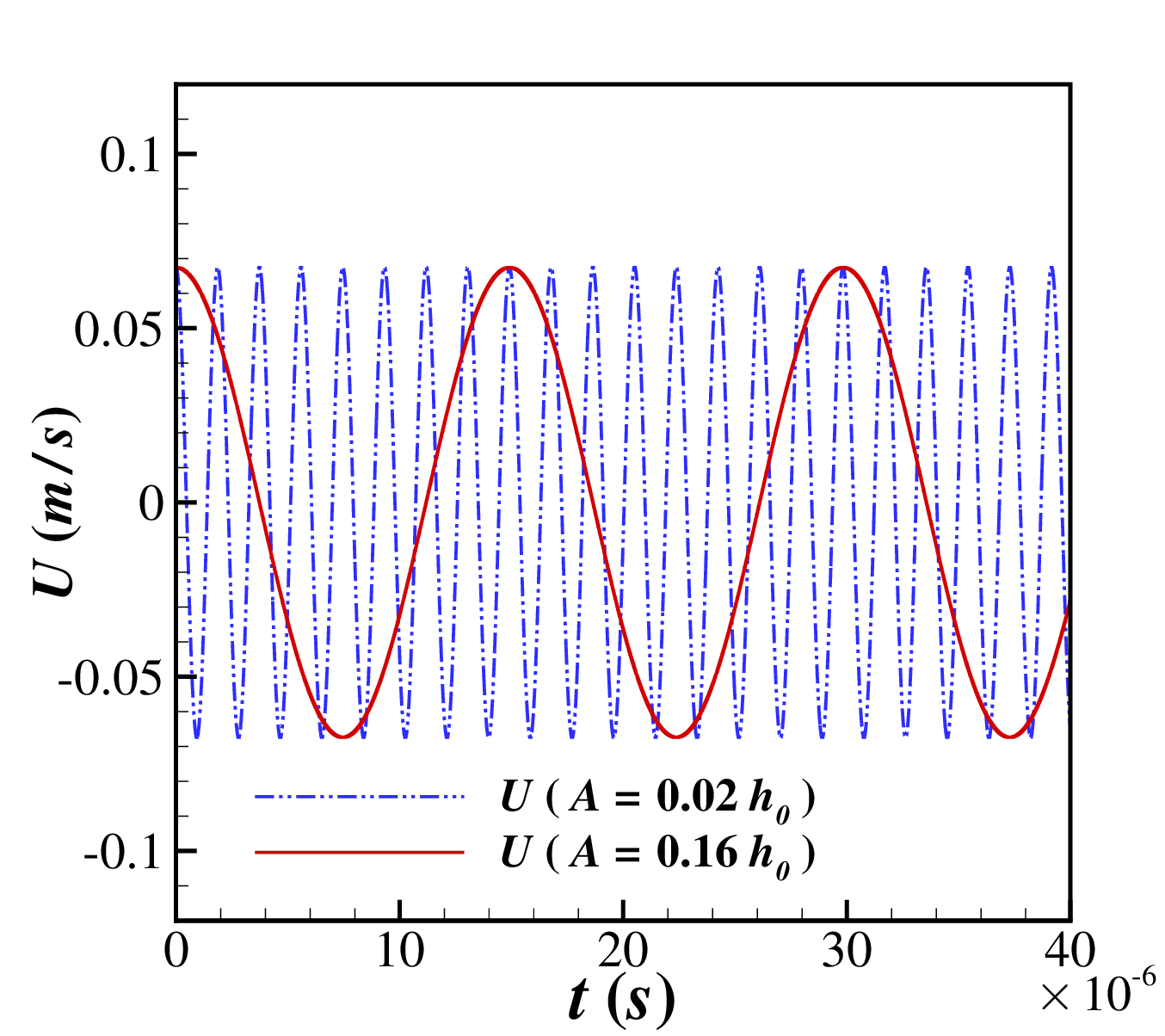}
		\label{guo-theory1}
	}
	\subfigure[]{
		\includegraphics[width=0.4 \textwidth]{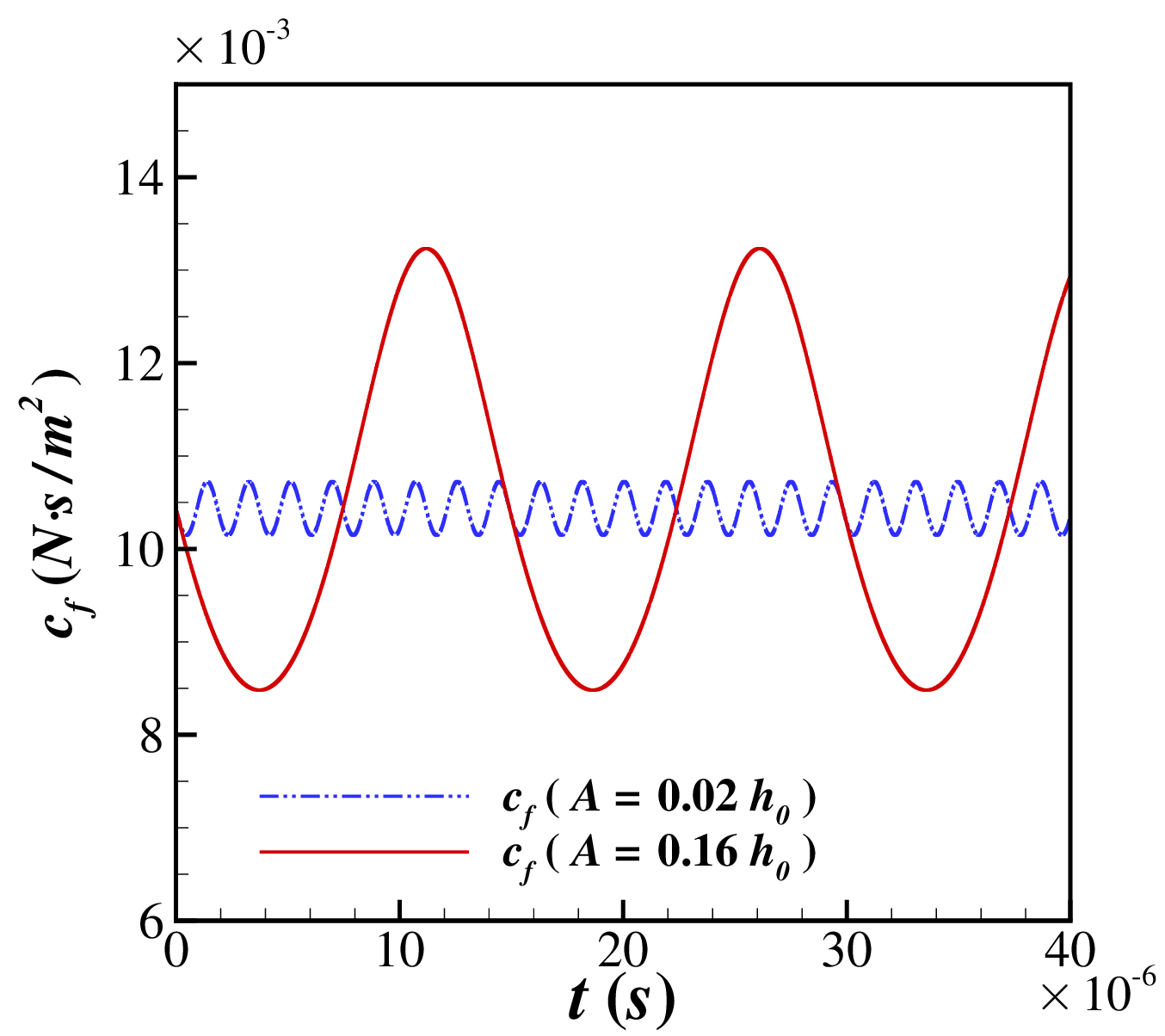}
		\label{guo-theory2}
	}
    \subfigure[]{
	    \includegraphics[width=0.4 \textwidth]{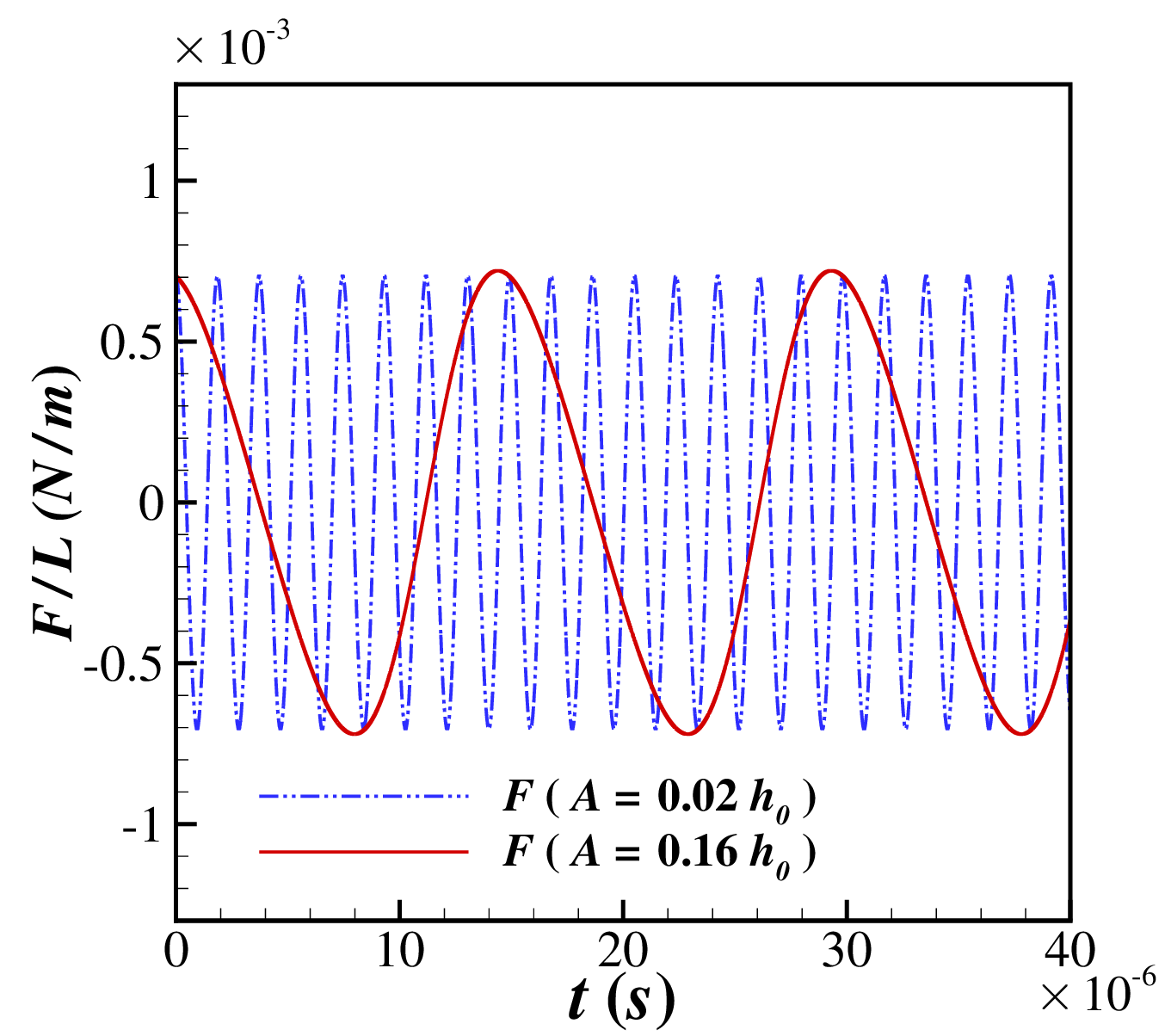}
	    \label{guo-theory3}
    }
	\caption{\label{guo_thoery} Comparisons of (a) motion velocities $U$, (b) damping coefficients $c_f$ and (c) damping forces $F$ at two oscillation amplitudes by Guo et al.'s compact model~\cite{Guo2009Compact} ($Kn^{(h)}=1.0$).}
\end{figure}

\begin{figure}
	\centering
	\subfigure[]{
		\includegraphics[width=0.55 \textwidth]{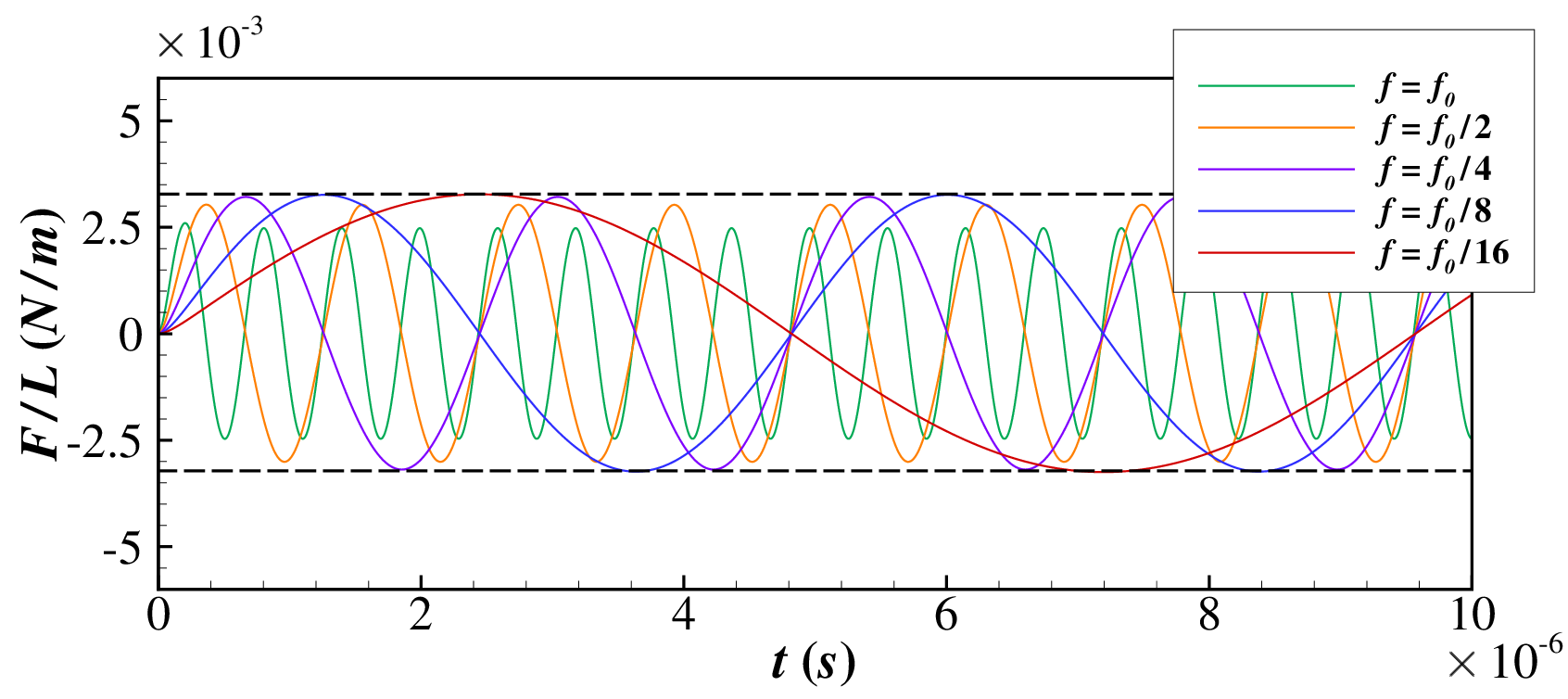}
		\includegraphics[width=0.55 \textwidth]{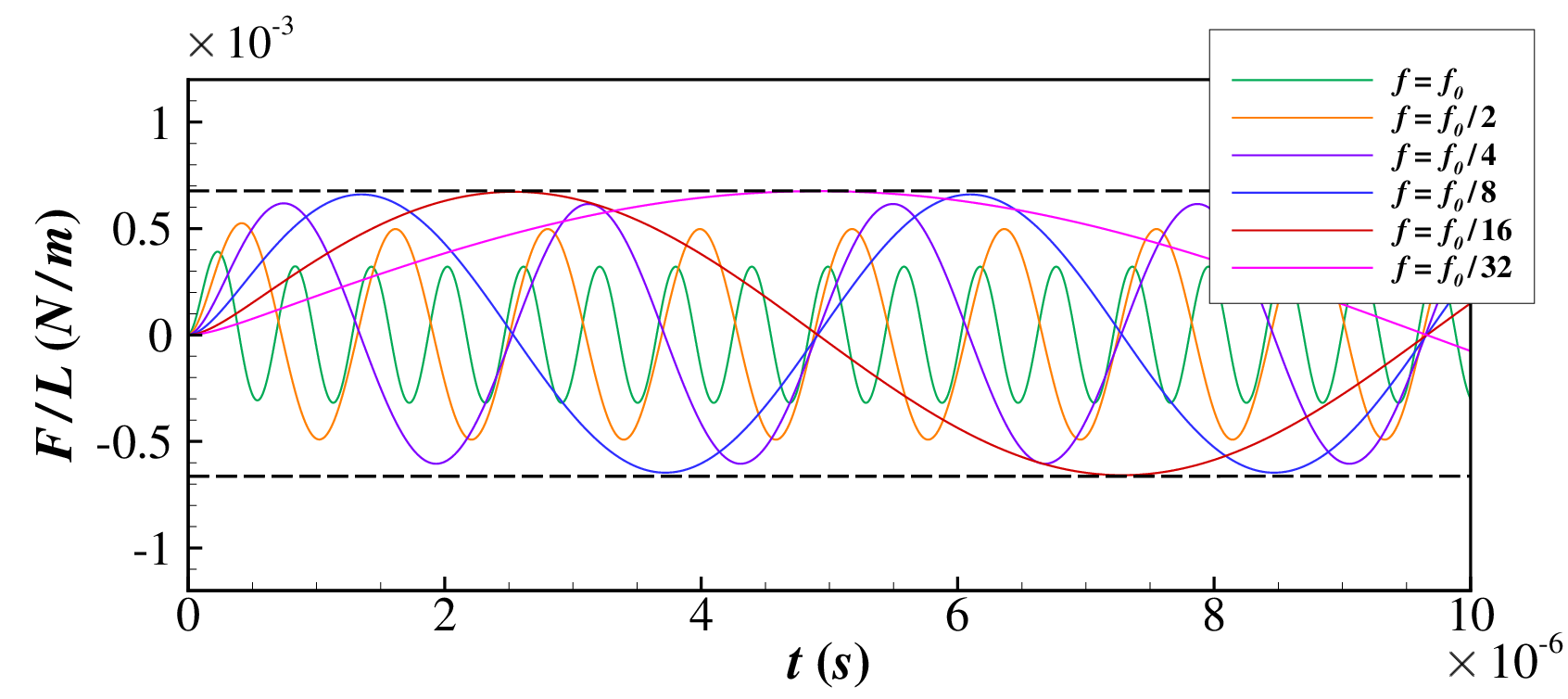}
	}
	\subfigure[]{
    	\includegraphics[width=0.55 \textwidth]{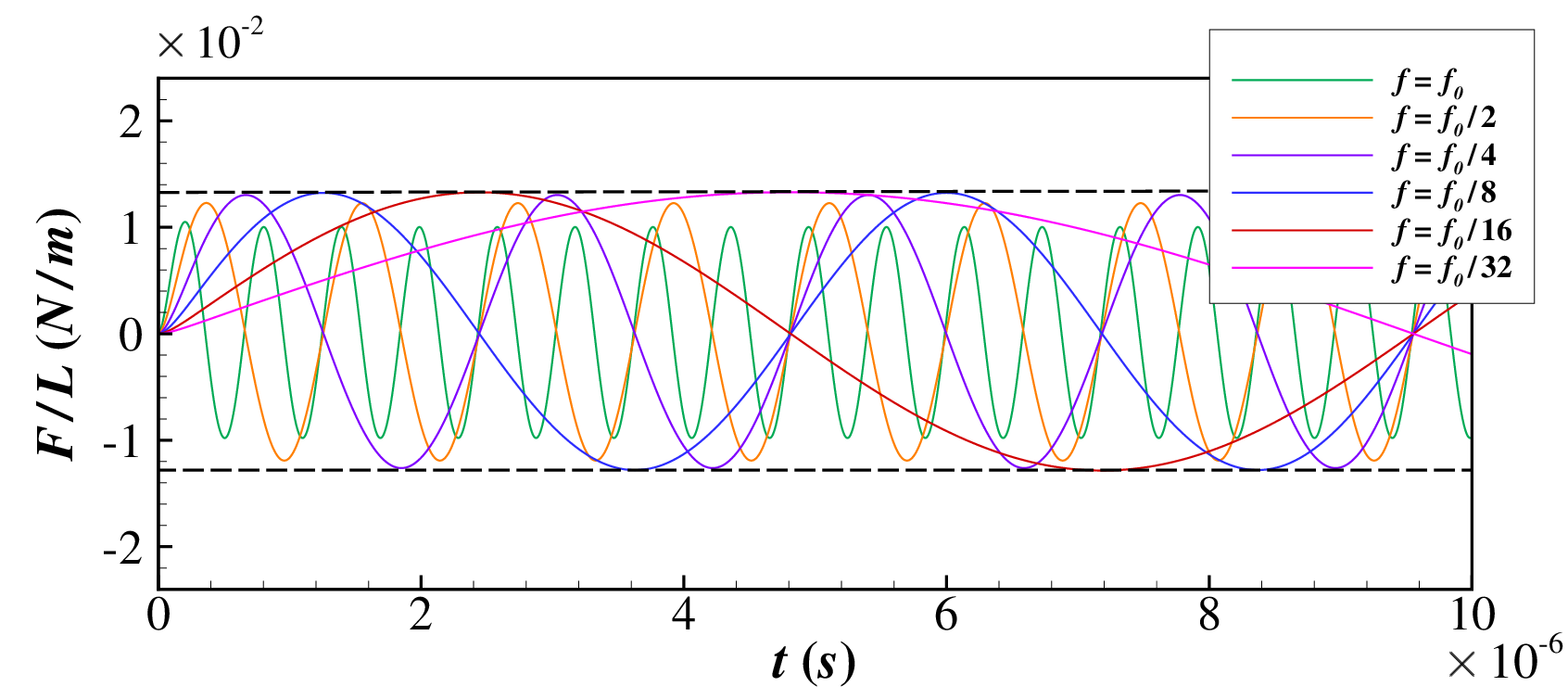}
	    \includegraphics[width=0.55 \textwidth]{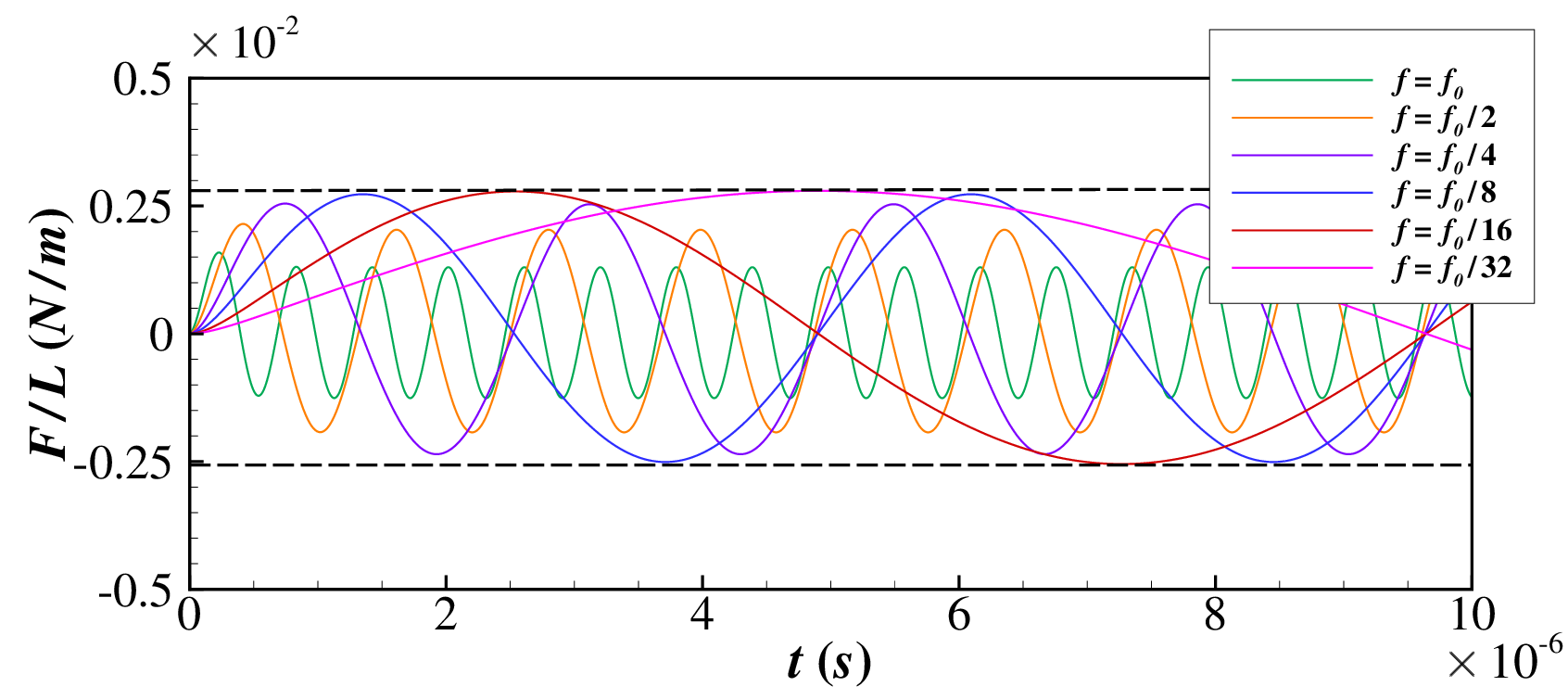}
    }
	\subfigure[]{
	    \includegraphics[width=0.55 \textwidth]{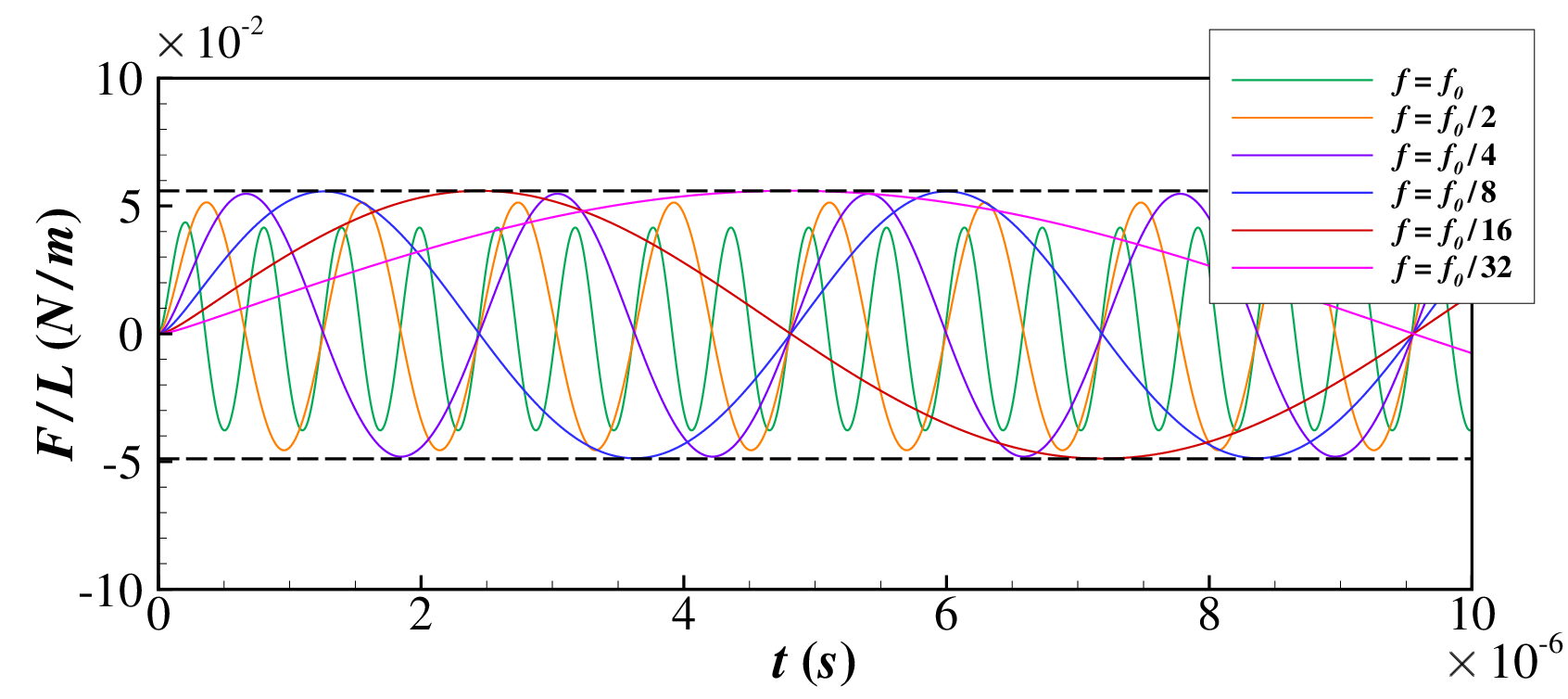}
	    \includegraphics[width=0.55 \textwidth]{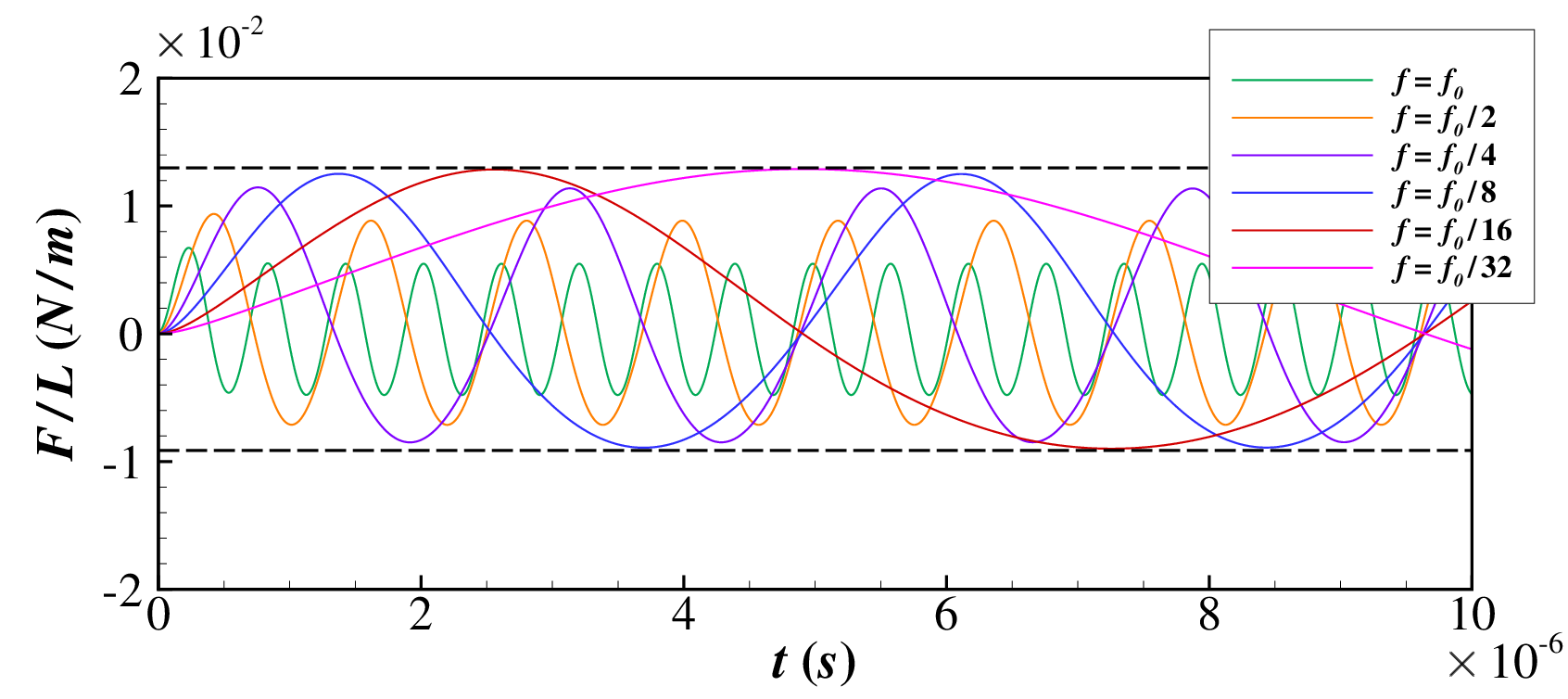}
    }	
	\caption{\label{Fcompare-df} Time evolutions of damping forces acting on a micro-beam with (a) $U_{max}=0.0674m/s$, (b) $U_{max}=0.27m/s$ and (c) $U_{max}=1.08m/s$. Two $Kn^{(h)}$ numbers, 0.1 (left) and 1.0 (right), are considered, $f_0$ is the initial test frequency with $f_{0}/(2\pi)=1.685MHz$, and the values of dashed lines shown in figures are obtained from Fig.~\ref{kn-press-velocity} at corresponding velocities.}
\end{figure}

\begin{figure}
	\centering
	\subfigure[]{
		\includegraphics[width=0.4 \textwidth]{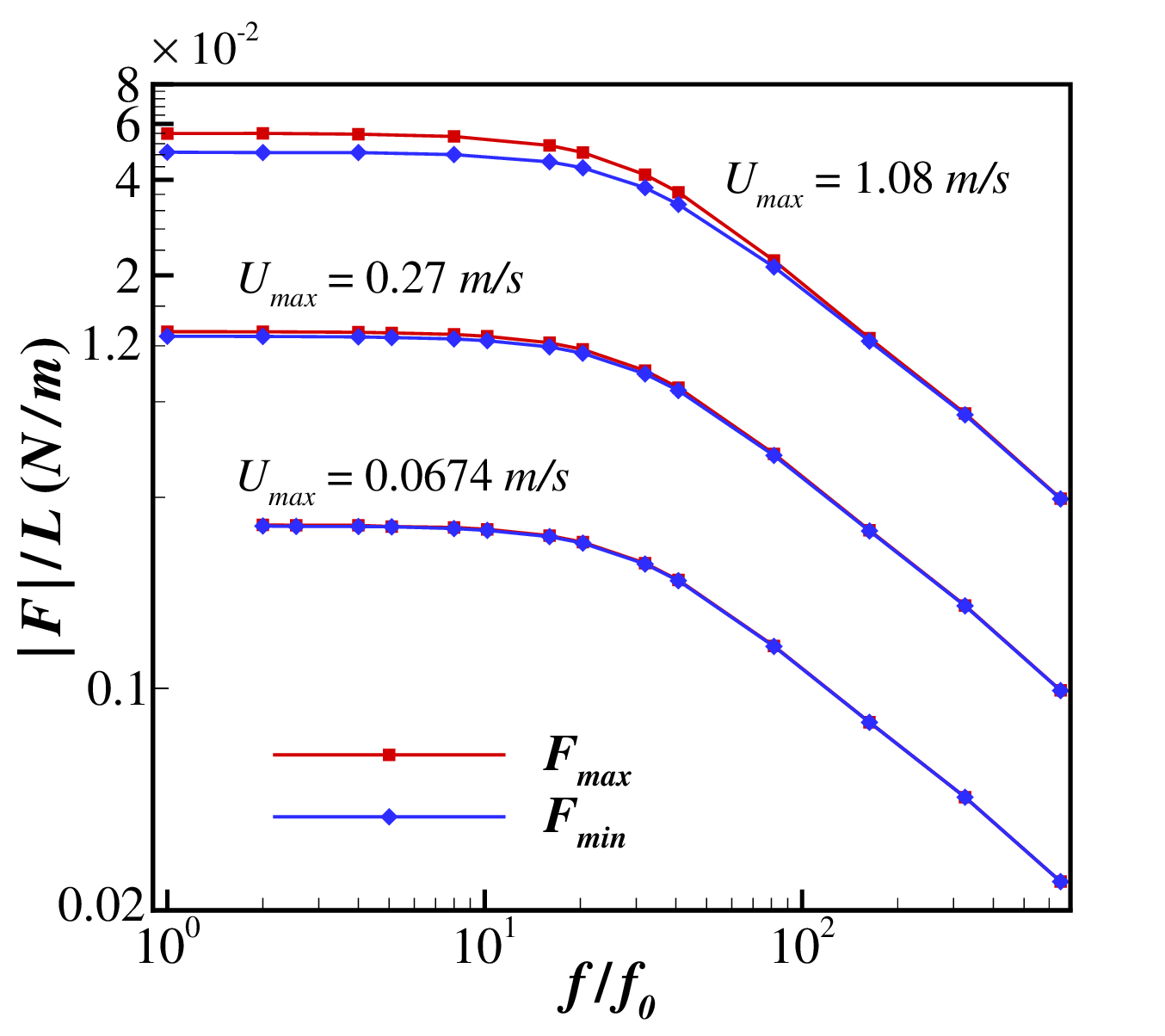}
	}
	\subfigure[]{
		\includegraphics[width=0.4 \textwidth]{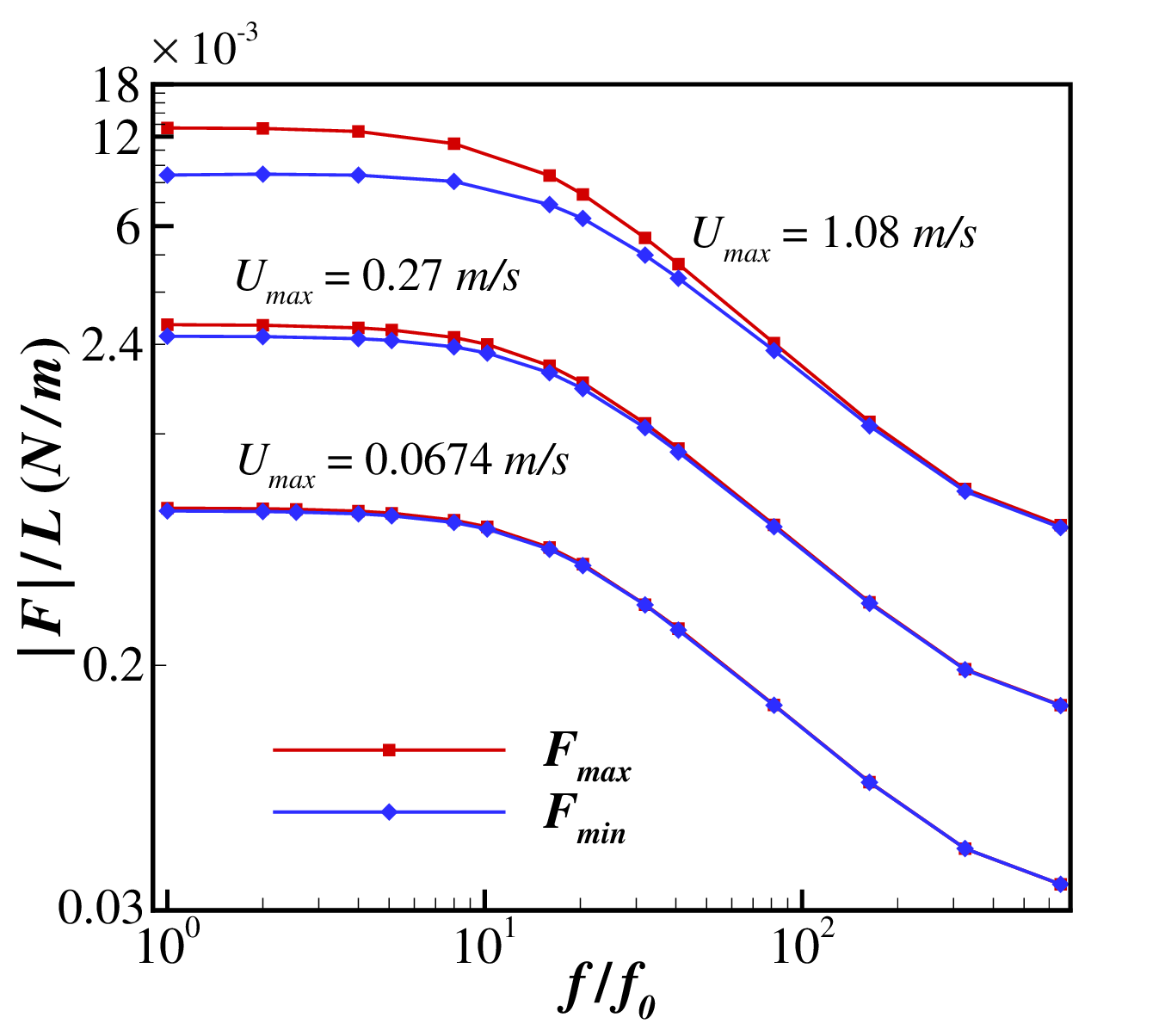}
	}	
	\caption{\label{kn-f-da} Comparisons of the largest damping force acting on a micro-beam at two $Kn^{(h)}$ numbers of (a) 0.1 and (b) 1.0. $f_0$ is the initial test frequency  with $f_0/(2\pi)=0.052656MHz$.}
\end{figure}

In this subsection, the influence of frequency on the damping force is further studied, which is not considered in Ref.~\cite{Chigullapalli2012Non}. By giving a maximum moving velocity $U_{max}$, a motion form of micro-beam is assumed:
\begin{equation}\label{oscillateing-form}
    y=Asin(\frac{U_{max}}{A}t),
\end{equation}
where $A$ is the oscillation amplitude of a moving micro-beam. Then with Eq.~\eqref{oscillateing-form}, the instantaneous relative height of gap $x_1$ in Eq.~\eqref{Guo_model} can be obtained. Shown in Fig.~\ref{guo_thoery}, with Guo et al.'s compact model~\cite{Guo2009Compact}, moving velocities $U$, damping coefficients $c_f$ and damping forces $F$ at two oscillation amplitudes, $0.02h_0$ and $0.16h_0$, are compared, where $h_0=1.0\times10^{-6}m$ is the initial gap height. Besides, a small value of $U_{max}$, $0.0674m/s$, is used; similar results will be obtained at high moving velocities due to the linear assumption of model. Although the damping coefficient has obvious variation during the oscillation at a large amplitude than that at a small one (Fig.~\ref{guo-theory2}), the maximum and minimum values of damping forces are almost the same (Fig.~\ref{guo-theory3}). So, the oscillation frequency does not influence the damping force in Guo et al.'s compact model. To further verify this model, a series of flows at $U_{max}$ of $0.0674m/s$, $0.27m/s$, and $1.08m/s$ is simulated, and the oscillating velocity is given by
\begin{equation}\label{ufunc}
	U=-U_{max}sin(ft),
\end{equation}
where $f$ is the oscillation frequency. Shown in Fig.~\ref{Fcompare-df}, the maximum damping forces at the initial test frequency $f_0$ ($f_{0}/(2\pi)=1.685MHz$) are lower than that obtained by the constant motion velocities (dashed lines shown in figures). And by decreasing the frequency $f$, the amplitudes of damping force gradually converges to the dashed lines, and the nonlinear phenomenons also can be observed as the absolute values of the maximum and minimum values of damping force are not equal to each others at a high oscillating velocity. So, Guo et al.'s compact model is only available at a low oscillation frequency. Finally, Fig.~\ref{kn-f-da} shows the largest damping forces at different oscillation frequencies. For the oscillation frequency lower than a threshold value, the largest damping forces will converge to a constant number, and the nonlinear phenomenon can be observed at a higher oscillating velocity. By increasing the oscillation frequency, the damping force decreases. Besides, when the oscillation frequency is larger than that threshold value, it seems that a linear relation can be found between the different oscillating velocities. So, further work can be continued to construct a modified compact model which is to consider the influence of oscillation frequency.

\subsection{\color{blue}Coupled framework: Squeeze-film damping force at the linear motion}\label{SecIII-II}
In this section, with the coupled FSI framework, the rarefied gas damping force acting on a micro-beam by the forced or free oscillation under the linear motion is studied. Some comparisons of result between decoupled method and coupled framework also are presented.
\subsubsection{Squeeze-film damping force at forced oscillation}
\begin{figure}
	\centering
	\subfigure[]{
		\includegraphics[width=0.55 \textwidth]{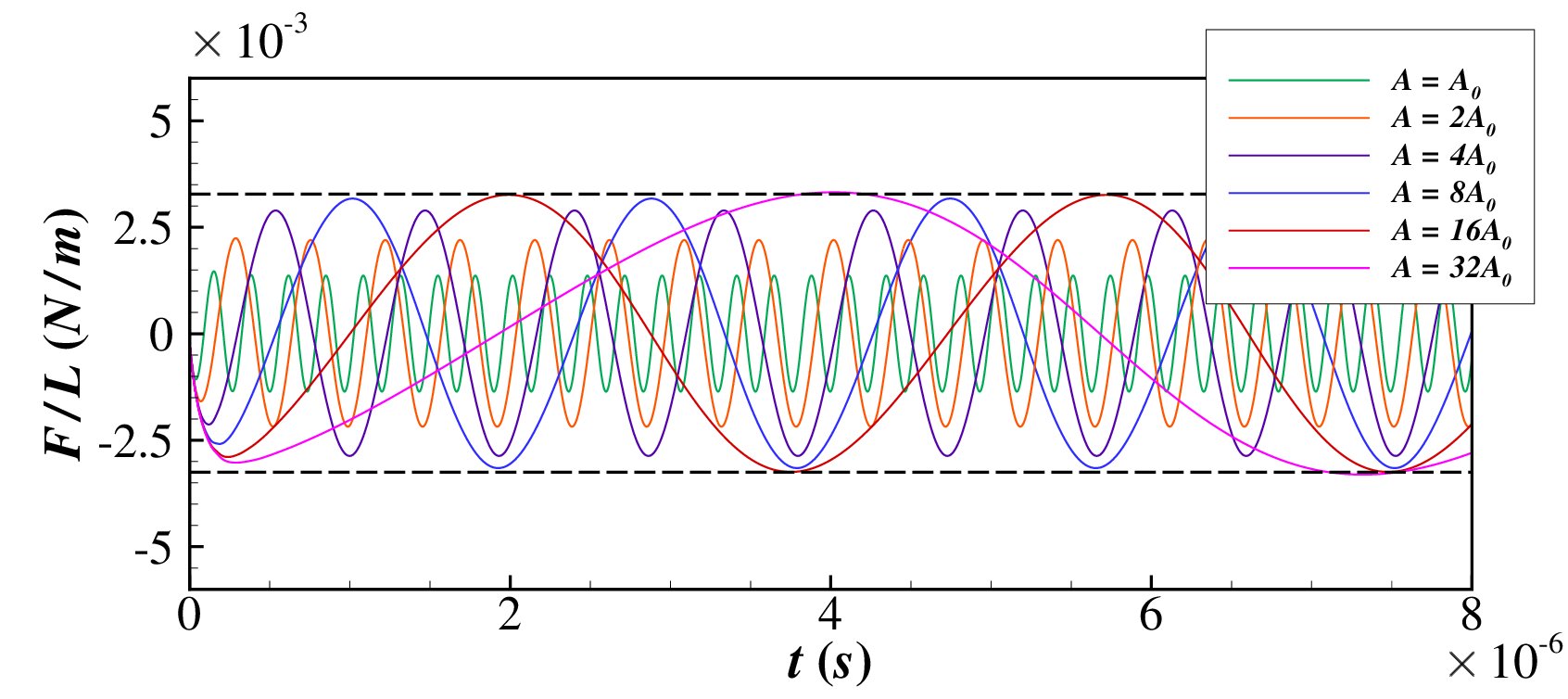}
		\includegraphics[width=0.55 \textwidth]{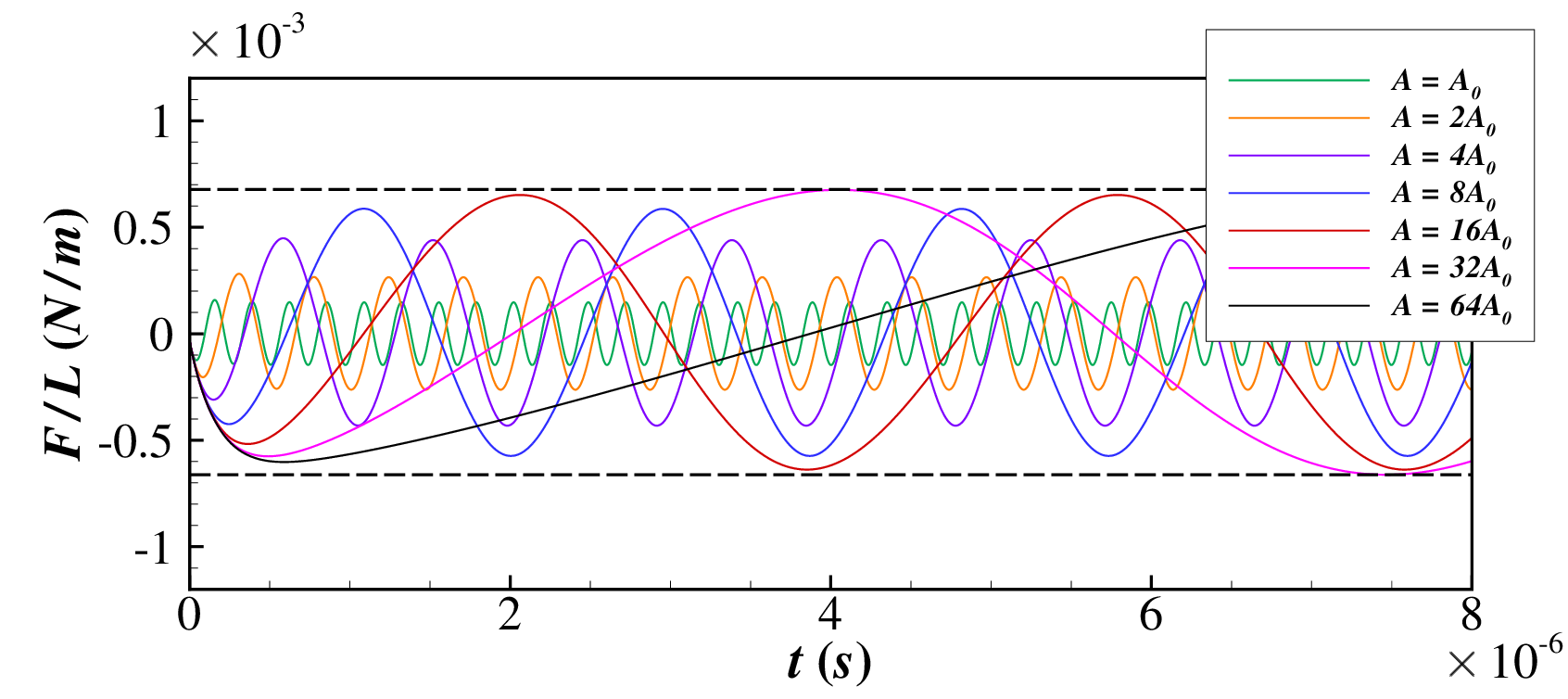}
	}
	\subfigure[]{
		\includegraphics[width=0.55 \textwidth]{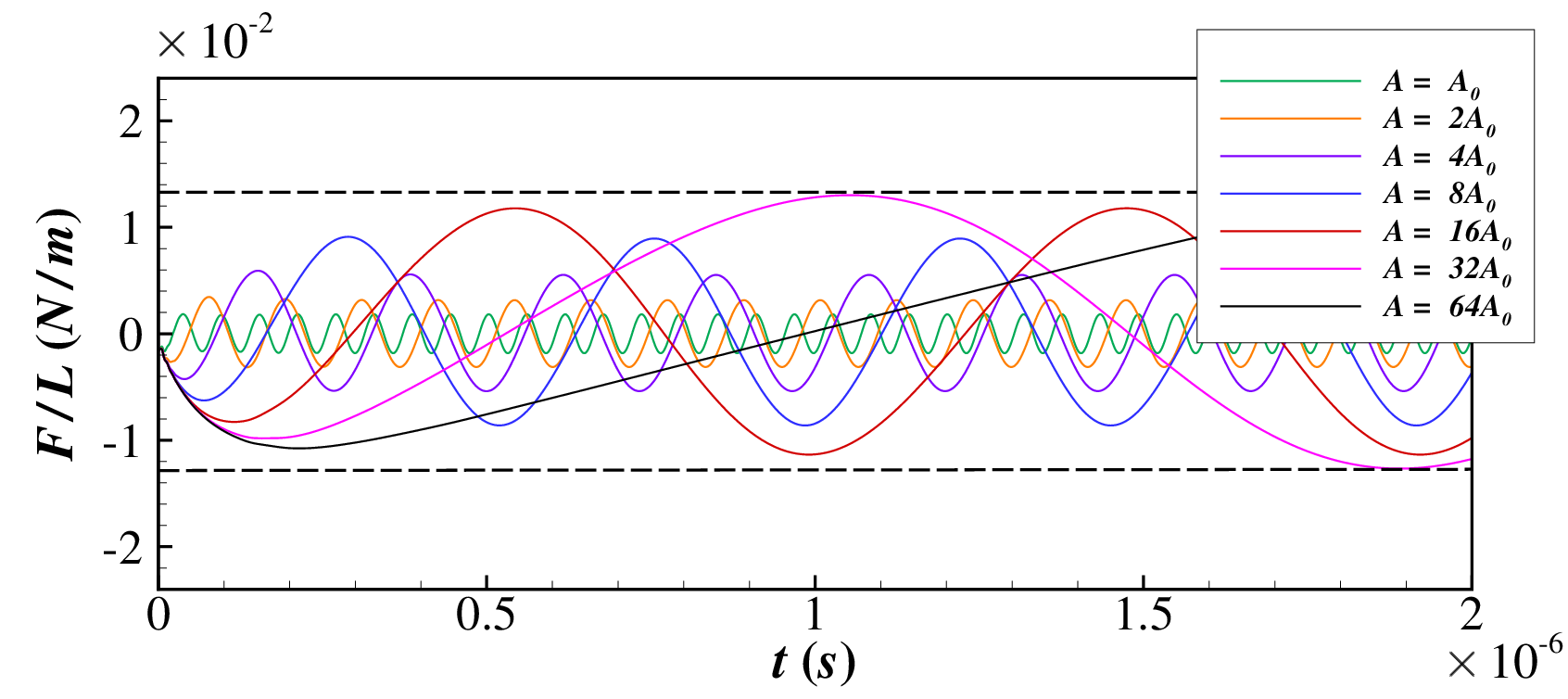}
		\includegraphics[width=0.55 \textwidth]{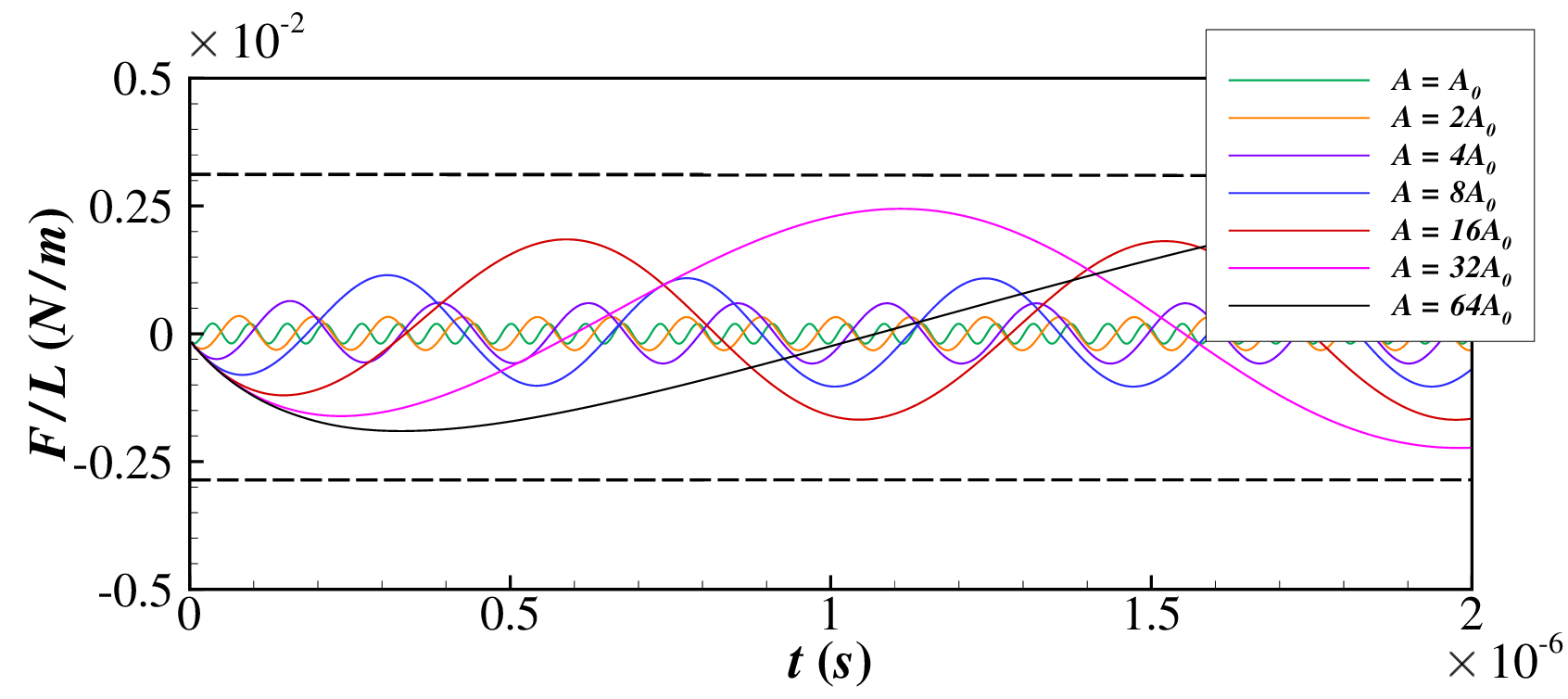}
	}
	\subfigure[]{
		\includegraphics[width=0.55 \textwidth]{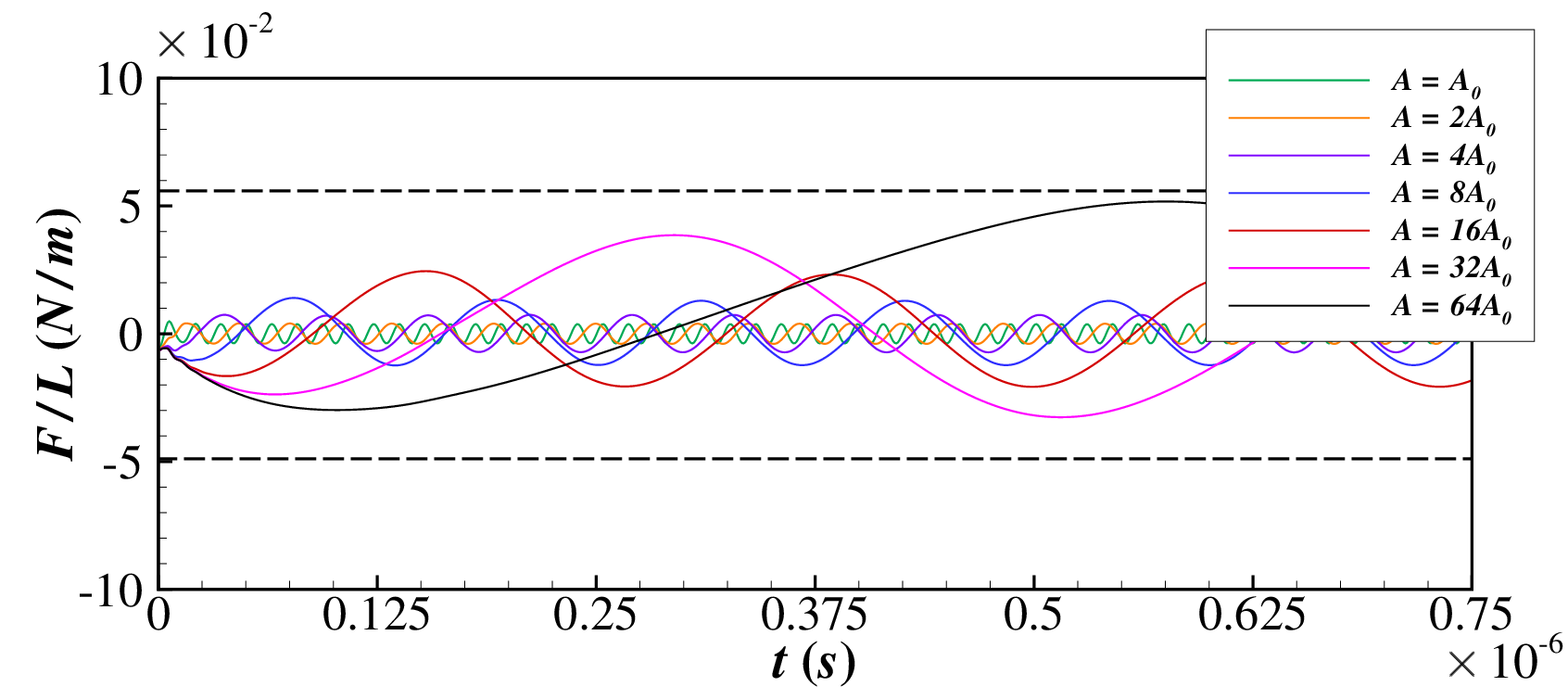}
		\includegraphics[width=0.55 \textwidth]{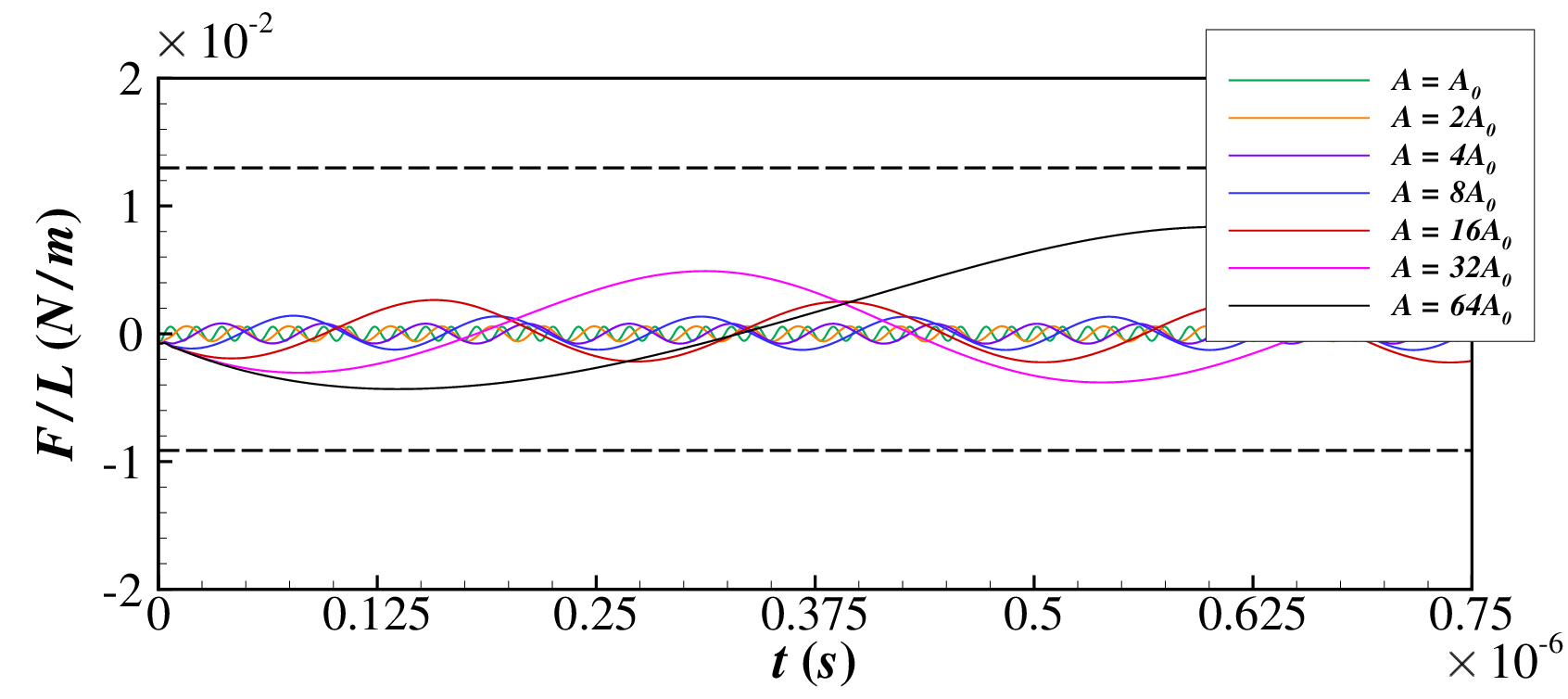}
	}	
	\caption{\label{Fcompare-da} Time evolutions of damping forces at (a) $U_{max}=0.0674m/s$, (b) $U_{max}=0.27m/s$ and (c) $U_{max}=1.08m/s$. Two $Kn^{(h_0)}$ numbers, 0.1 (left) and 1.0 (right), are considered; the values of dashed lines shown in figures are obtained from Fig.~\ref{kn-press-velocity} at corresponding velocities and the reference amplitude $A_0$ equals to $0.0025h_0$.}
\end{figure}

For the forced oscillation of a micro-beam, the motion form given by Eq.~\eqref{oscillateing-form} is also used here, and the minimum reference amplitude $A_0$ is $0.0025h_0$. With this equation, a smaller value of oscillation amplitude $A$ generates a higher oscillation frequency; and for the same oscillation amplitude, a larger maximum oscillating velocity $U_{max}$ also generates a higher oscillation frequency. Fig.~\ref{Fcompare-da} shows the time evolutions of damping forces at different maximum moving velocities and amplitudes. It is clear that for all the considered velocities, a smaller damping force will be generated by a high frequency and vice versa. And at a small value of moving velocity ($U_{max}=0.0674m/s$), the maximum and minimum values of damping force are almost the same. By increasing the oscillation amplitude, the corresponding values gradually converge to the results obtained by the decoupled method described in Sec.~\ref{SecIII-I}. Furthermore, there exists a threshold value that the largest damping force does not change when the oscillation frequency is lower than that value. For a higher oscillating velocity, $U_{max}=0.27m/s$ or $1.08m/s$, the nonlinear damping phenomenon can be observed as the maximum and minimum values of damping force are not equal to each others. Besides, the convergent values of damping forces will be much higher than the results obtained by the decoupled method, especially at a more higher oscillating velocity, $1.08m/s$. So, it proves again that the traditional compact model~\cite{Guo2009Compact} is only available at the low velocity and frequency of oscillation.

\begin{figure}
	\centering
	\subfigure[]{
		\includegraphics[width=0.4 \textwidth]{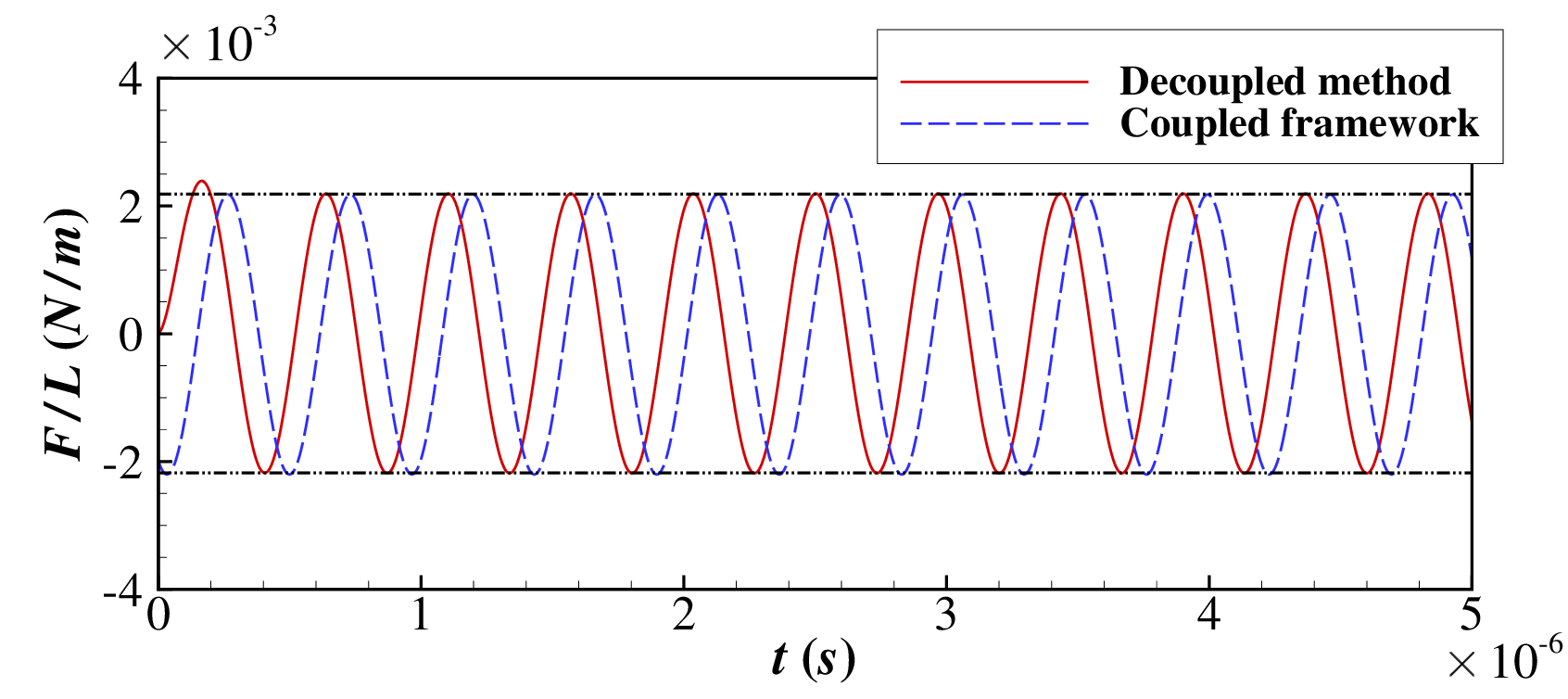}
		\includegraphics[width=0.4 \textwidth]{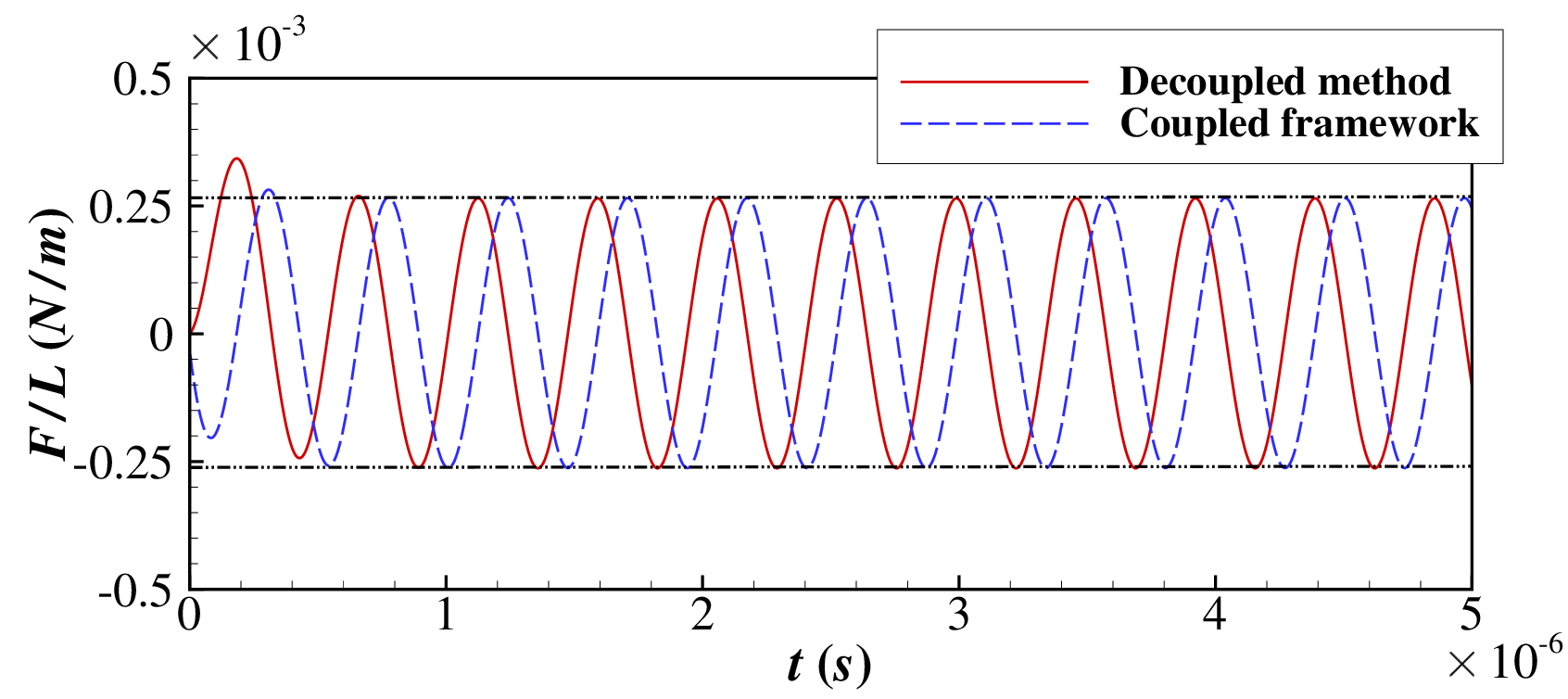}
		\label{fa-comp1}
	}
	\subfigure[]{
		\includegraphics[width=0.4 \textwidth]{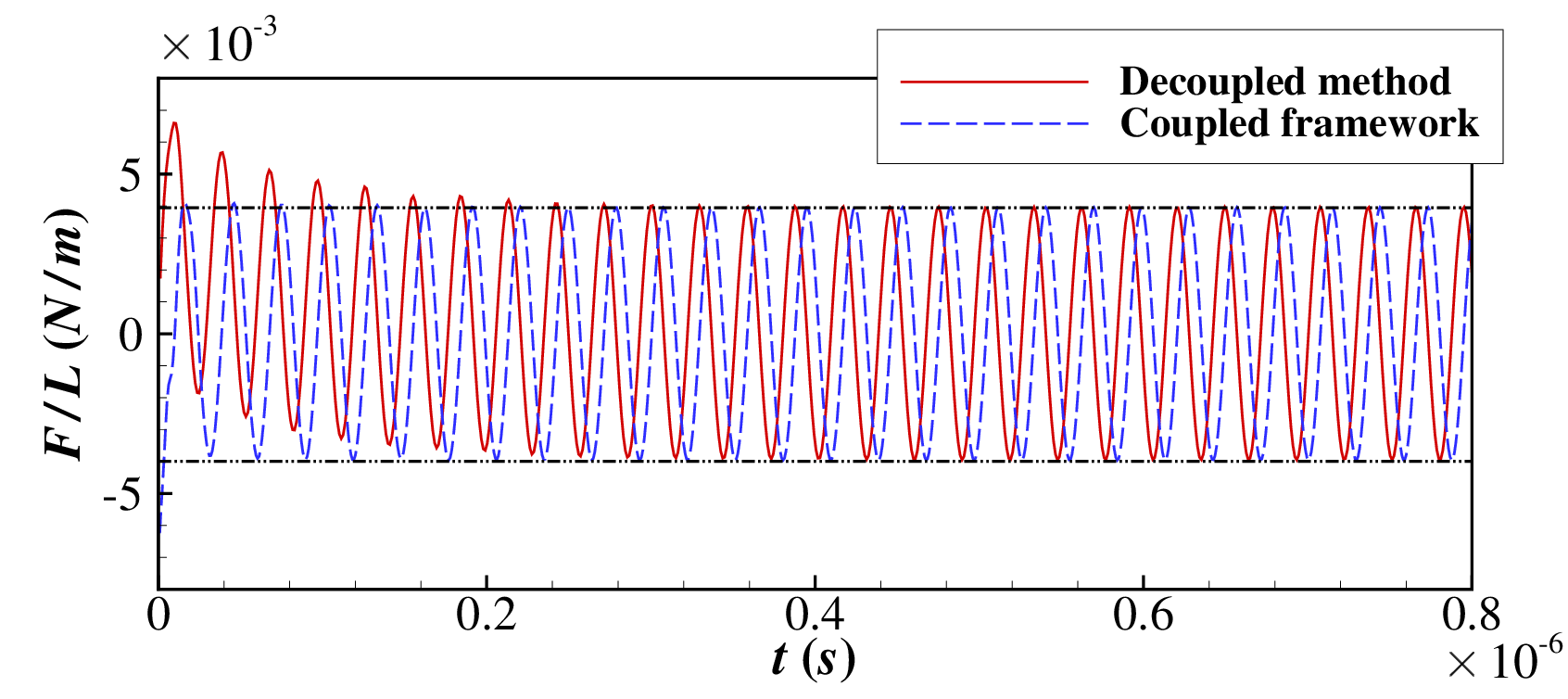}
		\includegraphics[width=0.4 \textwidth]{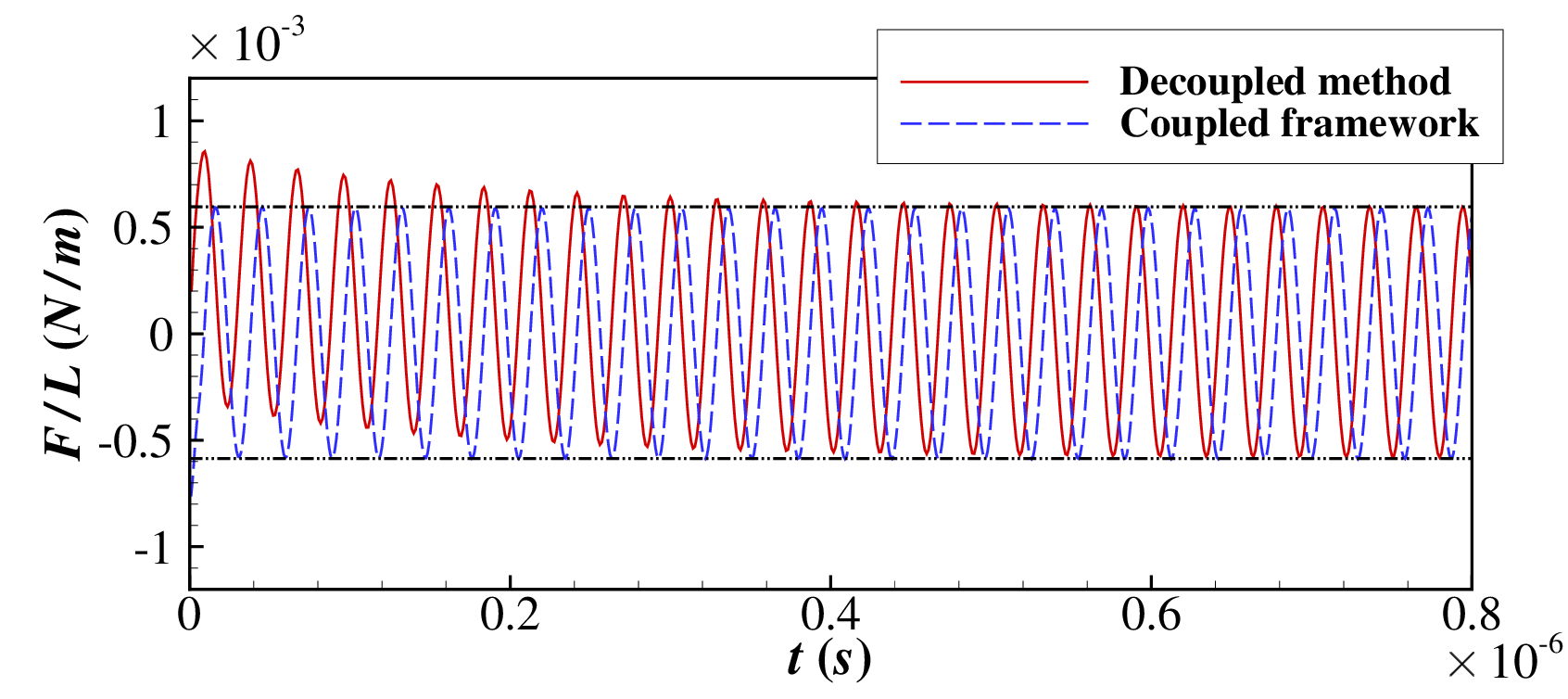}
		\label{fa-comp2}
	}
	\subfigure[]{
		\includegraphics[width=0.4 \textwidth]{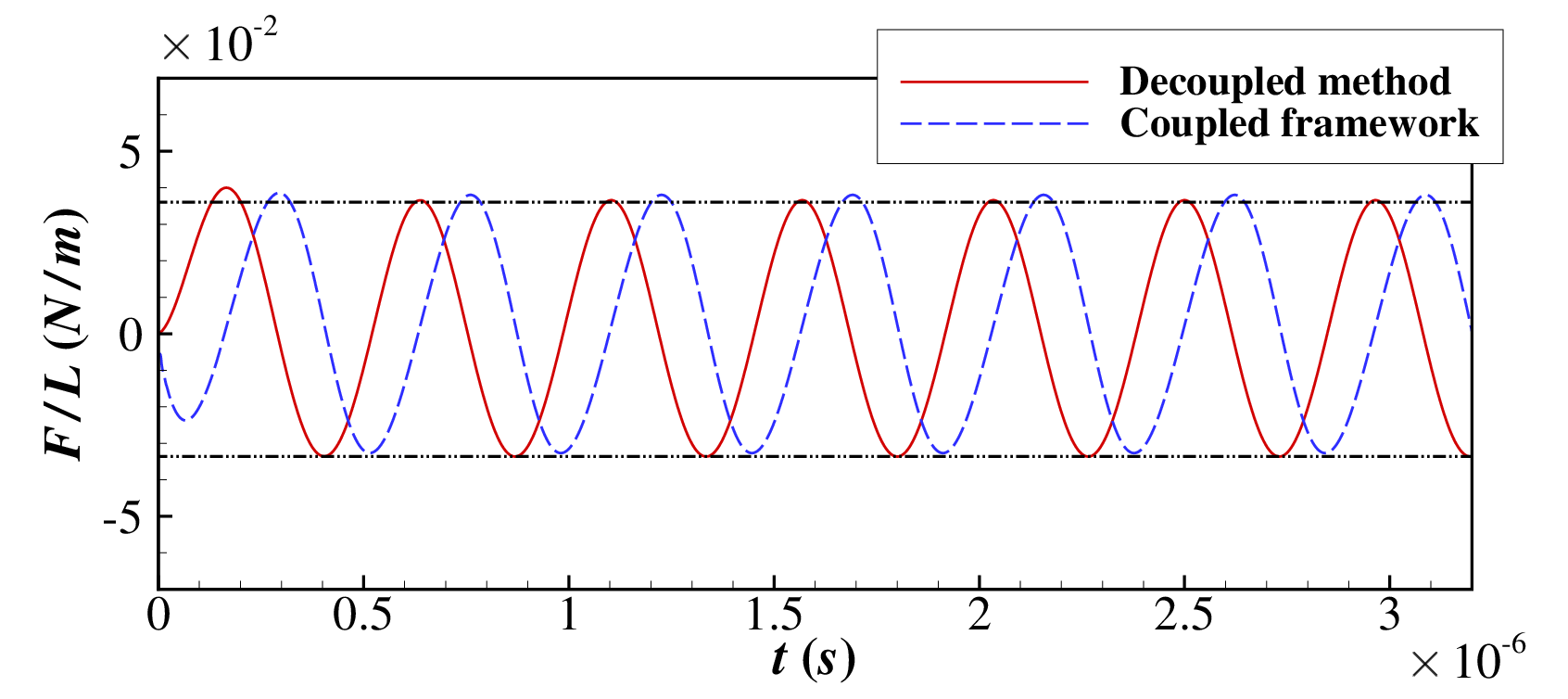}
		\includegraphics[width=0.4 \textwidth]{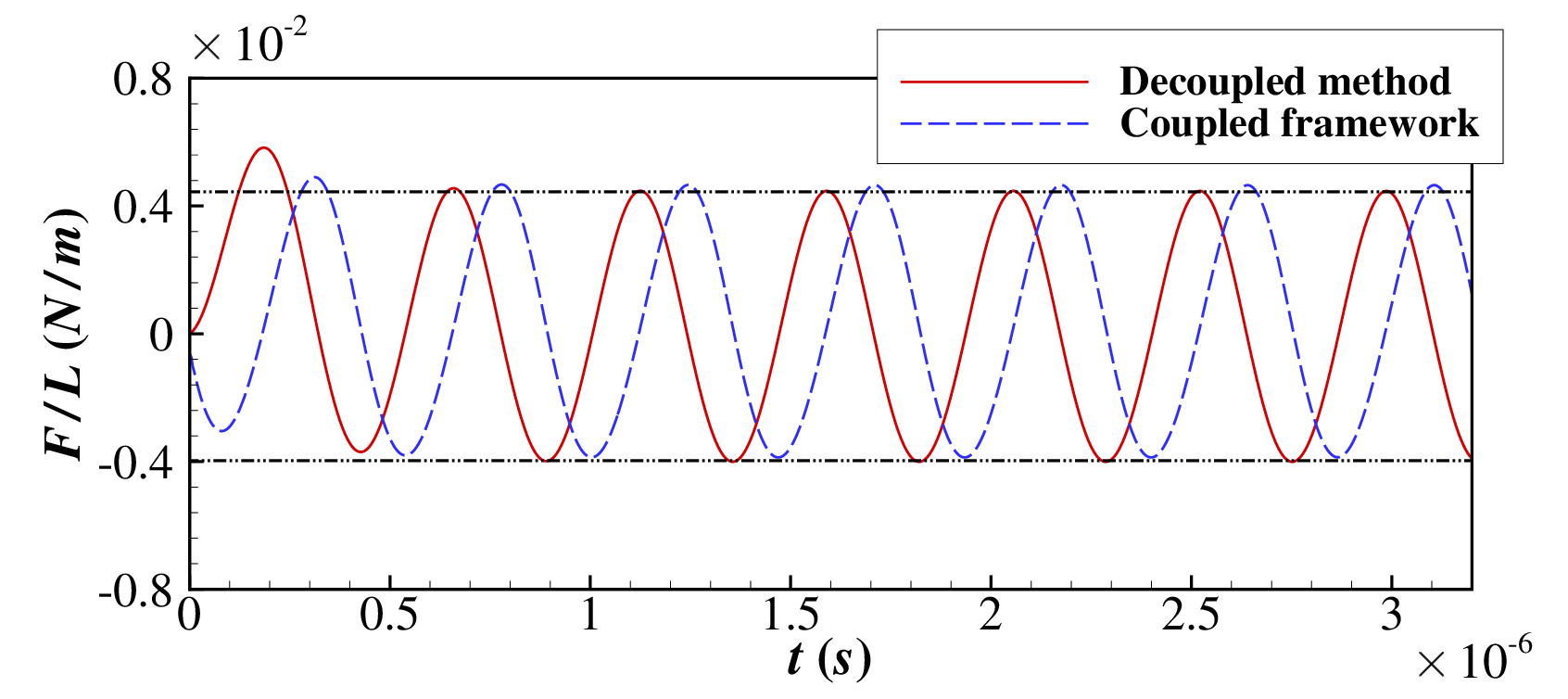}
		\label{fa-comp3}
	}
	\subfigure[]{
		\includegraphics[width=0.4 \textwidth]{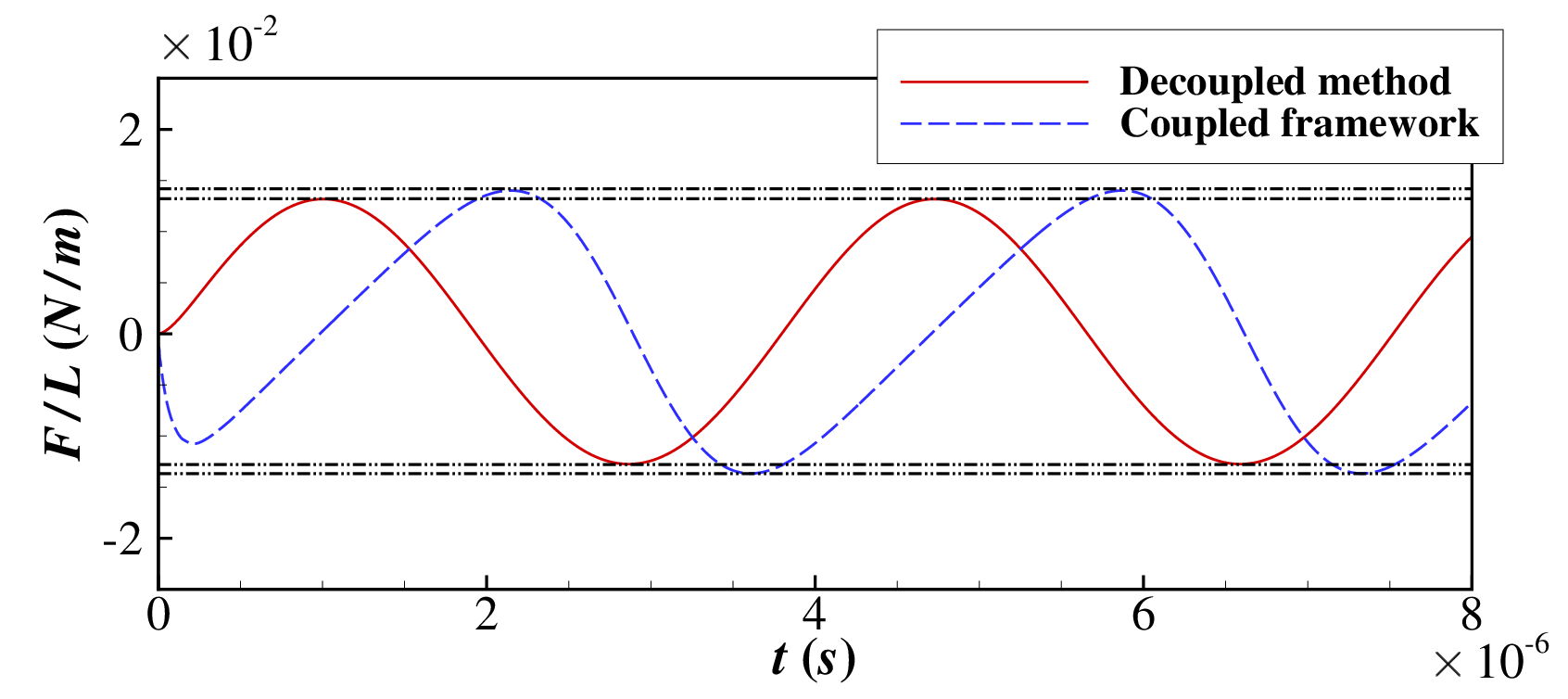}
		\includegraphics[width=0.4 \textwidth]{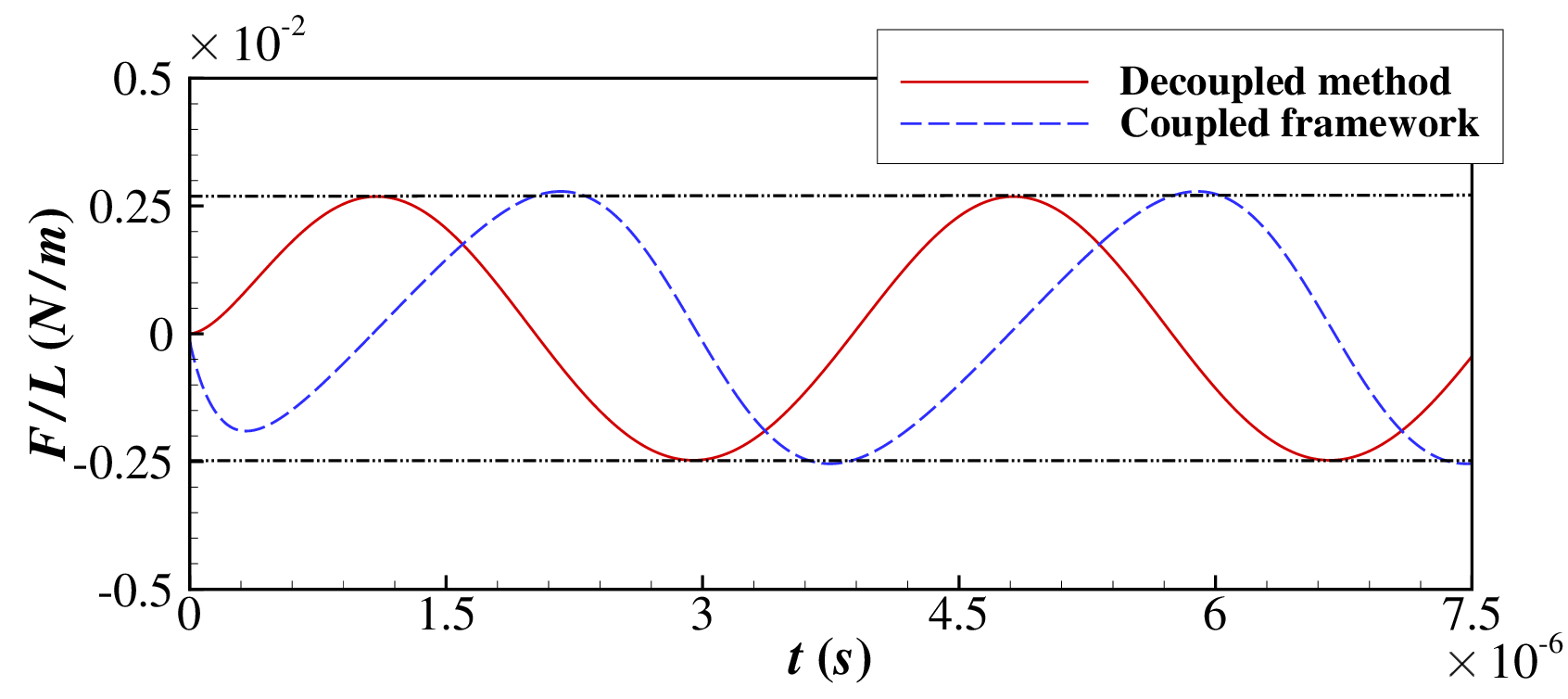}
		\label{fa-comp4}
	}
	\subfigure[]{
		\includegraphics[width=0.4 \textwidth]{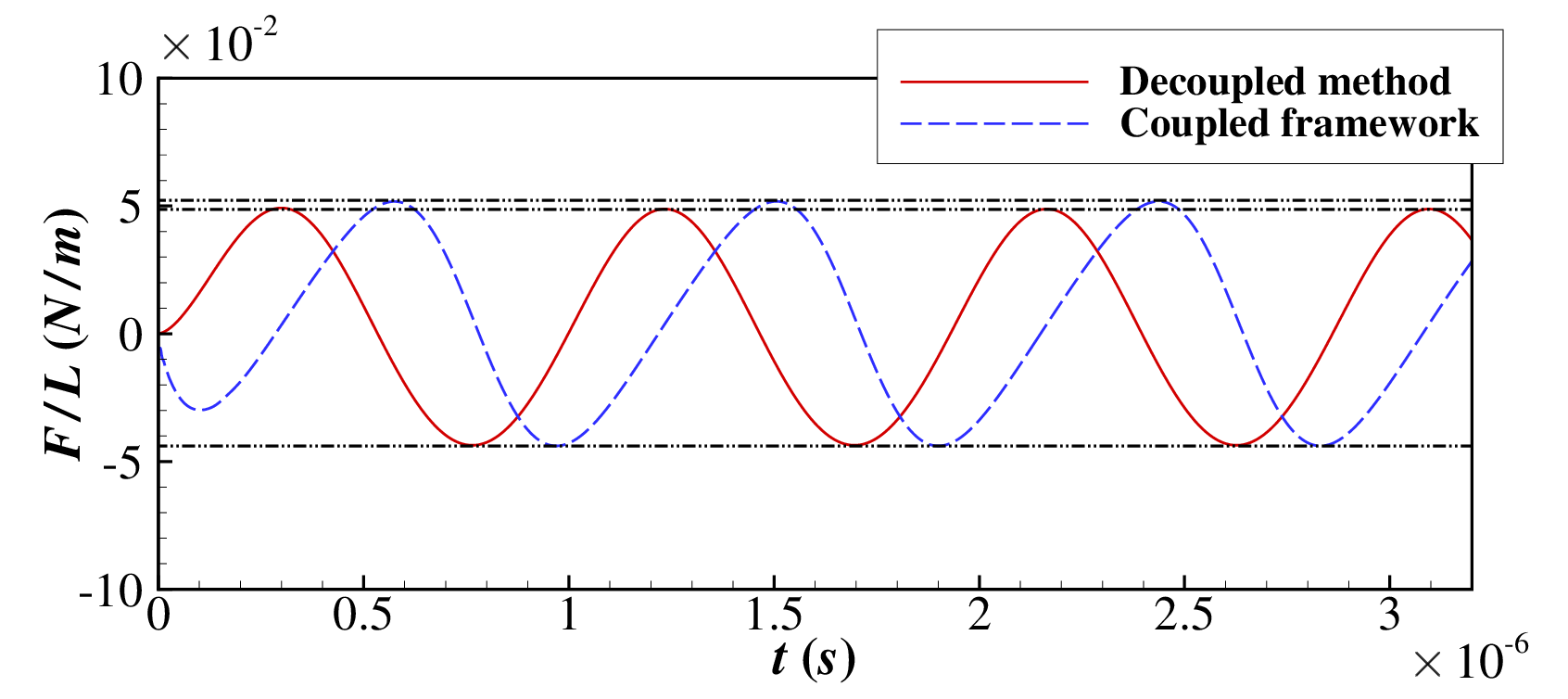}
		\includegraphics[width=0.4 \textwidth]{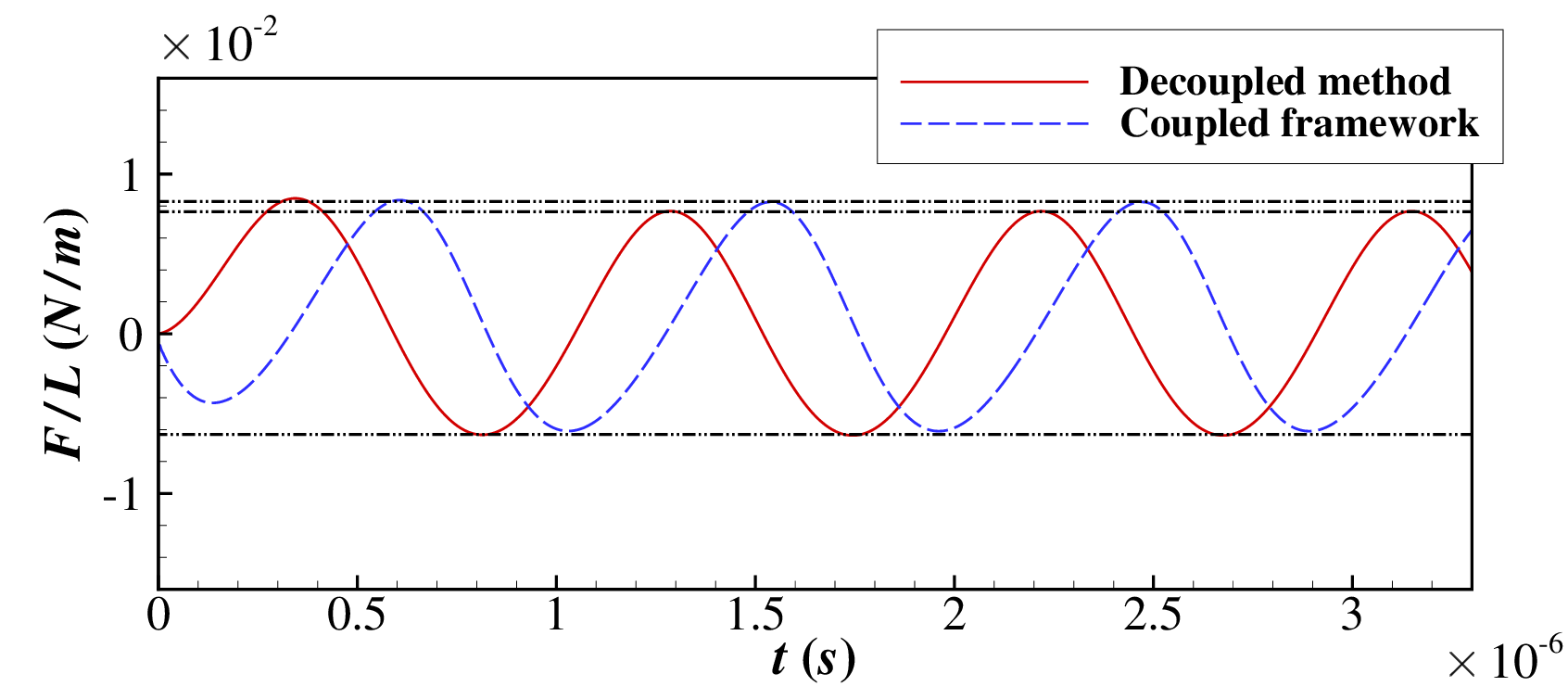}
		\label{fa-comp5}
	}
	\caption{\label{fa-comp} Comparisons of time evolutions of damping forces at two $Kn^{(h_0)}$ numbers, 0.1 (left) and 1.0 (right). The maximum oscillating velocities $U_{max}$ and amplitudes $A$ are set to (a) $U_{max}=0.0674m/s$ and $A=0.005h_0$, (b) $U_{max}=1.08m/s$ and $A=0.005h_0$, (c) $U_{max}=1.08m/s$ and $A=0.08h_0$, (d) $U_{max}=0.27m/s$ and $A=0.16h_0$, and (e) $U_{max}=1.08m/s$ and $A=0.16h_0$, respectively (the dash dot lines shown in figures are used for comparison).}
\end{figure}

\begin{figure}
	\centering
	\subfigure[]{
		\includegraphics[width=0.4 \textwidth]{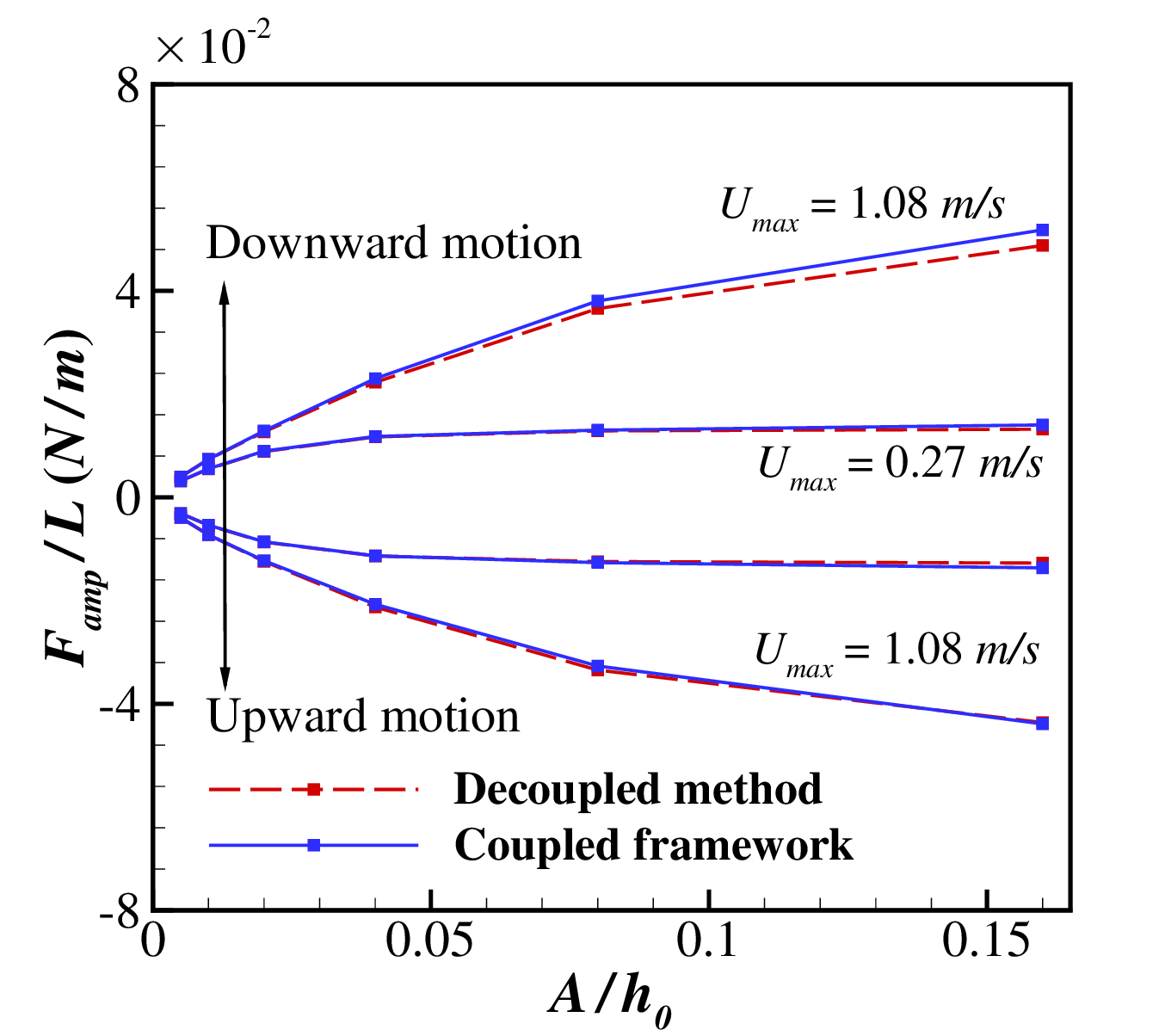}
	}
	\subfigure[]{
		\includegraphics[width=0.4 \textwidth]{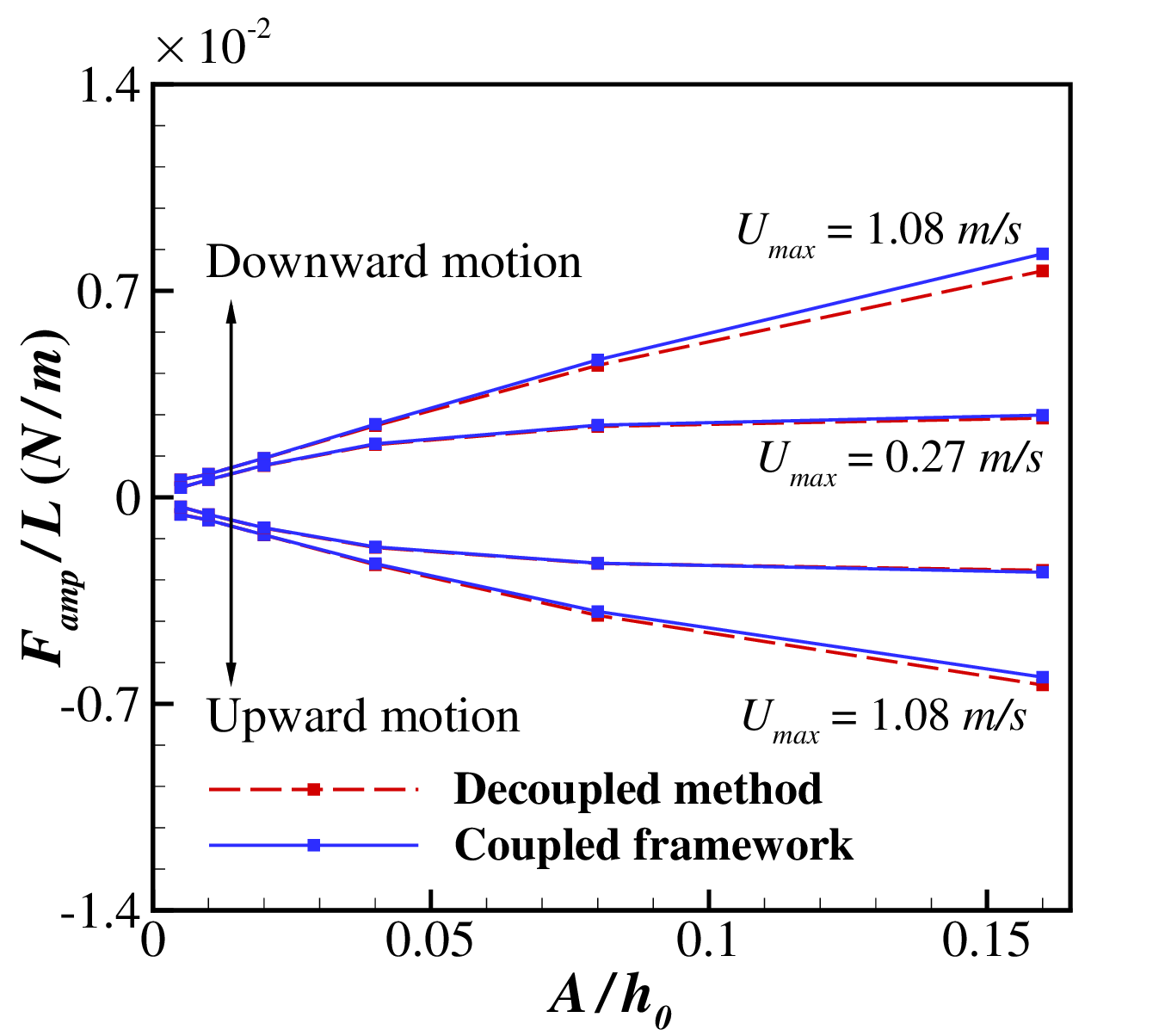}
	}	
	\caption{\label{kn-fa-da} Comparisons of the maximum and minimum values of damping force $F_{amp}$ at $Kn^{(h_0)}$ numbers of (a) 0.1 and (b) 1.0.}
\end{figure}

Then, the results obtained by decoupled method (Eq.~\eqref{ufunc}) and coupled framework (Eq.~\eqref{oscillateing-form}) are compared at the same oscillation frequency. Fig.~\ref{fa-comp} shows the comparisons of time evolutions of damping force. For the oscillation at a small value of amplitude (Figs.~\ref{fa-comp1} and~\ref{fa-comp2}), the largest damping forces are almost the same, so the influences of $U_{max}$ on the damping force are not obvious. As a result, for the high-frequency oscillation of a micro-beam, the decoupled method still exhibits a good performance to predict the damping force. For the oscillation at a high oscillating velocity $U_{max}=1.08m/s$ and a moderate oscillation amplitude $A=0.08h_0$ (Fig.~\ref{fa-comp3}), due to the influence of displacement of a micro-beam, the maximum of damping force obtained by the coupled framework is a little higher than that by the decoupled method (downward moving direction) and this tendency is inverse for the minimum one (upward moving direction). So, the advantage of the coupled framework is more accurate to predict the damping force at that computational condition. And for the oscillation at a high velocity and low frequency (Figs.~\ref{fa-comp4} and~\ref{fa-comp5}), the differences of results obtained by two methods are obvious. The sinusoidal shape of oscillation of damping force can not be maintained, and more larger damping force will be obtained by the coupled framework. So for the oscillation at a large displacement and low frequency, the decoupled method can not predict the damping force correctly. {\color{blue}Fig.~\ref{kn-fa-da} shows the comparisons of amplitude of damping force at different computational conditions. In consideration of the computational cost, a cost-effective framework can be adopted between these two methods in the practical application.}

\subsubsection{Squeeze-film damping force at free oscillation}\label{linear-free-oscillation}
Next, the free oscillation problem of a micro-beam is considered. By giving an external excitation force, Eq.~\eqref{linear-struct-coulped} is rewritten as
\begin{equation}\label{linear-struct2}
   m\ddot{y}(t)+ky=F_{ext}+F=F_0cos(\omega_{n}t)+F,
\end{equation}
where $F_0$ is the amplitude of external excitation force and $\omega_{n}$ is the frequency of $F_0$. If $\omega_{n}$ is set equal to the natural frequency of structure $\omega$ ($\omega=\sqrt{k/m}$), the resonance phenomenon is excited. Due to the gas damping force, the displacement of the structure is no divergent. So, Eq.~\eqref{linear-struct2} is used in the simulation. Furthermore, to make a comparison, a theoretical solution can be obtained for a low-speed oscillation. Based on the theory of ordinary differential equation, the theoretical solution of a structure dynamic equation:
\begin{equation}\label{linear-struct3}
    m\ddot{y}(t)+c\dot{y}(t)+ky=F_0cos(\omega_{n}t),
\end{equation}
can be obtained in the resonance regime:
\begin{equation}\label{linear-theory}
  y(t)=\frac{F_0}{c\omega_{n}}[sin(\omega_{n}t)-\frac{1}{\sqrt{1-\zeta^2}}e^{-\zeta{\omega_{n}}t}sin(\sqrt{1-\zeta^2}\omega_{n}t)],
\end{equation}
where $\zeta=c/2m\omega$ is the damping ratio. With Eq.~\eqref{linear-theory}, the gas damping force can be verified by the present coupled framework. As the maximum oscillation velocity is $F_0/c$, by giving a damping coefficient, $F_0$ can be determined. Then with the maximum displacement of structure $F_0/c\omega_{n}$, the frequency of external excitation force $\omega_{n}$ can also be determined. Finally, $k$ is obtained by assuming a mass of structure $m$.

\begin{figure}
	\centering
	\subfigure[]{
		\includegraphics[width=0.55 \textwidth]{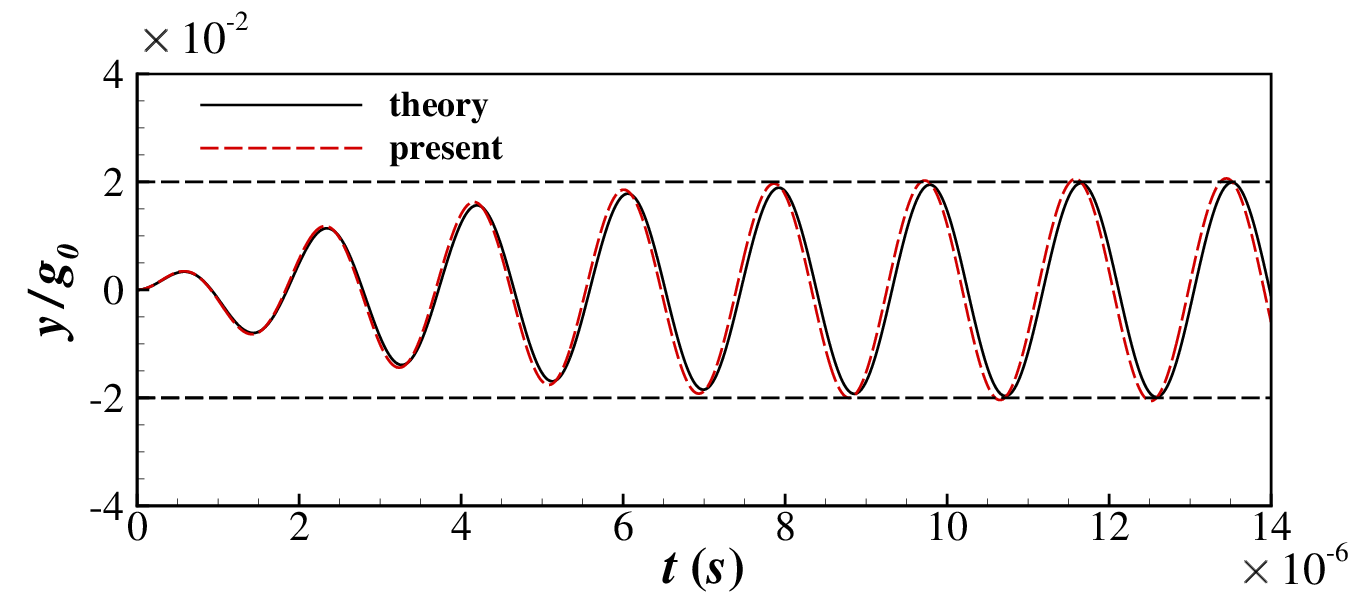}
		\includegraphics[width=0.55 \textwidth]{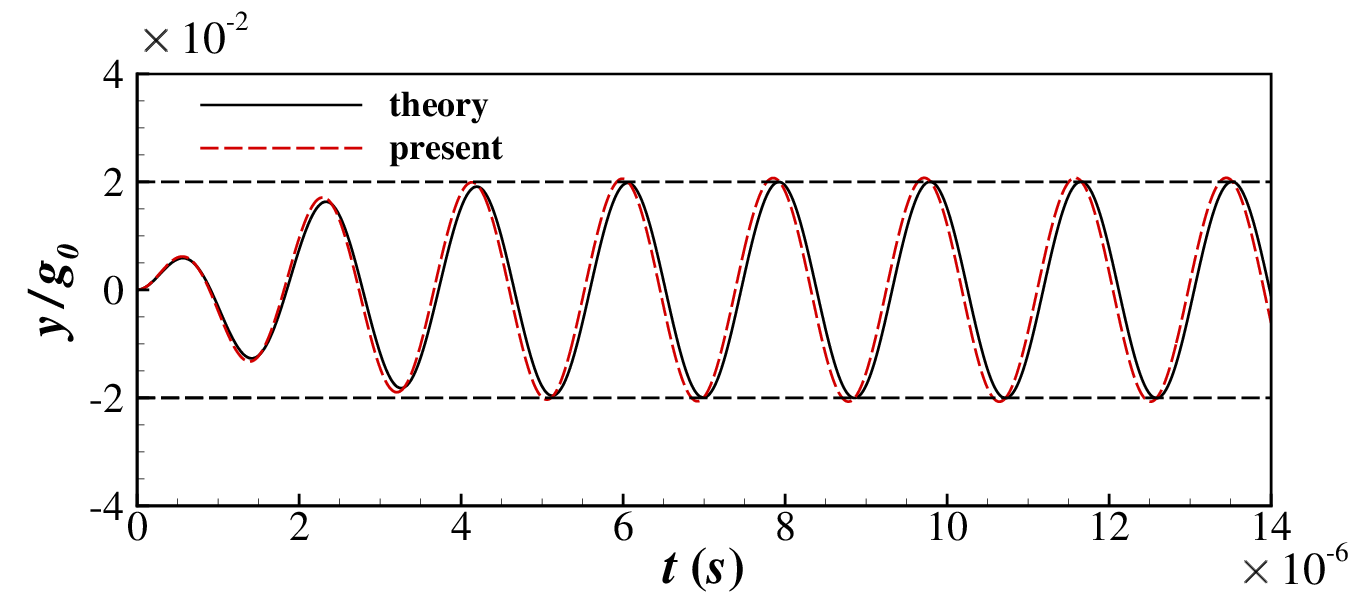}
	}
	\subfigure[]{
		\includegraphics[width=0.55 \textwidth]{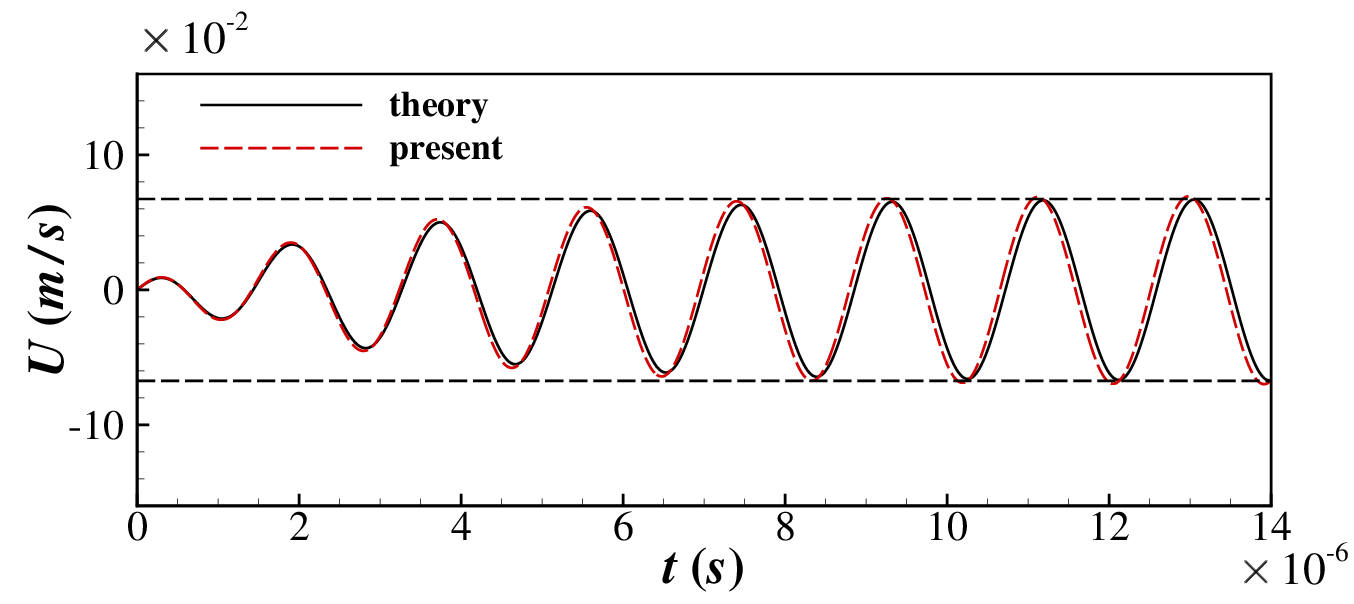}
		\includegraphics[width=0.55 \textwidth]{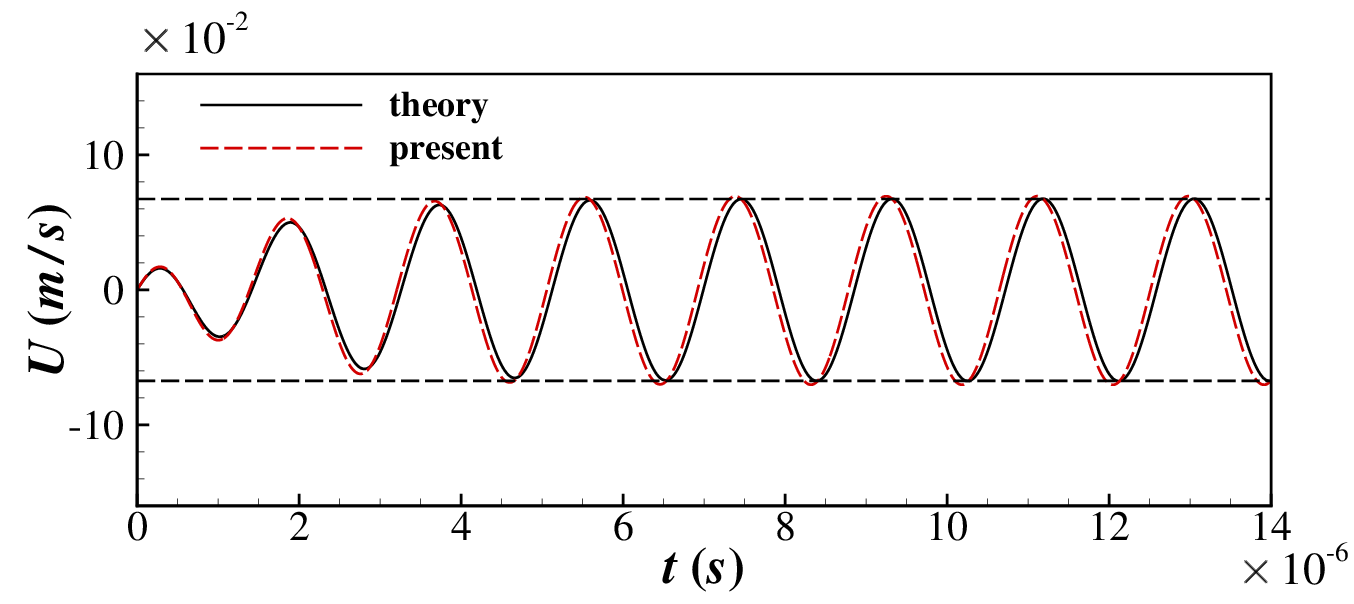}
	}
	\subfigure[]{
	    \includegraphics[width=0.55 \textwidth]{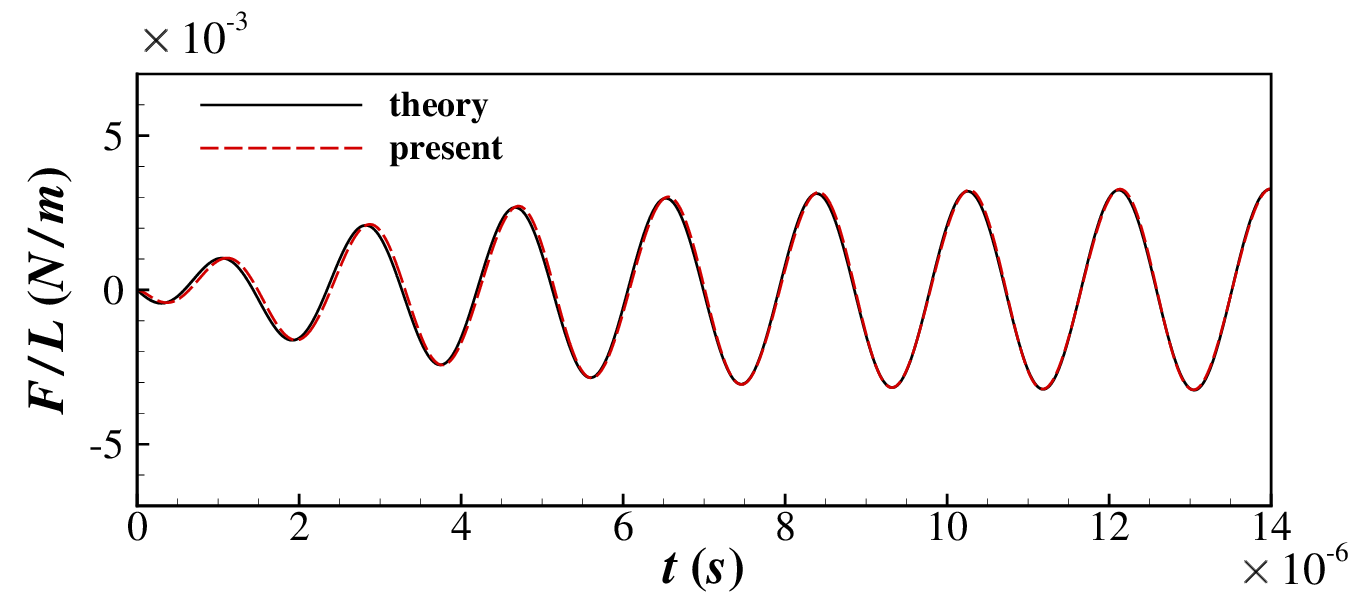}
	    \includegraphics[width=0.55 \textwidth]{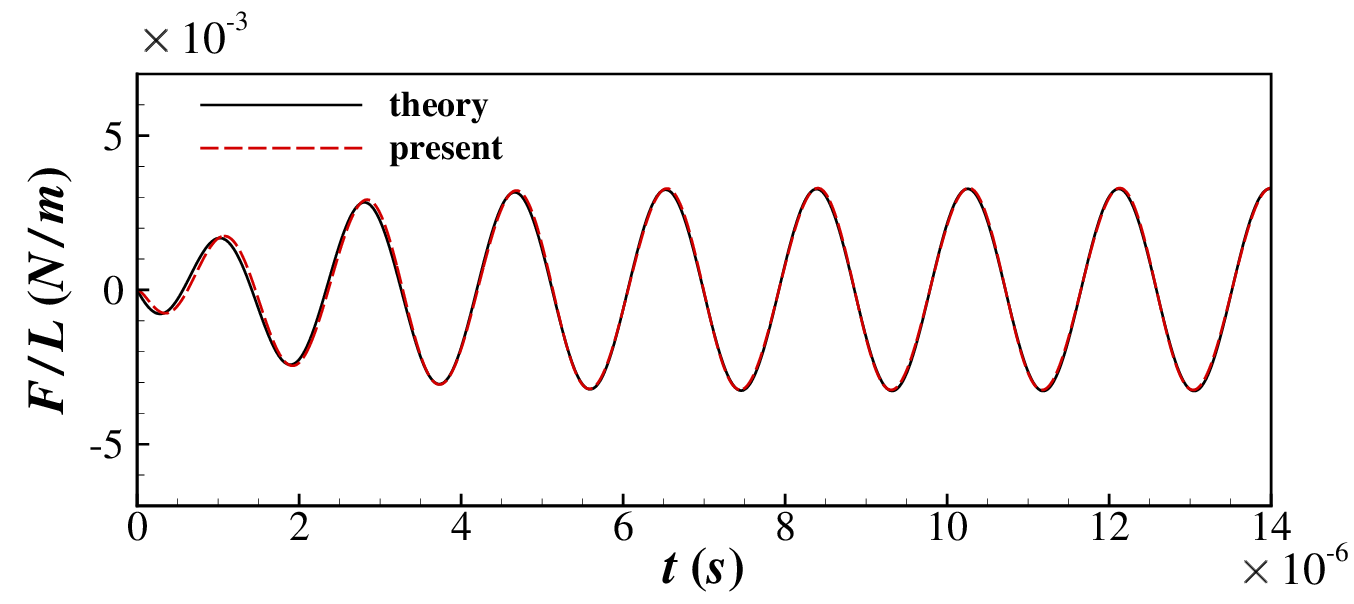}
    }	
	\caption{\label{fsi-u0p0002-kn0p1} Time evolutions of (a) displacement, (b) moving velocity and (c) damping force of a micro-beam with linear free oscillation at $Kn^{(h_0)}=0.1$. The maximum moving velocity is $0.0674m/s$, and two non-dimensional mass of micro-beam $M^*$, 2769.5 (left figures) and 1384.7 (right figures) are considered (the dash lines shown in the figures are the maximum theoretical values used for comparison).}
\end{figure}

\begin{figure}
	\centering
	\subfigure[]{
		\includegraphics[width=0.55 \textwidth]{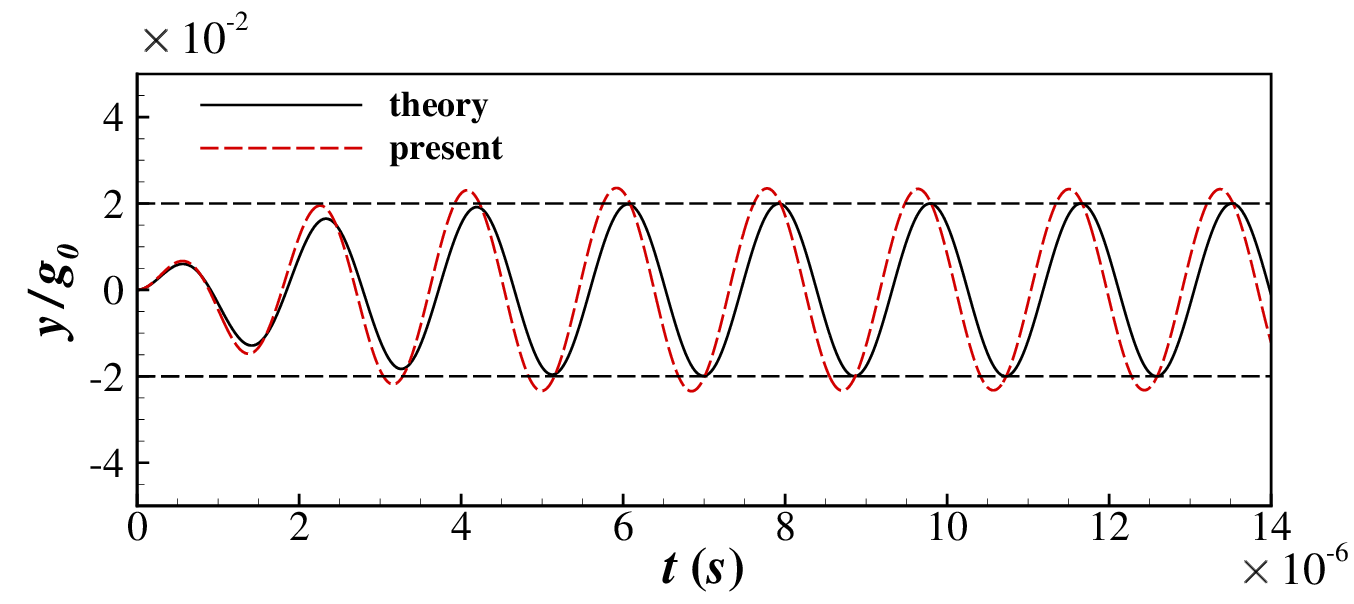}
	}
	\subfigure[]{
		\includegraphics[width=0.55 \textwidth]{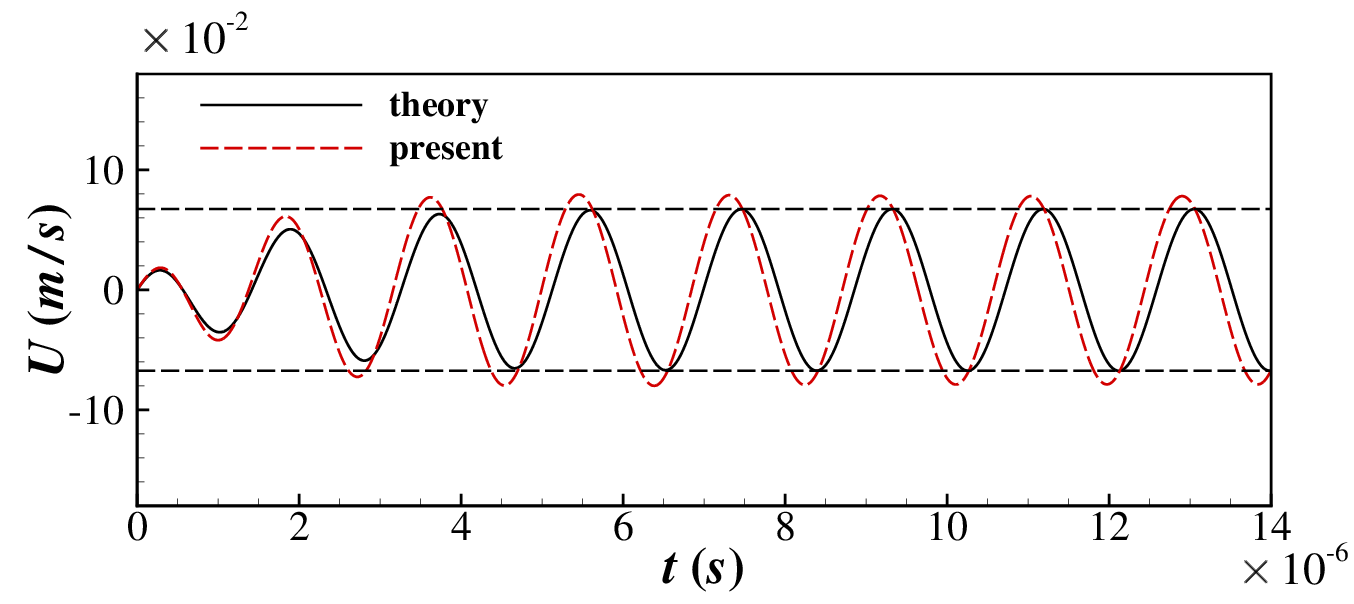}
	}
	\subfigure[]{
		\includegraphics[width=0.55 \textwidth]{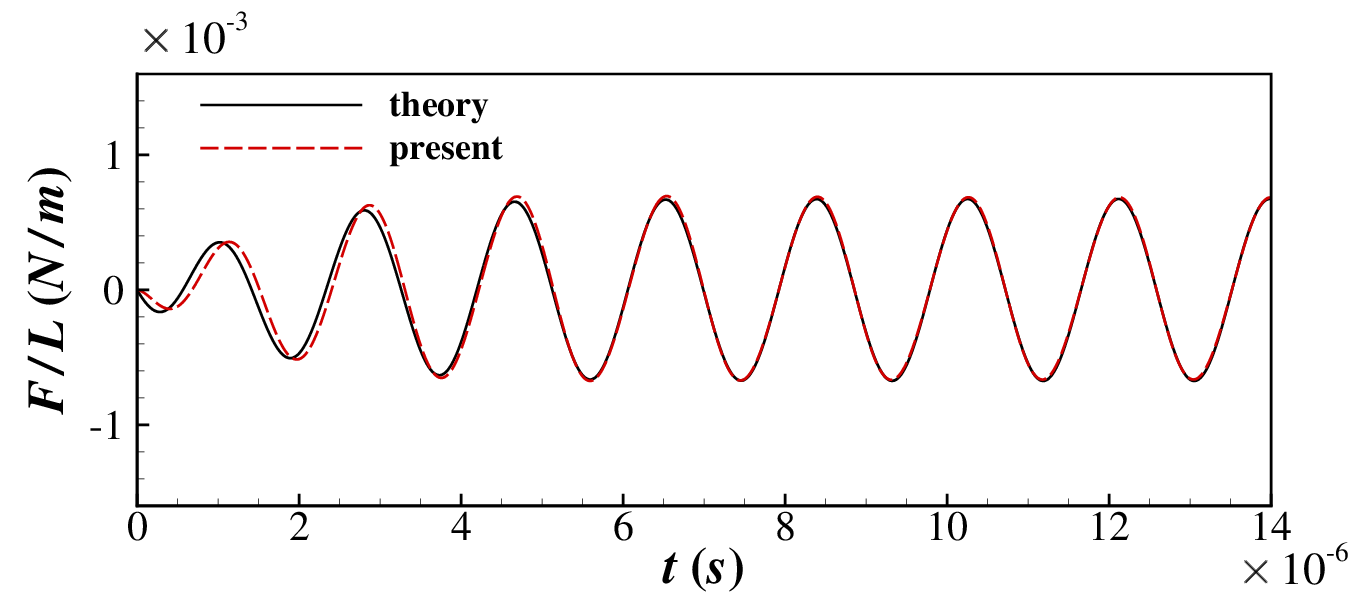}
	}	
	\caption{\label{fsi-u0p0002-kn1} Time evolutions of (a) displacement, (b) moving velocity and (c) damping force of a micro-beam with linear free oscillation at $Kn^{(h_0)}=1.0$. The maximum moving velocity is $0.0674m/s$, and the non-dimensional mass of micro-beam $M^*$ is 2769.5 (the dash lines shown in the figures are the maximum theoretical values used for comparison).}
\end{figure}

Firstly, the free oscillation at a low-speed $U_{max}=0.0674m/s$ and a small amplitude $A=0.02h_0$ is studied. For flow at $Kn^{(h_0)}=0.1$, the equivalent structure damping coefficient $c$ ($c=c_fL$) is $0.0485L$, and the value of $c_f$ is obtained from Fig.~\ref{Cf-dkn}.  For flow at $Kn^{(h_0)}=1.0$, $c$ equals to $0.01L$. Fig.~\ref{fsi-u0p0002-kn0p1} shows the time evolutions of displacement, moving velocity and damping force of a micro-beam. Here, a non-dimensional mass of micro-beam $M^*$ is used with $M^*=m/\rho{LD}$, where $m$ the actual mass of micro-beam and $\rho$ is the density of rarefied gas. In our simulations, two values of $M^*$, 2769.5 and 1384.7, are considered. For flow at the continuum flow regime, those values are 10 and 5, respectively. Generally, the results of numerical simulation agree well with the theoretical solution. And the convergence time of displacement of a micro-beam developing to its maximum value with a heavier mass is slower than that with a lighter one. Shown in Fig.~\ref{Fcompare-da}, for the forced oscillation at $Kn^{(h_0)}=0.1$ and $A=0.02h_0$, the maximum damping force is a little lower than that obtained by the decoupled method (about $3\%$). So $c$ used to calculate the theoretical solution in Eq.~\eqref{linear-theory} is slightly larger than the real one in the simulation, and the numerical result is also slightly larger than that of the theoretical one. For flow at $Kn^{(h_0)}=1.0$ shown in Fig.~\ref{fsi-u0p0002-kn1}, as that difference increases to about $10\%$, the numerical results are much larger than the theoretical solutions. Consequently, it illustrates again that the influence of oscillation frequency must be introduced into the damping model.

\begin{figure}
	\centering
	\subfigure[]{
		\includegraphics[width=0.55 \textwidth]{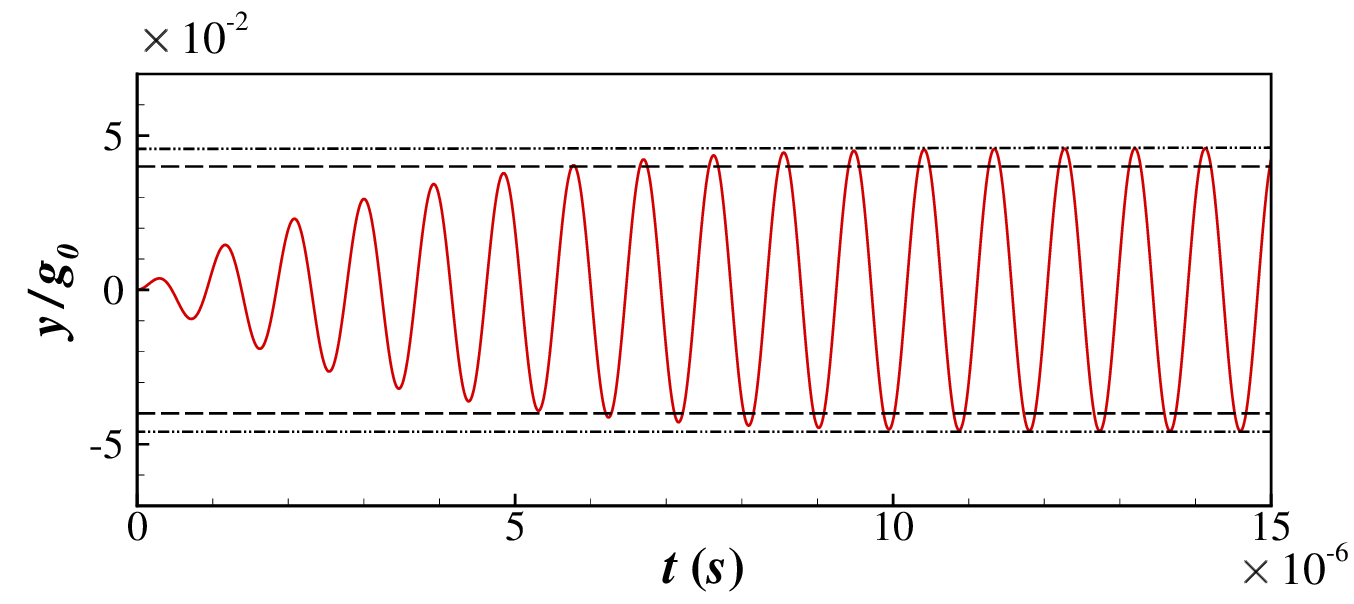}
	}
	\subfigure[]{
		\includegraphics[width=0.55 \textwidth]{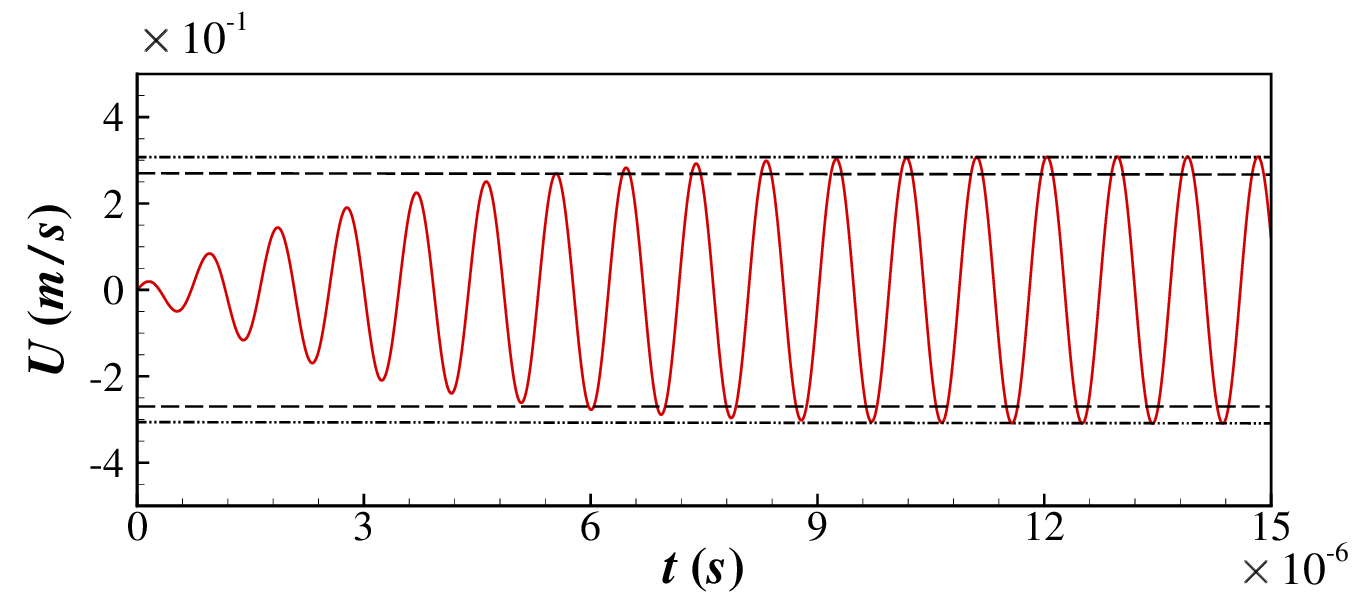}
	}
	\subfigure[]{
		\includegraphics[width=0.55 \textwidth]{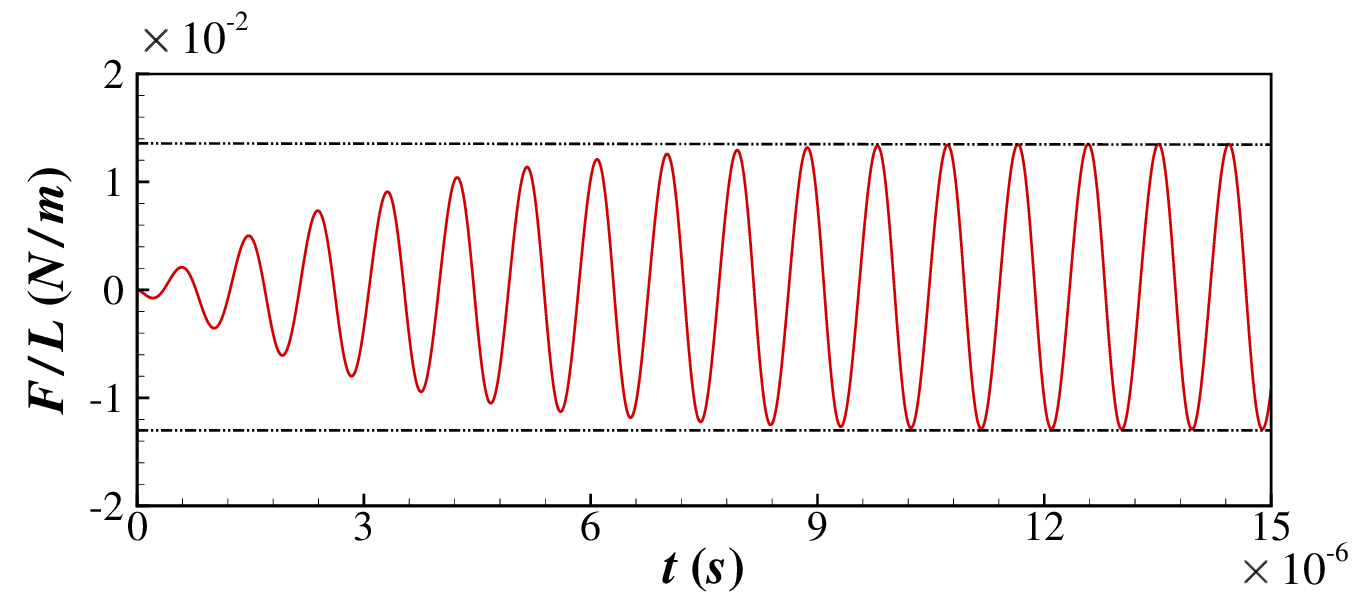}
	}	
	\caption{\label{fsi-u0p0008-kn0p1} Time evolutions of (a) displacement, (b) moving velocity and (c) damping force of a micro-beam with linear free oscillation at $Kn^{(h_0)}=0.1$. In the simulation, $U_{max}=0.27m/s$, $M^*=2769.5$, and $A=0.04h_0$ are used. (the dash lines shown in the figures are the maximum theoretical values, and the dash-dot lines are the maximum values obtained by numerical simulation).}
\end{figure}

\begin{figure}
	\centering
	\subfigure[]{
		\includegraphics[width=0.55 \textwidth]{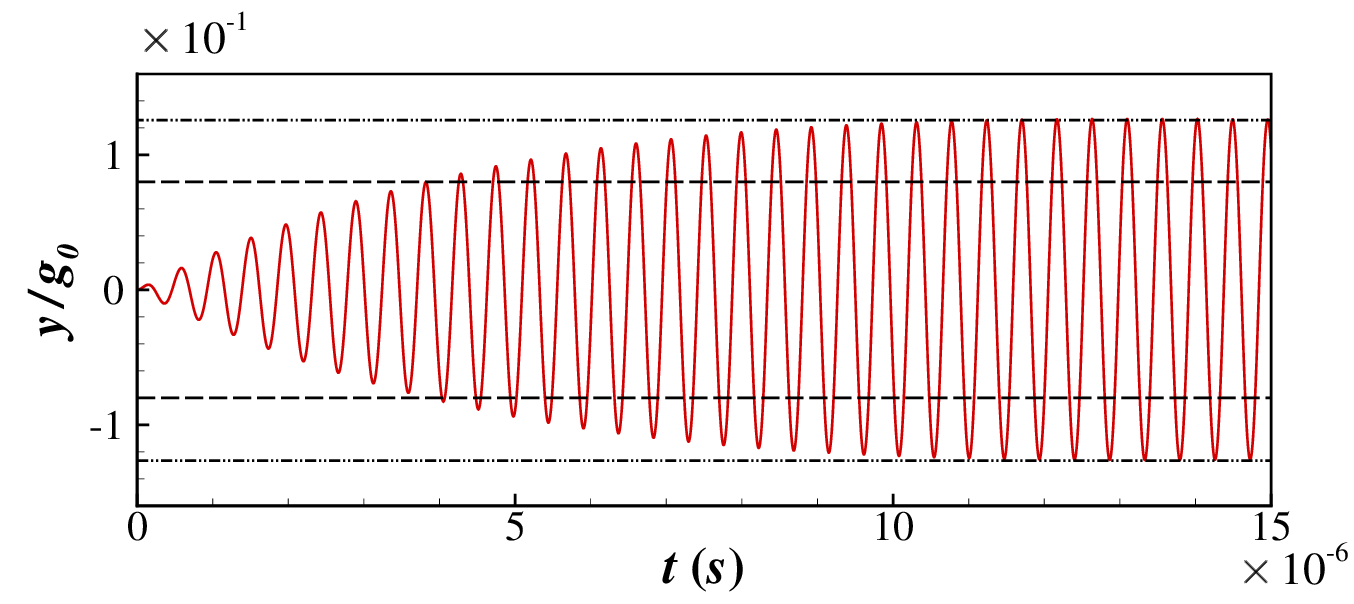}
	}
	\subfigure[]{
		\includegraphics[width=0.55 \textwidth]{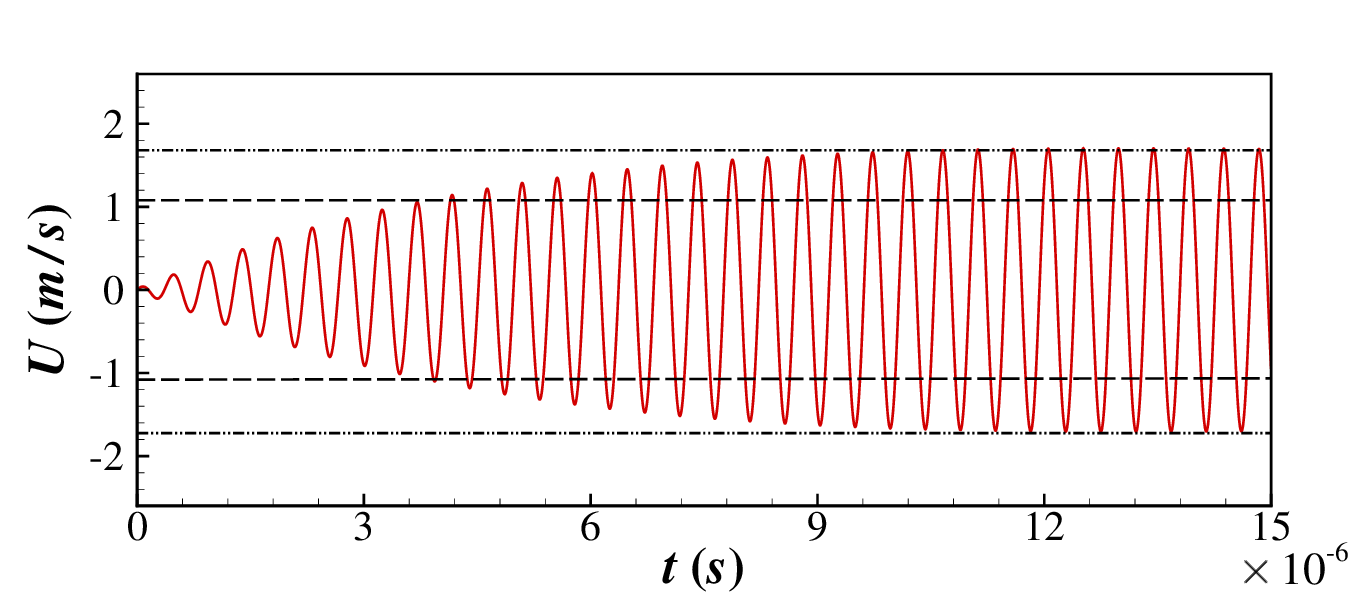}
	}
	\subfigure[]{
		\includegraphics[width=0.55 \textwidth]{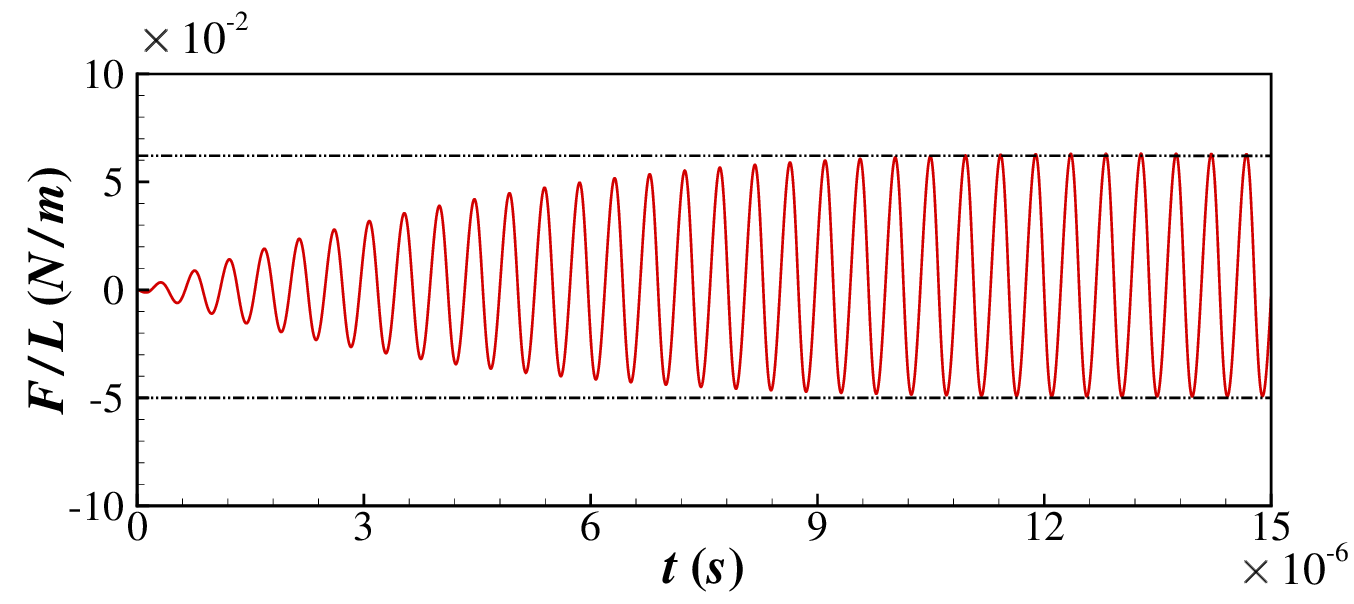}
	}	
	\caption{\label{fsi-u0p0032-kn0p1} Time evolutions of (a) displacement, (b) moving velocity and (c) damping force of a micro-beam with linear free oscillation at $Kn^{(h_0)}=0.1$. In the simulation, $U_{max}=1.08m/s$, $M^*=2769.5$, and $A=0.08h_0$ are used. (the dash lines shown in the figures are the maximum theoretical values, and the dash-dot lines are the maximum values obtained by numerical simulation).}
\end{figure}

Secondly, two free oscillation cases at higher velocities are simulated, with the computational conditions are set to $U_{max}=0.27m/s$, $A=0.04h_0$ and $U_{max}=1.08m/s$, $A=0.08h_0$, respectively. And $Kn^{(h_0)}$ is set to 0.1. Due to the nonlinear phenomenon of damping force, the damping coefficient $c$ used for a theoretical solution is difficult to construct and is also set to $0.0485L$. Shown in Figs.~\ref{fsi-u0p0008-kn0p1} and~\ref{fsi-u0p0032-kn0p1}, for the displacements and moving velocities of a micro-beam, as $c$ obtained from a low-frequency simulation can not reflect the real damping at a high-frequency oscillation, the numerical results are much higher than the theoretical solutions. Besides, although the nonlinear phenomenon of damping forces also can be observed, the maximum and minimum values of displacement and moving velocity of a micro-beam are almost the same. For this reason, in the practical computation, an empirical parameter $c$ maybe be constructed and used for high velocity oscillation.

\subsection{\color{blue}Decoupled method and coupled framework: squeeze-film damping torque at the tilting motion}\label{SecIII-III}
In this section, with the coupled FSI framework, the rarefied gas damping torque acting on a micro-beam by the forced or free oscillation under the tilting motion is studied. The initial height of gap $h_0$ also is set to $1.0\times10^{-6}m$.

\subsubsection{Decoupled method: squeeze-film damping torque}
\begin{figure}
	\centering
	\subfigure[]{
		\includegraphics[width=0.5 \textwidth]{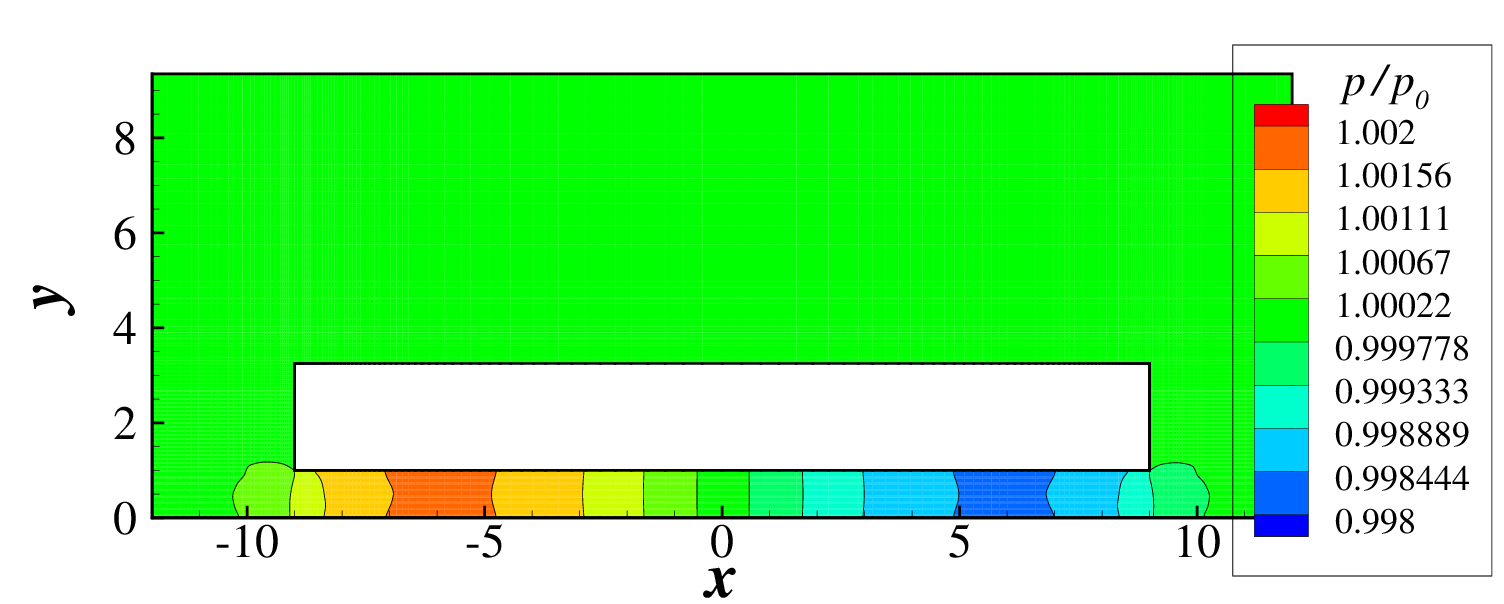}
	}
	\subfigure[]{
		\includegraphics[width=0.5 \textwidth]{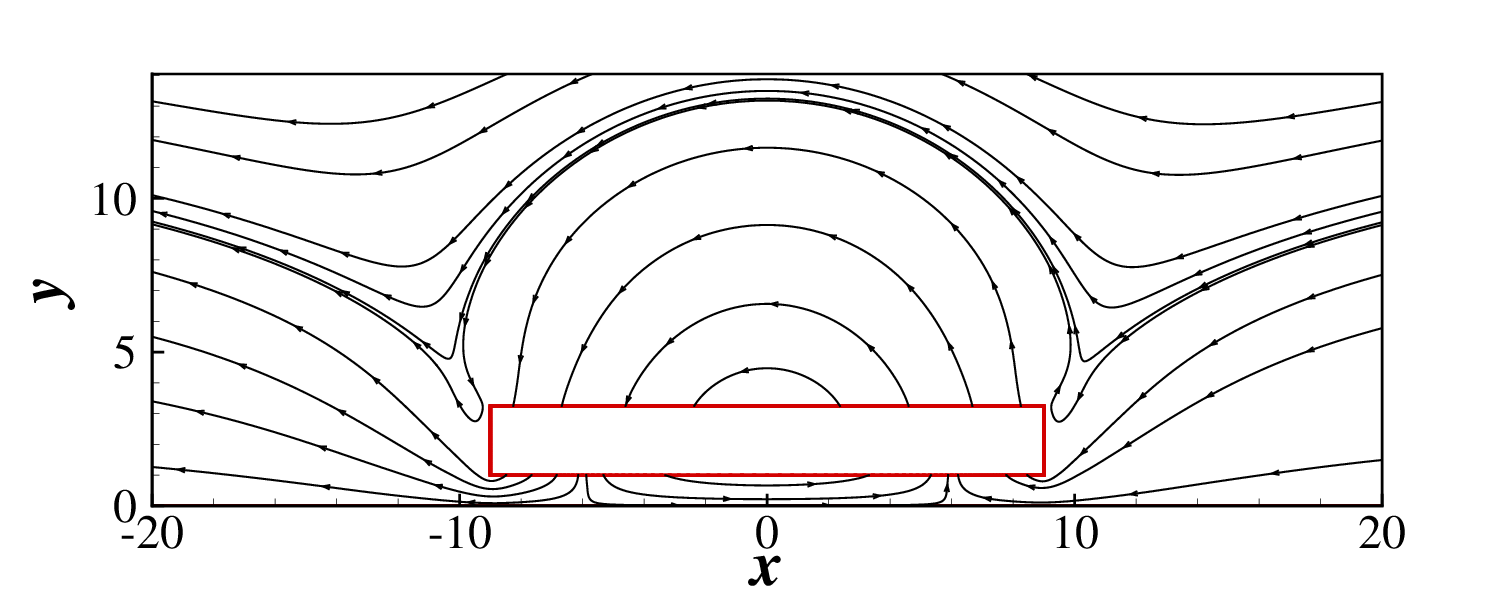}
	}	
	\caption{\label{tilt-pre} (a) Pressure contours and (b) streamlines for flow around a micro-beam in the rarefied gas at $Kn^{(h_0)}=0.1$. $\theta_0$ and $f$ in Eq.~\eqref{tilt-motion2} are set to $\theta_0=0.5^\circ$ and $f/(2\pi)=0.2106MHz$, respectively.}
\end{figure}

\begin{figure}
	\centering
	\subfigure[]{
		\includegraphics[width=0.4 \textwidth]{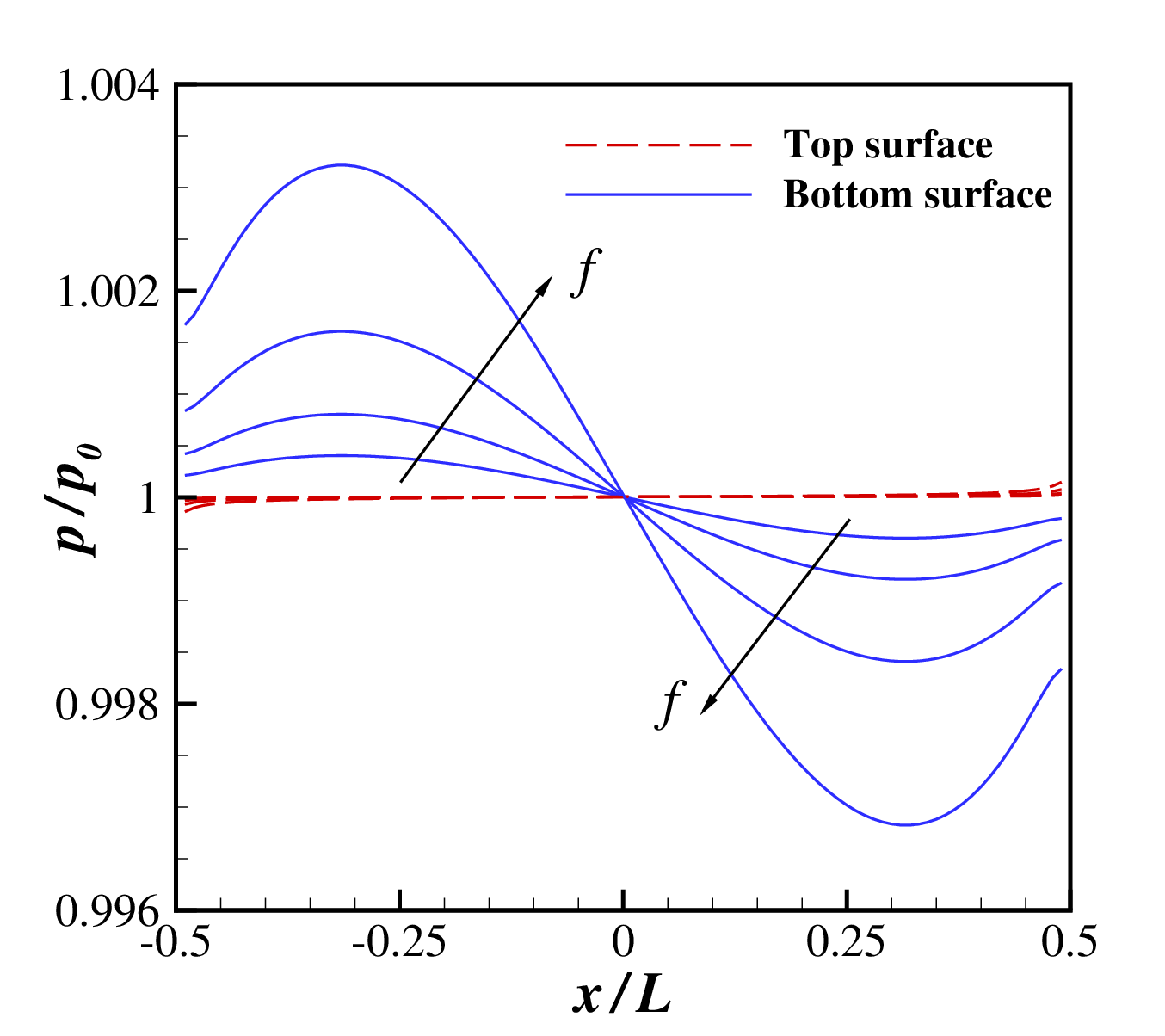}
	}
	\subfigure[]{
		\includegraphics[width=0.4 \textwidth]{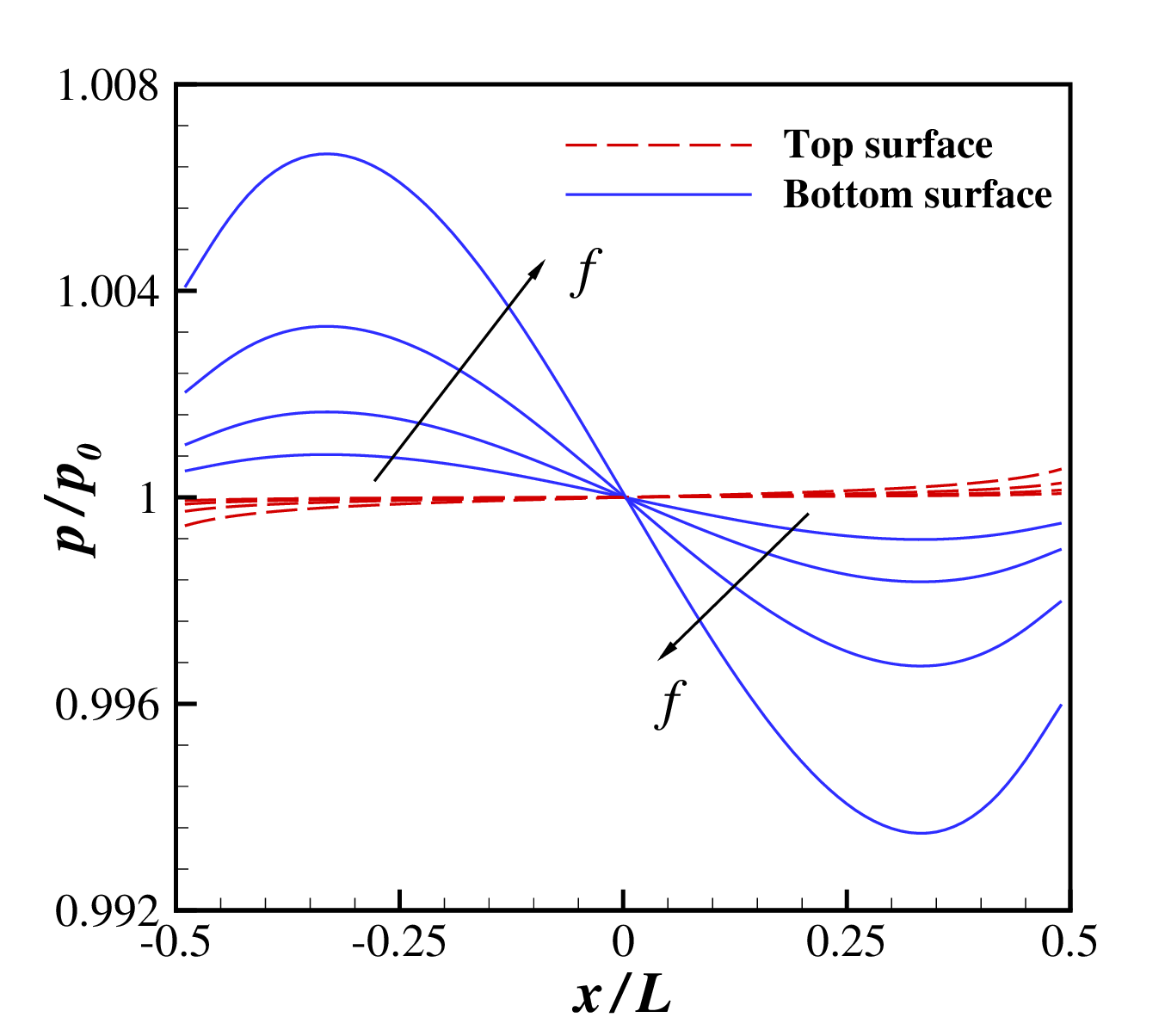}
	}	
	\caption{\label{tilt-press-line} Pressure distributions along the top and bottom surfaces of micro-beam with different tilting frequencies at (a) $Kn^{(h_0)}=0.1$ and (b) $Kn^{(h_0)}=1.0$. Four tilting frequencies ($f/(2\pi)$), $0.05265MHz$, $0.1053MHz$, $0.2106MHz$, and $0.4213MHz$ at $\theta_0=0.5^{\circ}$ are shown in figures.}
\end{figure}

\begin{figure}
	\centering
	\subfigure[]{
		\includegraphics[width=0.4 \textwidth]{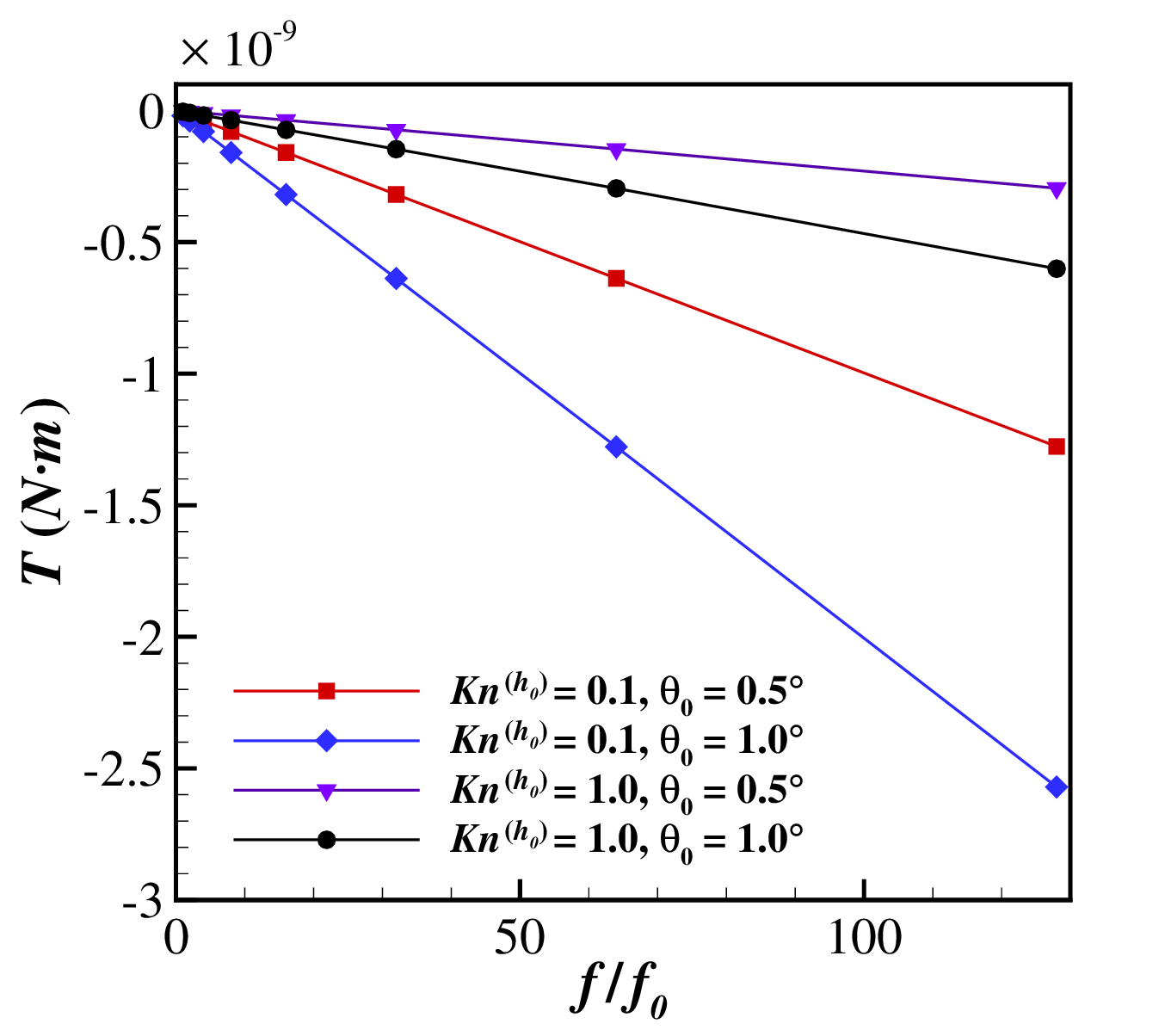}
	}
	\subfigure[]{
		\includegraphics[width=0.4 \textwidth]{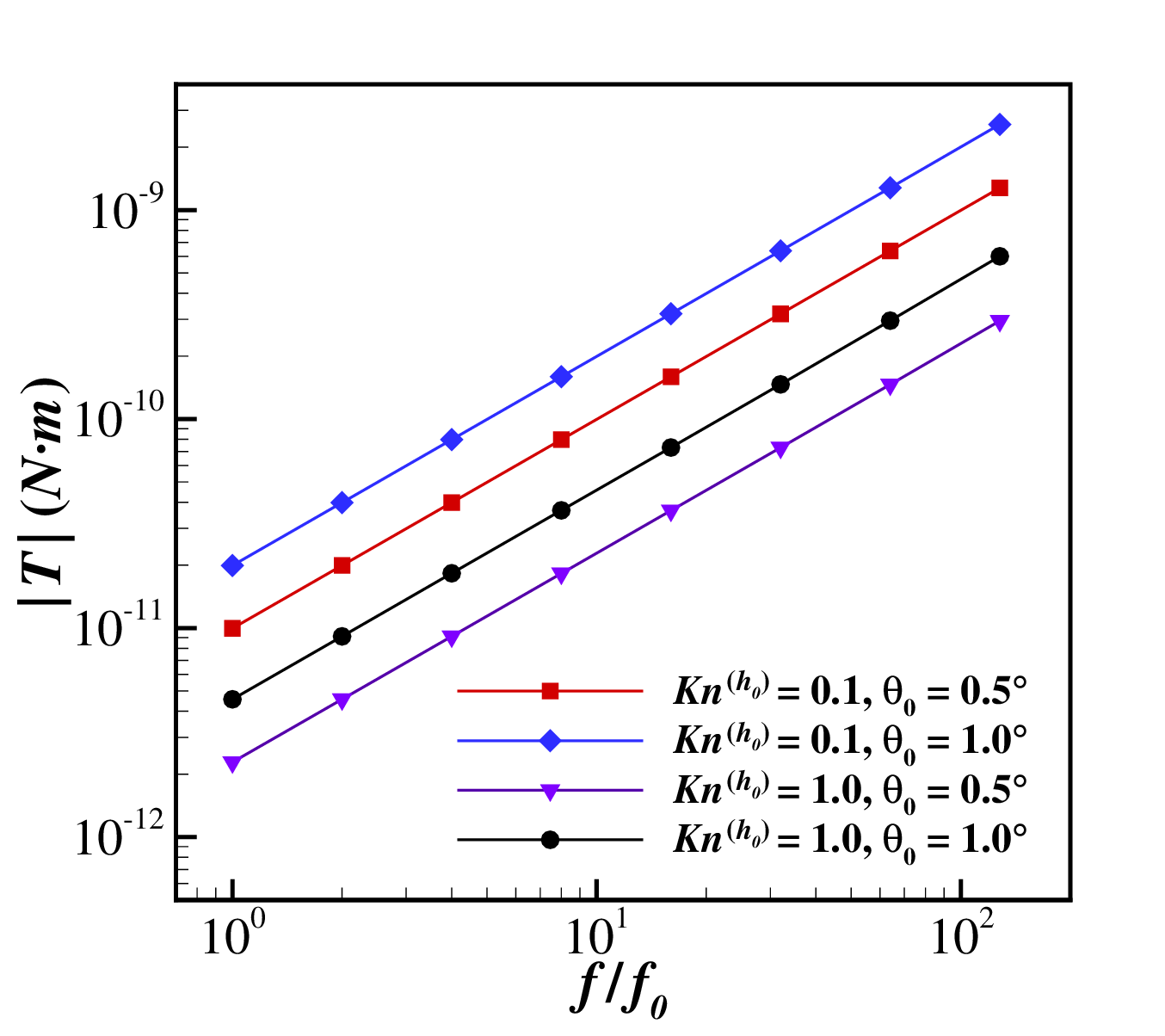}
	}	
	\caption{\label{tilt-cm-stat} Comparisons of gas damping torque at different tilting frequencies with (a) original coordinates and (b) logarithm coordinates. The initial test frequency $f_0$ equals to $f_0/(2\pi)=0.052656MHz$.}
\end{figure}

For the forced tilting oscillation, the motion form is given as
\begin{equation}\label{tilt-motion}
	\theta=\theta_0sin(ft),
\end{equation}
where $\theta_0$ is the maximum tilting angle; and two values, $0.5^\circ$ and $1.0^\circ$, are considered. With Eq.~\eqref{tilt-motion}, assuming the maximum damping torque is obtained at the maximum angular velocity, the tilting angle of a micro-beam used in the simulation is $0^\circ$. Shown in Fig.~\ref{plate-tilt-decoupled}, the velocity profile imposed on the surface of wall is given by
\begin{equation}\label{tilt-motion2}
	V=R\theta_0f\frac{\pi}{180},
\end{equation}
where $V$ is the magnitude of velocity vector, and $R$ is the length from the surface of micro-beam to its center. Fig.~\ref{tilt-pre} shows the pressure contours and streamlines near the micro-beam at $\theta_0=0.5^\circ$ and $f/(2\pi)=0.2106MHz$. In this study, similar to the definition of the Strouhal number, a parameter of $fh_0/V_{max}$ is used to nondimensionalize the tilting frequency, where $V_{max}$ is the maximum velocity at the surface of micro-beam. Clearly, due to the effect of tilt, the pressure in one side of the gap increases, and that at the other side decreases; then it generates the damping torque. Figs.~\ref{tilt-press-line} and~\ref{tilt-cm-stat} show the pressure distributions along the surface of a micro-beam, and the damping torque acting on it, respectively. Here, a coefficient $T/(0.25\rho{V_{max}^2}h_0L)$ is used to nondimensionalize the damping torque $T$. Similar to the linear motion described in Sec.~\ref{SecIII-II}, due to the influence of the substrate, the variation of pressure at the bottom surface is much more significant than that at the top surface, and the pressure distributions show some kind of linear relation between the different frequencies. Further, the variation of damping torque also shows a linear relation between the oscillation frequency $f$, the maximum tilting angle $\theta_0$ and the Knudsen number $Kn^{(h_0)}$. Although, the maximum moving velocity on the surface of a micro-beam is about $6.65m/s$ at $\theta_0=1.0^\circ$ and $f/(2\pi)=6.74MHz$, the nonlinear phenomenon can not be observed.

\subsubsection{Coupled method: squeeze-film damping torque at forced oscillation}
\begin{figure}
	\centering
	\subfigure[]{
		\includegraphics[width=0.55 \textwidth]{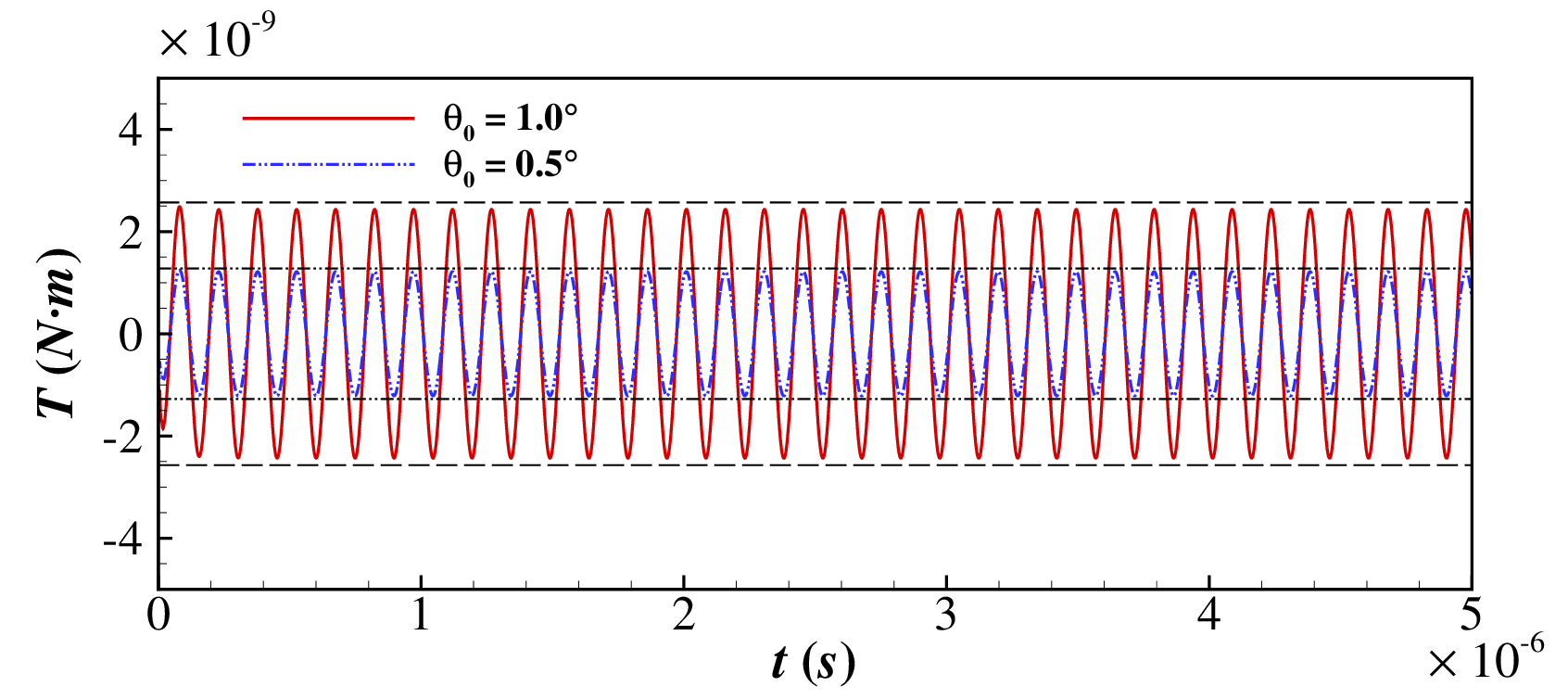}
		\includegraphics[width=0.55 \textwidth]{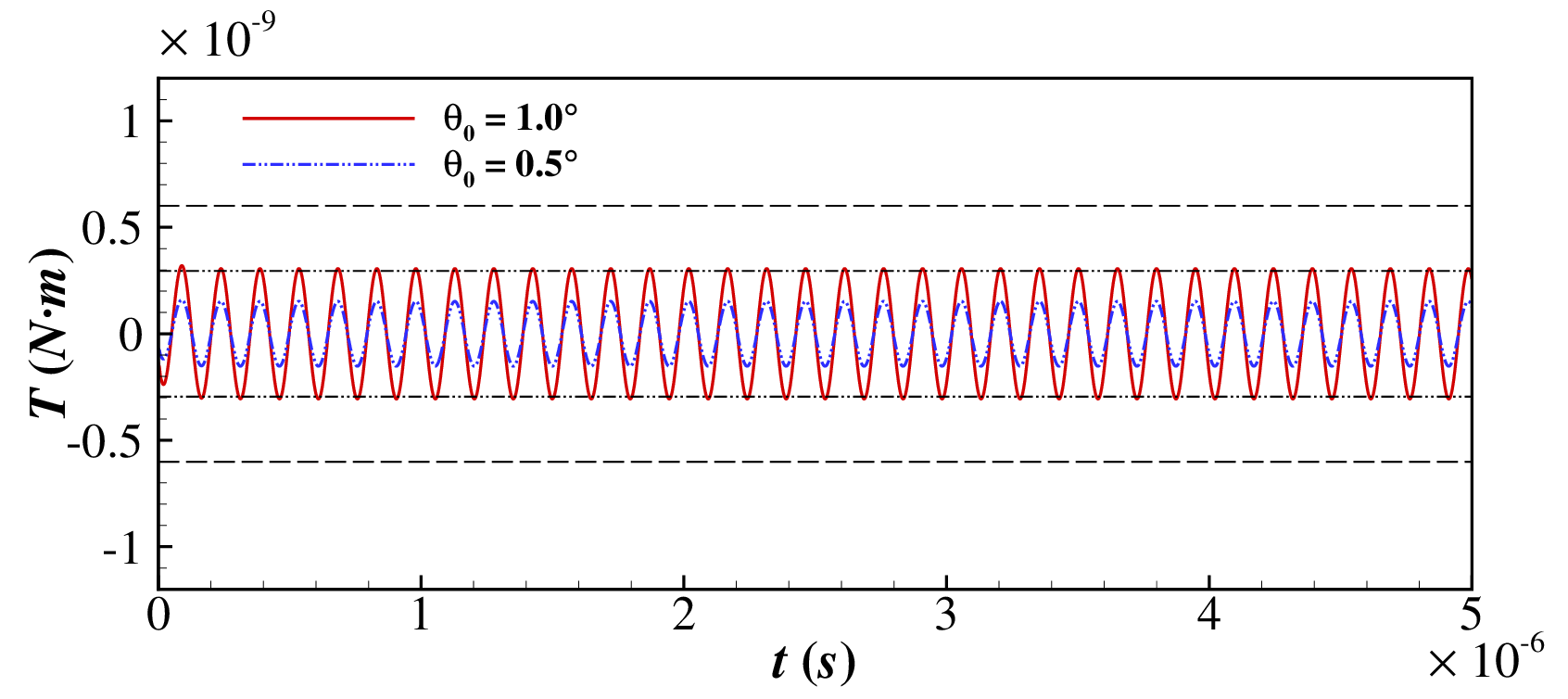}
	}
	\subfigure[]{
		\includegraphics[width=0.55 \textwidth]{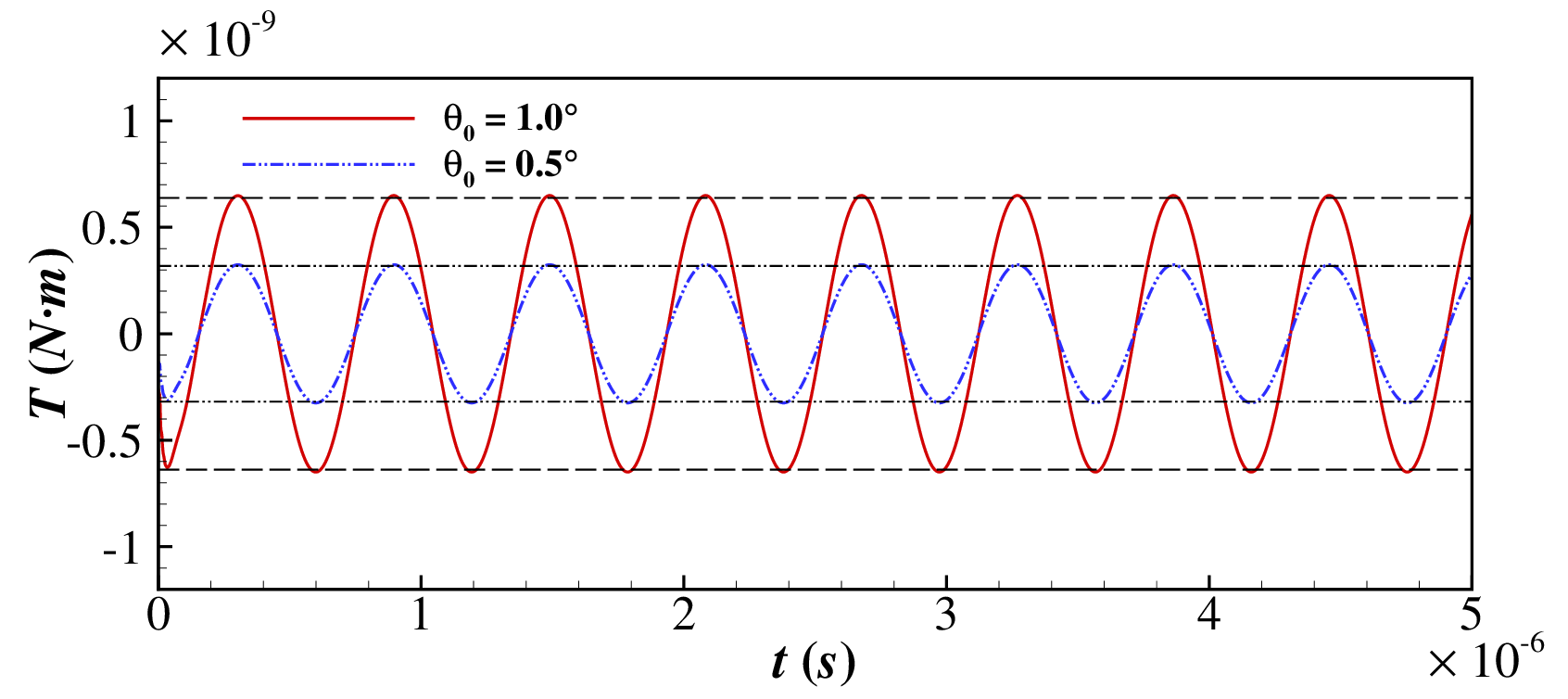}
		\includegraphics[width=0.55 \textwidth]{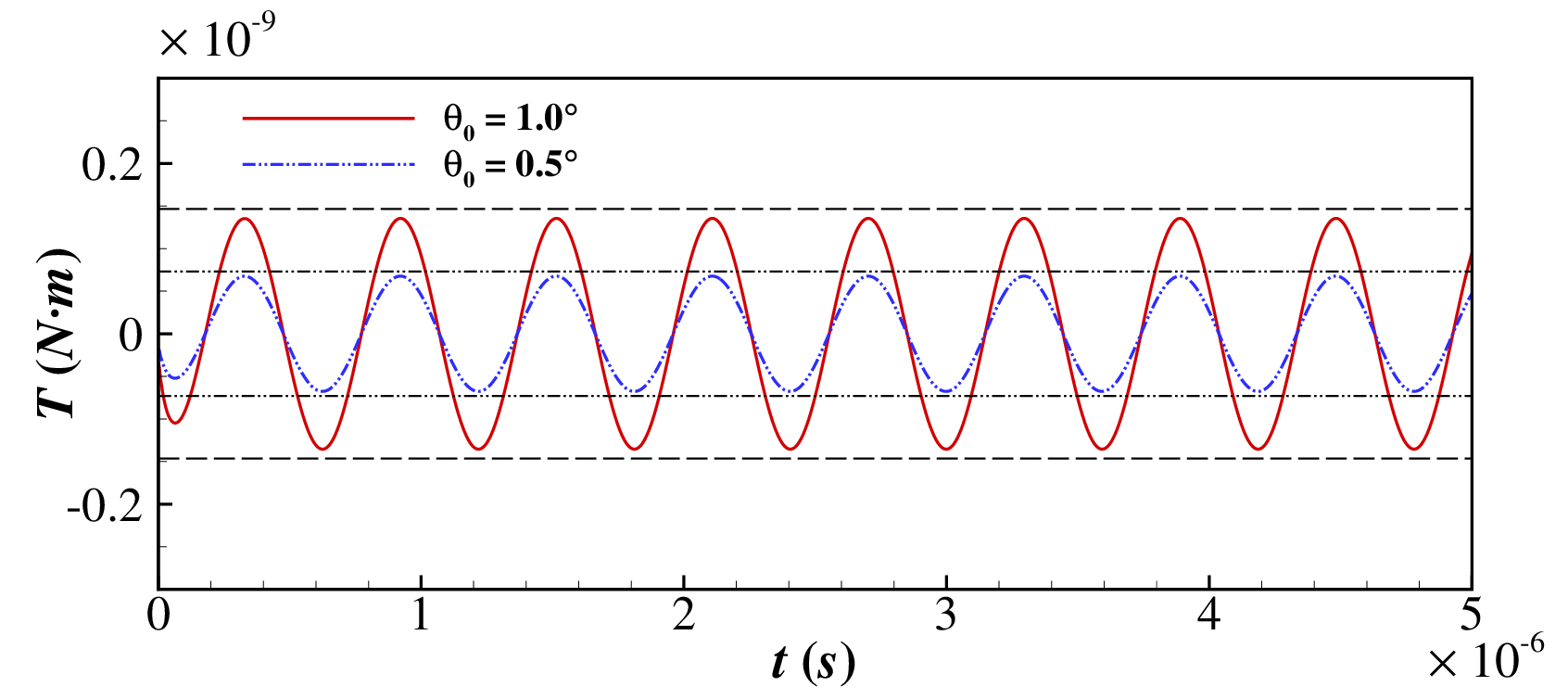}
	}
	\subfigure[]{
		\includegraphics[width=0.55 \textwidth]{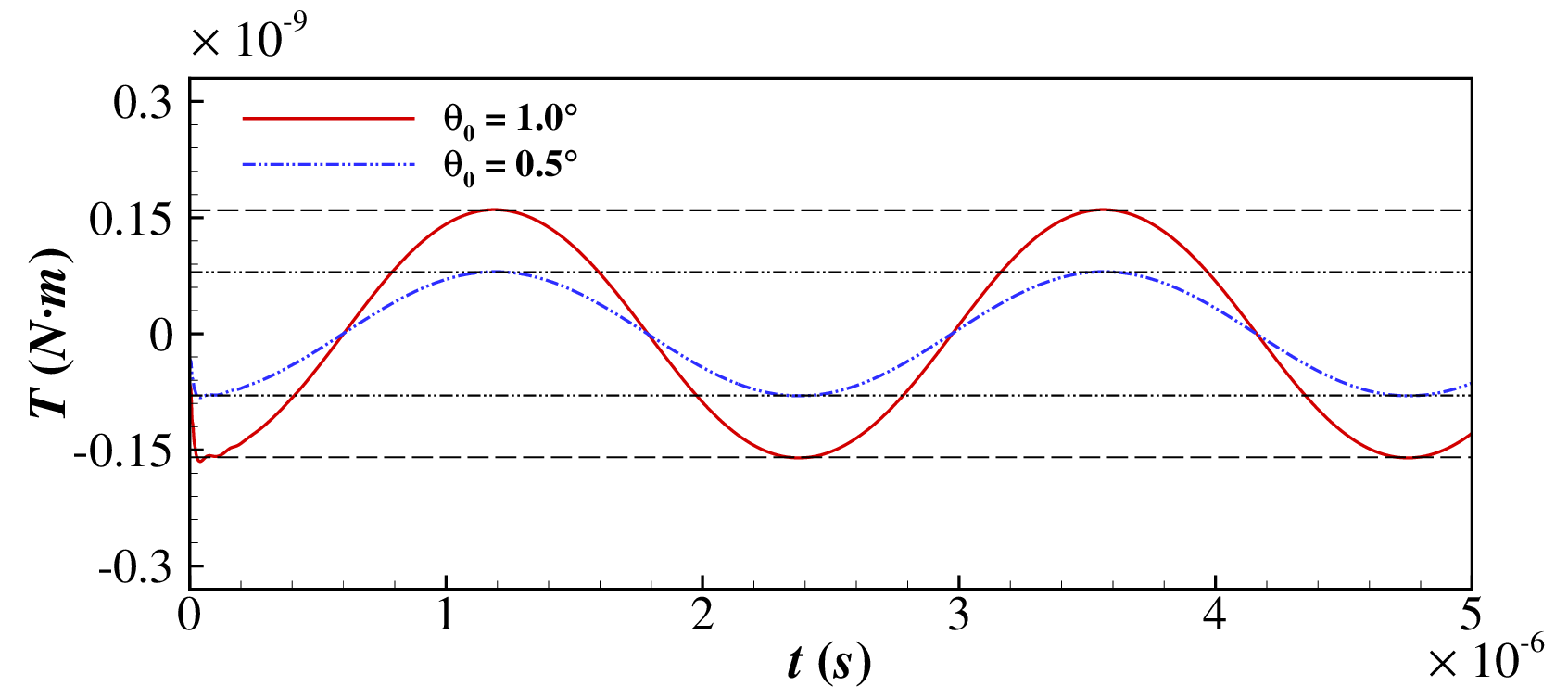}
		\includegraphics[width=0.55 \textwidth]{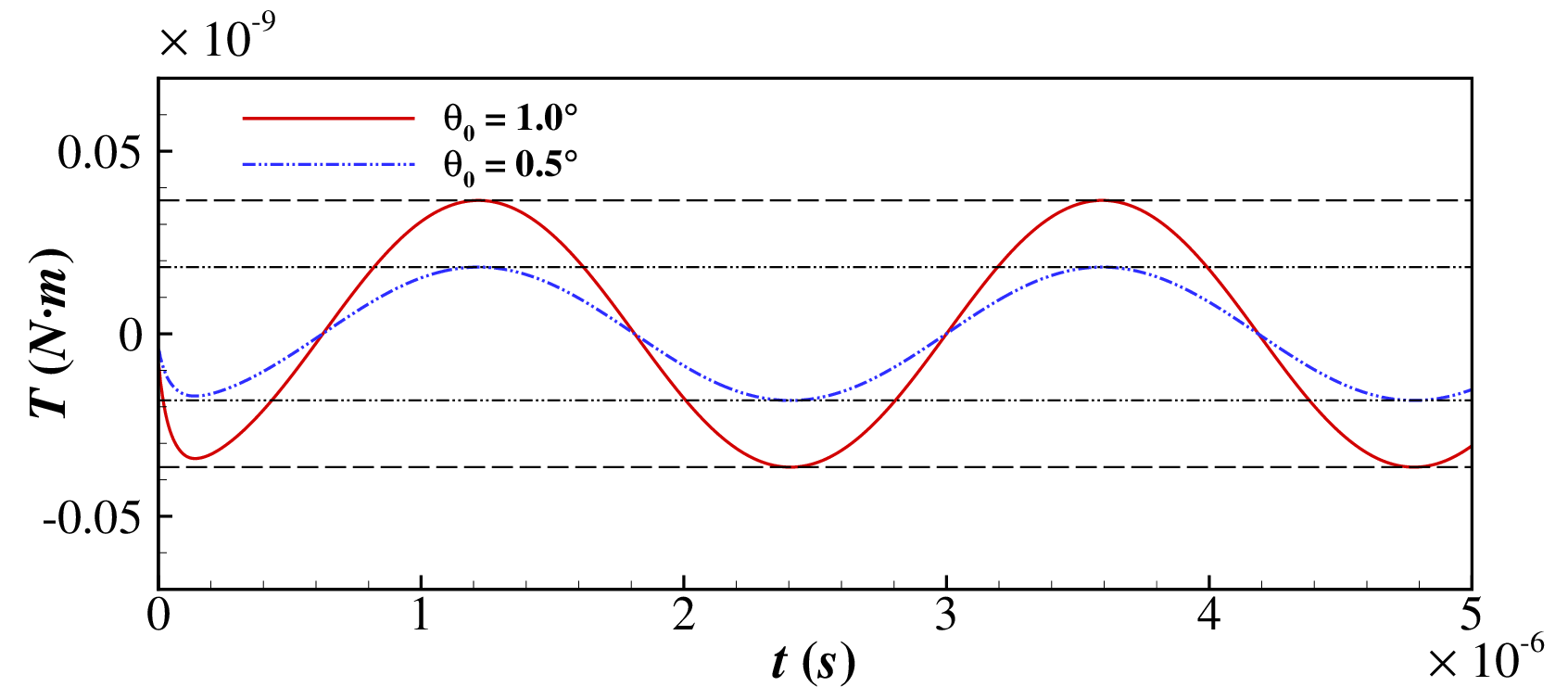}
	}	
	\caption{\label{tilt-force-kn} Time evolutions of damping torque at two $Kn^{(h_0)}$ numbers, 0.1 (left) and 1.0 (right), and the tilting oscillation frequencies ($f/(2\pi)$) are set to (a) $6.74MHz$, (b) $1.685MHz$ and (c) $0.42125MHz$ (the values of dash lines and dash dots lines shown in figures are obtained from Fig.~\ref{tilt-cm-stat} for comparison).}
\end{figure}

\begin{figure}
	\centering
	\subfigure[]{
		\includegraphics[width=0.4 \textwidth]{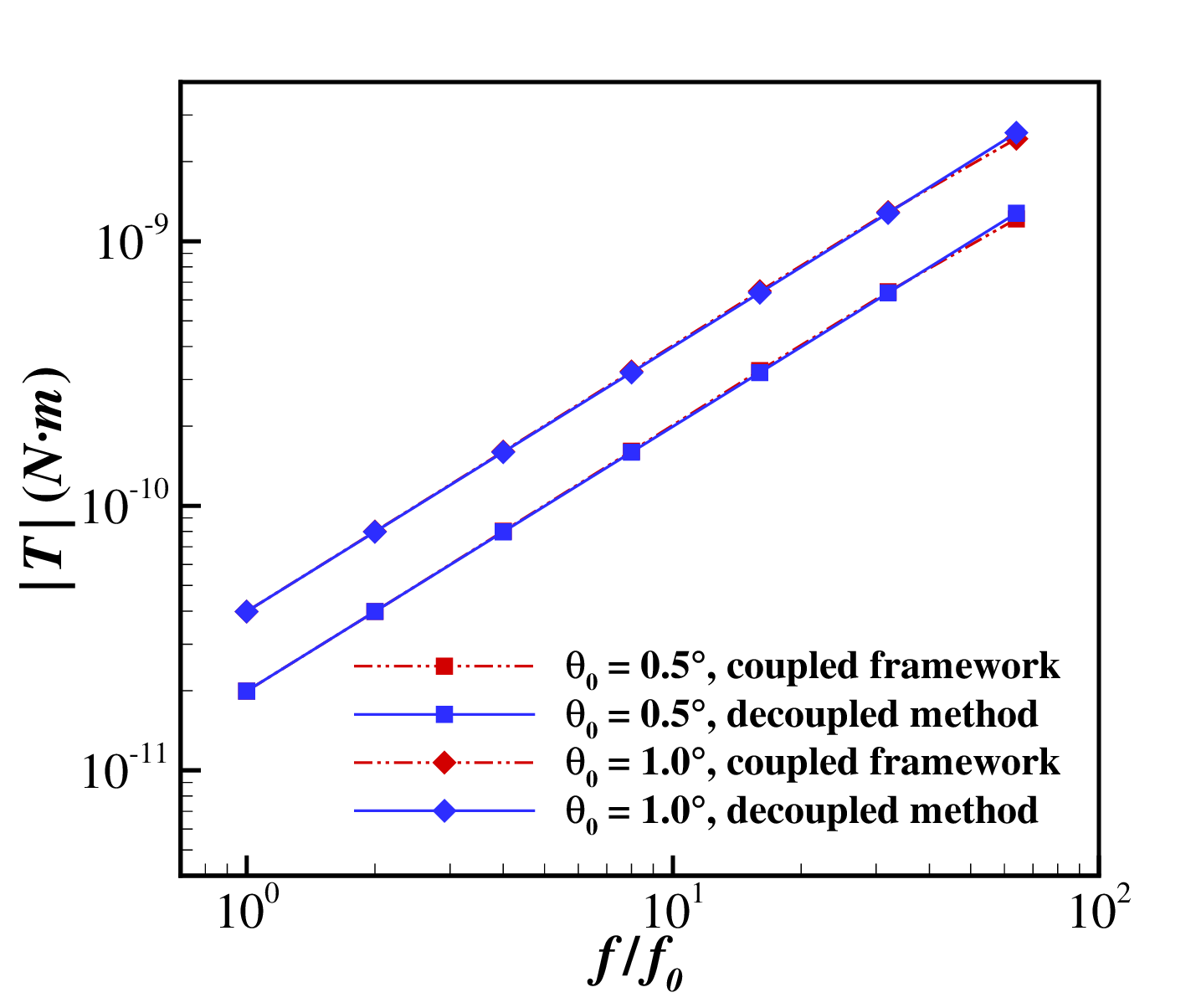}
	}
	\subfigure[]{
		\includegraphics[width=0.4 \textwidth]{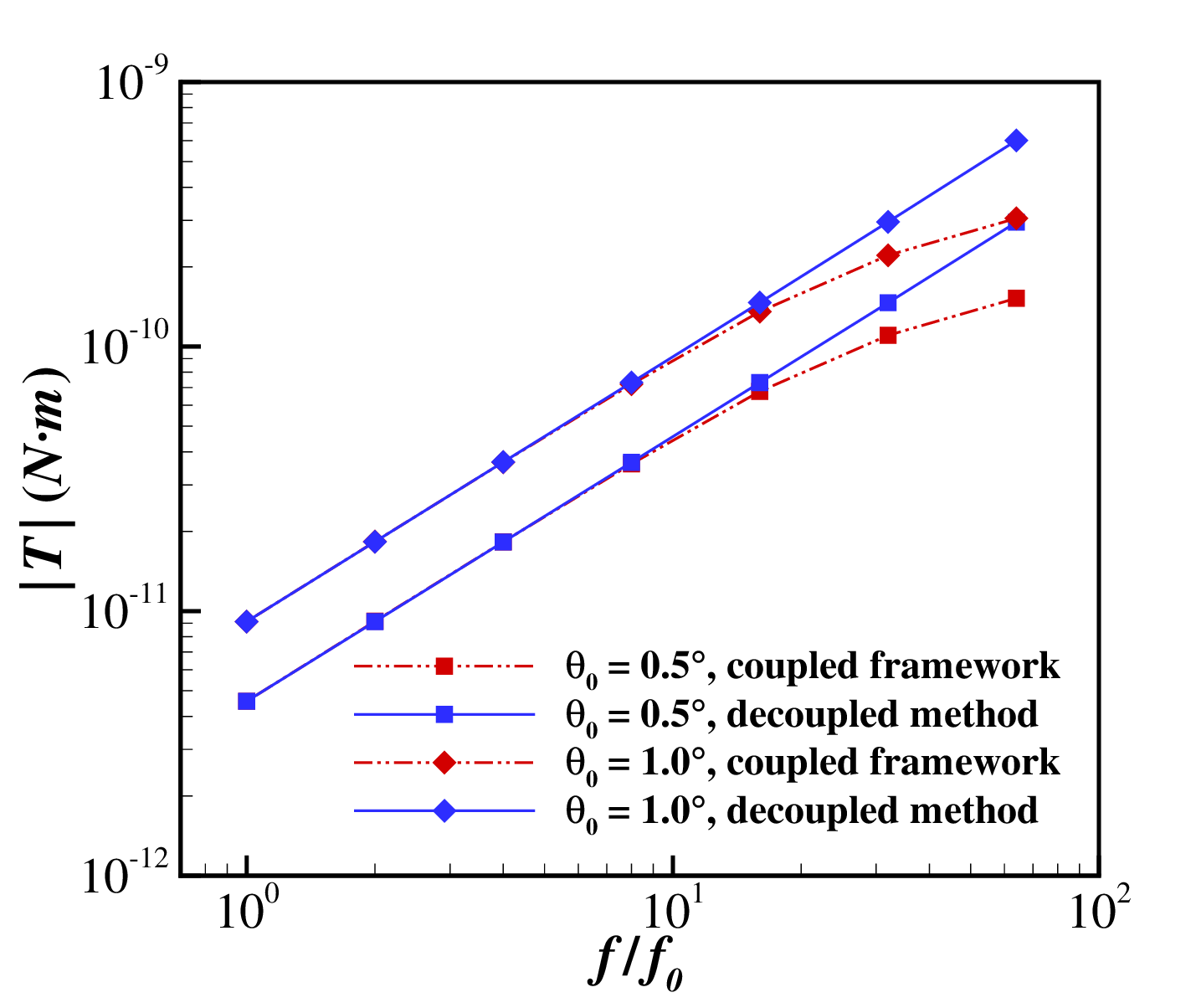}
	}	
	\caption{\label{tilt-torque-max} Comparisons of the maximum gas damping torque at different tilting frequencies $f$ and angles $\theta_0$ with (a) $Kn^{(h_0)}=0.1$ and (b) $Kn^{(h_0)}=1.0$, where $f_0$ is the initial test frequency and equals to $f_0/(2\pi)=0.1053MHz$.}
\end{figure}

\begin{figure}
	\centering
	\subfigure[]{
		\includegraphics[width=0.55 \textwidth]{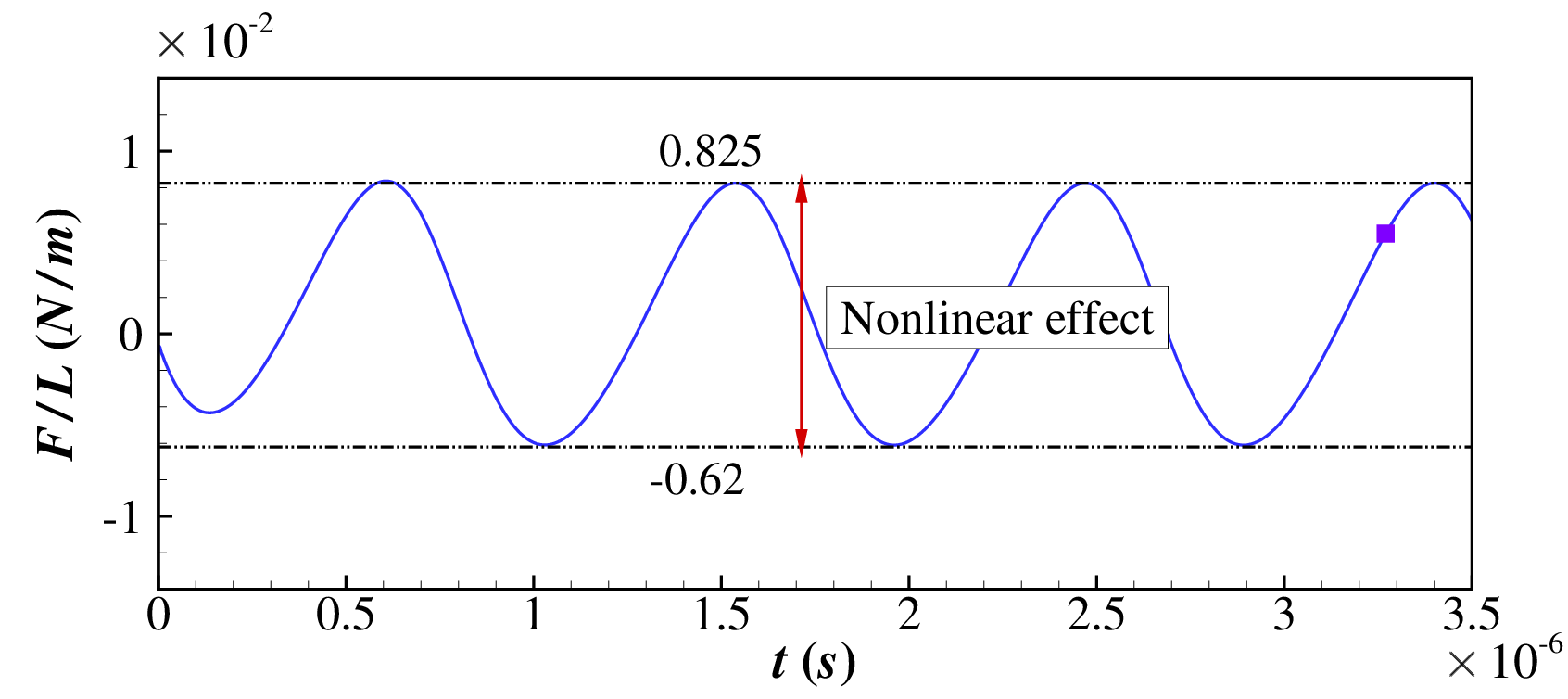}
		\includegraphics[width=0.55 \textwidth]{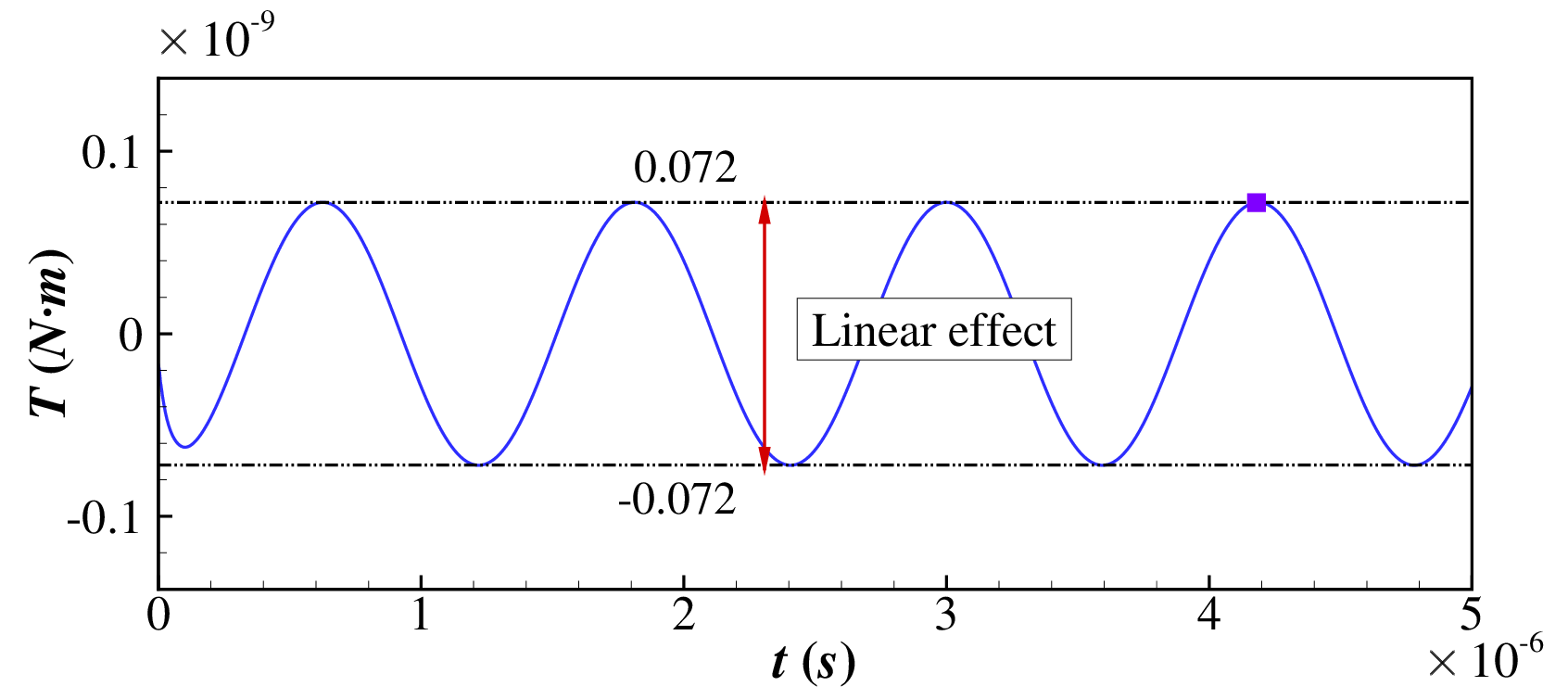}
	}
	\subfigure[]{
		\includegraphics[width=0.55 \textwidth]{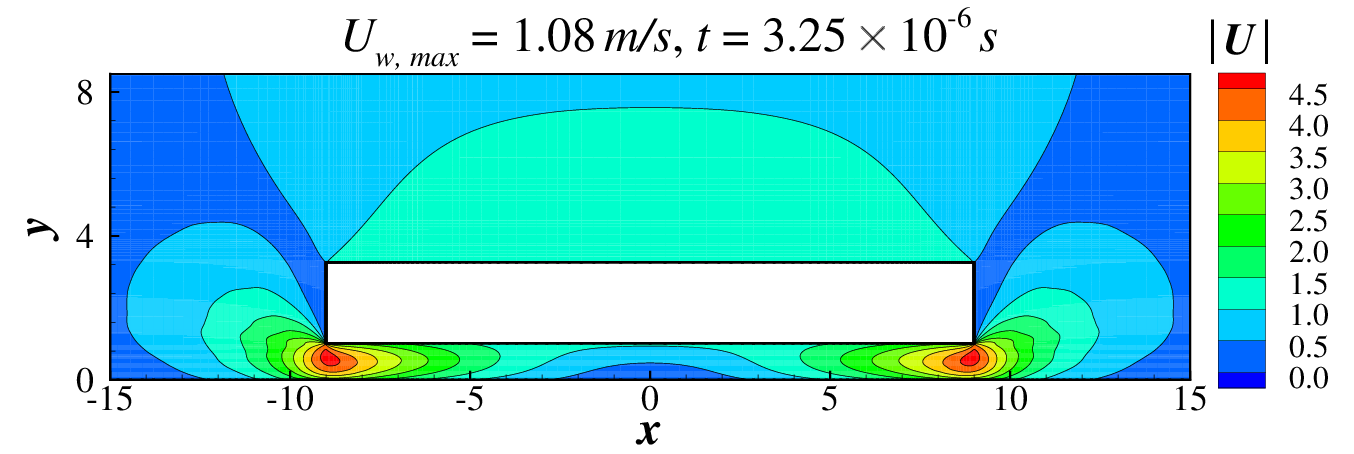}
		\includegraphics[width=0.55 \textwidth]{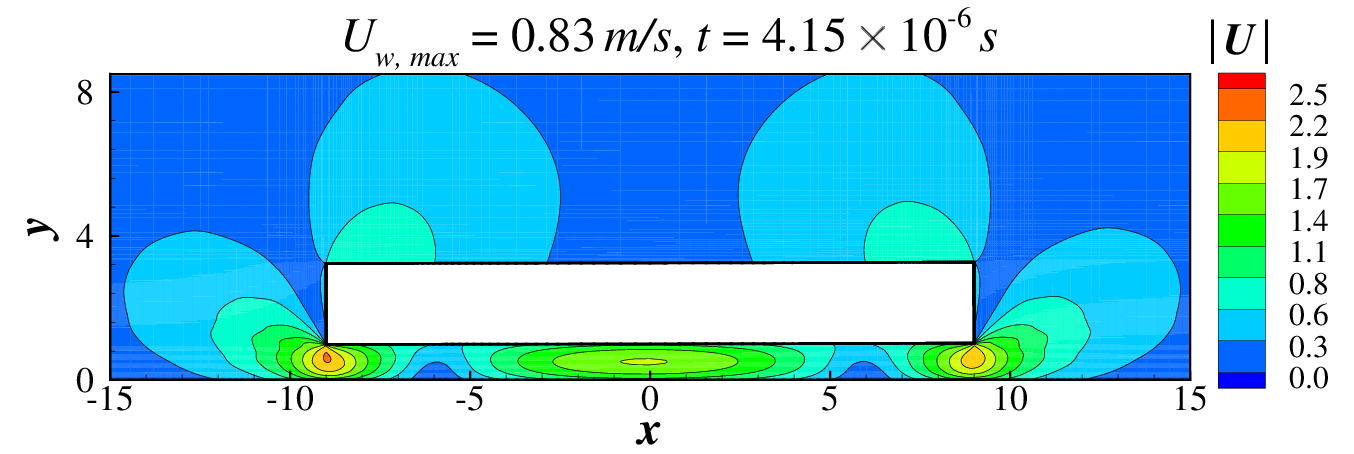}
	}
	\subfigure[]{
		\includegraphics[width=0.53 \textwidth]{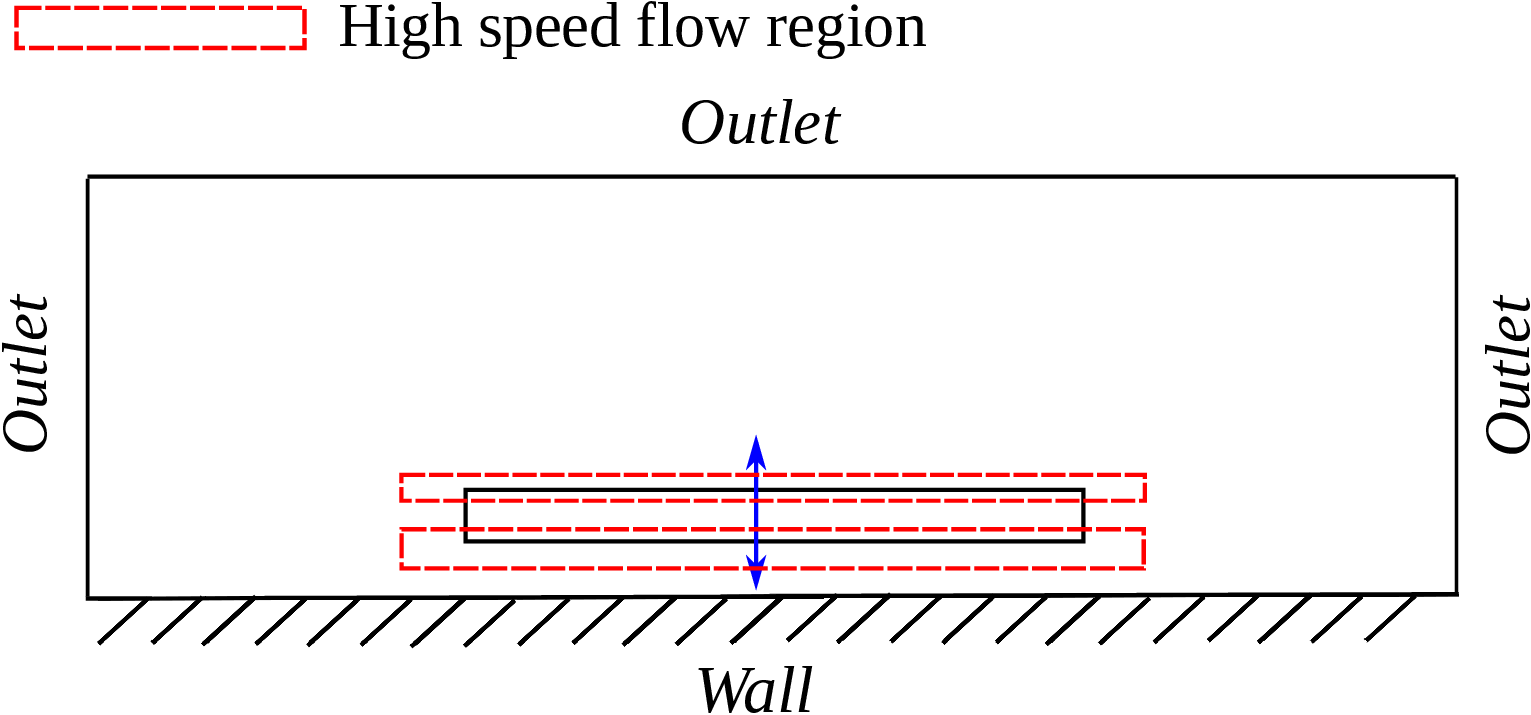}
		\includegraphics[width=0.53 \textwidth]{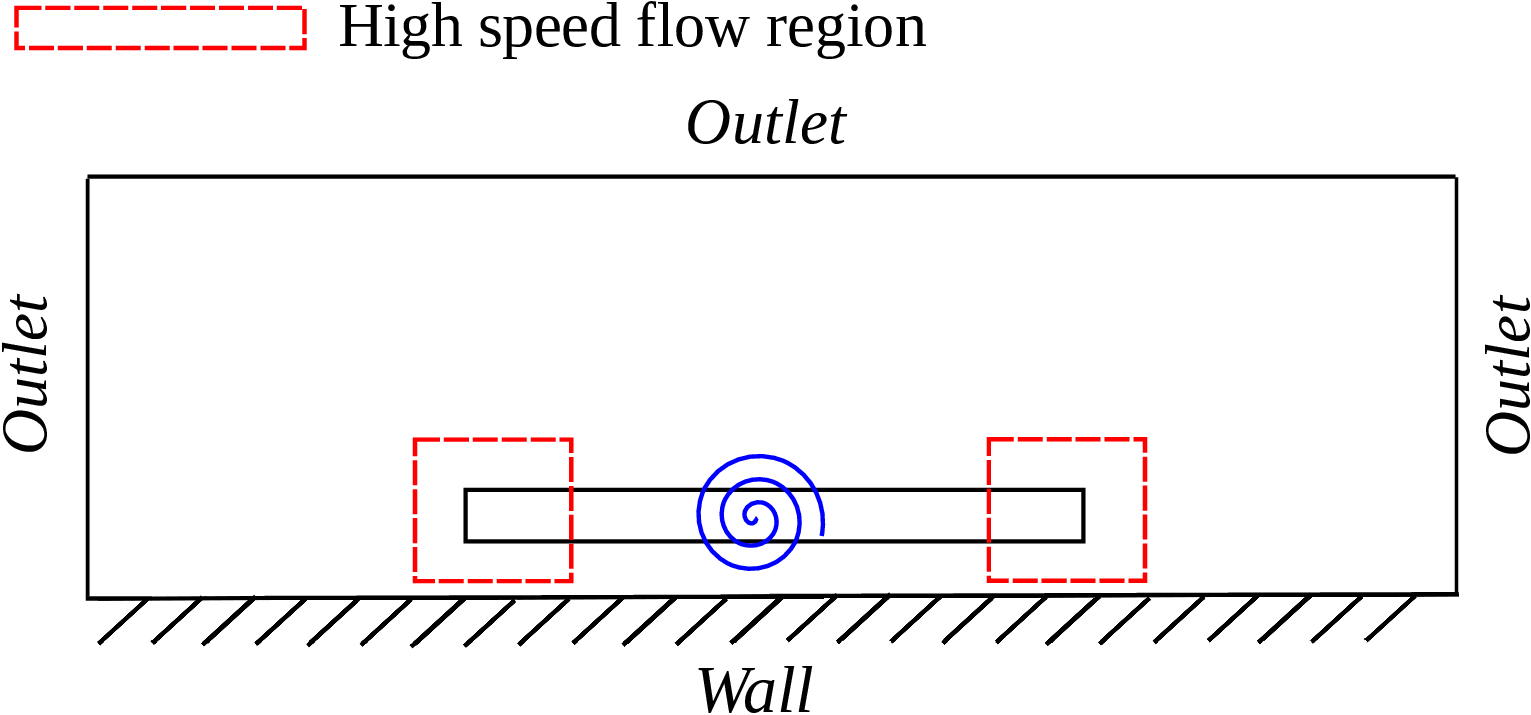}
	}	
	\caption{\label{linear-effect} (a) Time evolutions of force and torque, (b) velocity magnitude contours $|\bm{U}|$ and (c) sketches of high speed flow region at forced linear oscillation (left) and tilting oscillation (right) with $Kn^{(h_0)}=1.0$, where $U_{w, max}$ is maximum moving velocity at the surface of a micro-beam. For the forced linear oscillation, $U_{max}$ and $A$ in Eq.~\eqref{oscillateing-form} are $1.08m/s$ and $0.16h_0$, respectively. So, the corresponding oscillation frequency is $1.074MHz$. And for the forced tilting oscillation, the frequency $f$ in Eq.~\eqref{tilt-motion} is $f/(2\pi)=0.8425MHz$, and the maximum tilting angle $\theta_0$ is $1.0^{\circ}$.}
\end{figure}

For the forced tilting oscillation, Eq.~\eqref{tilt-motion} is used to control the variation of tilting angle. Fig.~\ref{tilt-force-kn} shows the time evolutions of damping torque at three tilting oscillation frequencies. Fig.~\ref{tilt-torque-max} shows the comparisons of the largest damping torque acting on a micro-beam by two different methods. For flow at $Kn^{(h_0)}=0.1$, the largest damping torques obtained by two different methods are almost the same, so the traditional decoupled method still exhibits a good performance to predict the rarefied gas damping torque coefficient $\eta$ in Eq.~\eqref{tilt-struct}. And for flow at $Kn^{(h_0)}=1.0$, due to the rarefaction effect, the differences of results are obvious at a high tilting oscillation frequency. For example, the result obtained by the decoupled method is about twice larger than that by the coupled framework at $f/(2\pi)=6.74MHz$. So, the influence of tilting oscillation frequency must be introduced to build the gas damping torque model at a high tilting oscillation frequency and a high Knudsen number. Besides, the nonlinear phenomenon also can not be observed as the absolute values of maximum and minimum values of damping torque are almost the same. Different from the linear motion at a high velocity, the high moving velocity regime only focuses on the left and right sides of a tilting micro-beam. So, the cause of the nonlinear phenomenon may be a low oscillation frequency and a large contact area of high velocity between rarefied gas and micro-beam (see Fig.~\ref{linear-effect}).

\subsubsection{Coupled method: squeeze-film damping torque at free oscillation}
\begin{figure}
	\centering
	\subfigure[]{
		\includegraphics[width=0.55 \textwidth]{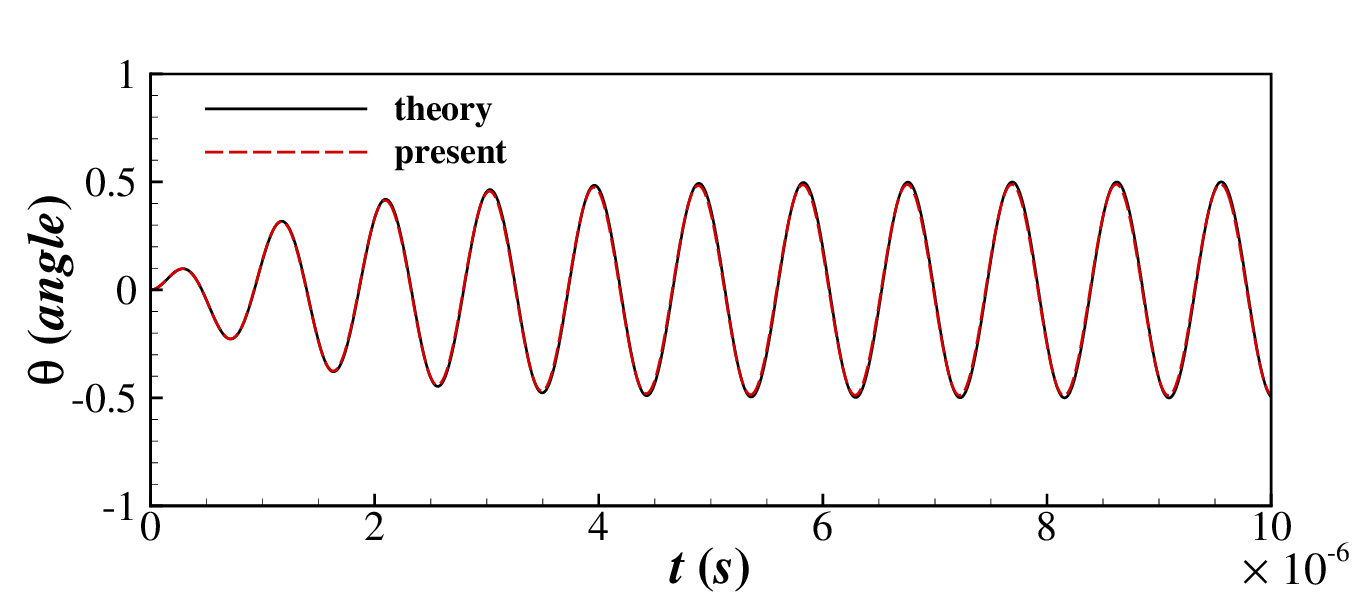}
	}
	\subfigure[]{
		\includegraphics[width=0.55 \textwidth]{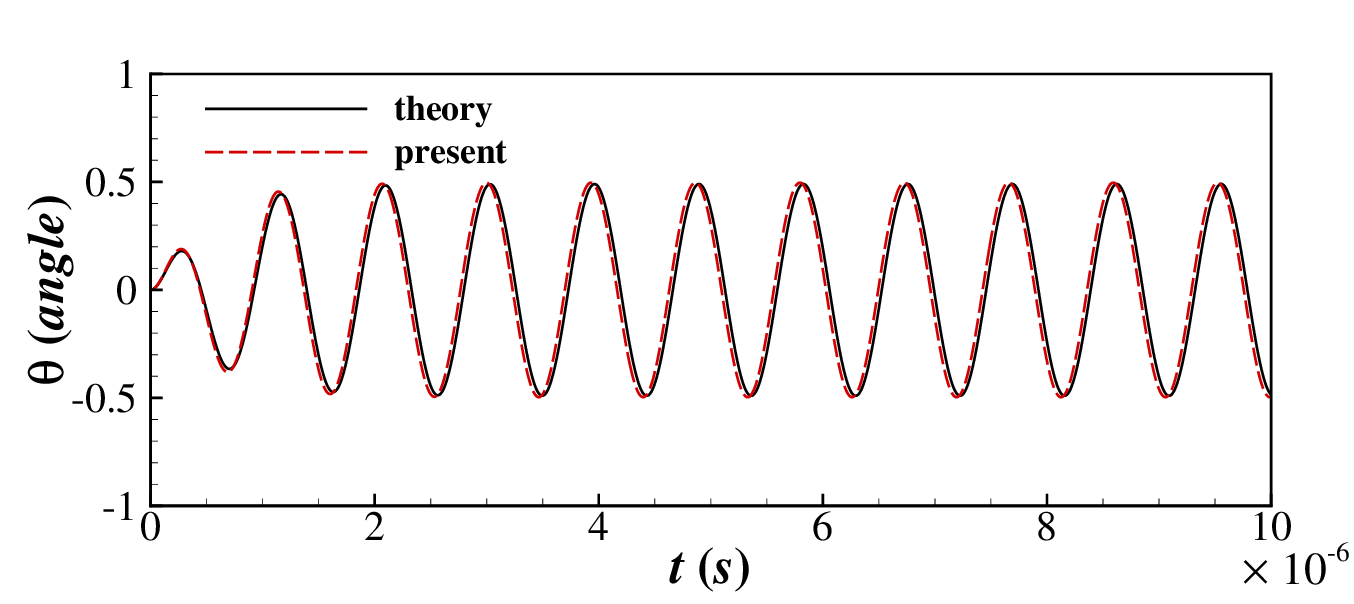}
	}
	\caption{\label{fsi-tilt} Time evolutions of tilting angle of a micro-beam with tilting oscillation at (a) $Kn^{(h_0)}=0.1$ and (b) $Kn^{(h_0)}=1.0$. The maximum tilting angle $\theta_0$ is $0.5^{\circ}$, and the non-dimensional mass of micro-beam $M^*$ is 2769.5.}
\end{figure}

For the free tilting oscillation, similar to linear motion described in Sec.~\ref{linear-free-oscillation}, an external excitation torque is also introduced, and Eq.~\eqref{tilt-struct-coupled} is modified as
\begin{equation}\label{tilt-struct2}
   I\ddot{\theta}(t)+K\theta=T_{ext}+T=T_0cos(\omega_{n}t)+T,
\end{equation}
where $T$ is the gas damping torque, $T_0$ is the amplitude of external excitation torque, and $\omega_{n}$ is the frequency of $T_0$. If $y$ and $c$ in Eq.~\eqref{linear-theory} are replaced by $\theta$ and $\eta$, respectively, the theoretical solution of the time evolution of tilting angle $\theta$ also can be obtained. So, Eq.~\eqref{tilt-struct2} is used in the numerical simulation, and the modified form of Eq.~\eqref{linear-theory} is used to make a comparison. Then assuming the maximum tilting angle $\theta_0$ equals to $0.5^\circ$, the tilting oscillation frequency equals to $\omega_n/(2\pi)=1.073MHz$, and the torsional damping coefficient $\eta$ is calculated from Fig.~\ref{tilt-cm-stat} at the corresponding frequency, $T_0$ can be obtained. In this case, the non-dimensional mass of micro-beam $M^*$ is 2769.5, and the polar moment of inertia $I$ is $M^*(L^2+D^2)/12$ for a rigid plate tilted around its center. Fig.~\ref{fsi-tilt} shows the numerical results at two $Kn^{(h_0)}$ numbers, 0.1 and 1.0. Generally, numerical results agree well with the theoretical solution. For flow at $Kn^{(h_0)}=1.0$, due to the damping torque coefficient used in the theoretical solution is a little different from the real one in the numerical simulation, a little difference between the two results can be observed. {\color{blue}Consequently, with Fig.~\ref{tilt-torque-max}, for predicting the response of a micro-beam by free tilting oscillation, the traditional decoupled method can be used for low-frequency oscillation, and the coupled framework must be used for high-frequency oscillation due the rarefied gas effect.}

\section{Conclusion}\label{conclusion}
In the present study, a decoupled method based on the DUGKS and a coupled framework based on the ALE-DUGKS are used for studying the squeeze-film damping in MEMS. For the decoupled method, based an Eulerian-scheme, the damping force is calculated by imposing a velocity profile on the stationary wall. And for the implementation of coupled framework, a loosely-coupled algorithm is used, in which the fluid and structure dynamic solvers are used alternately in each time iteration step. To validate these two methods, a micro-Couette flow in rarefied gas and an elastically mounted square cylinder oscillating in continuum flow are simulated. Results of both test cases agree well with existing numerical results. For the SFD problems in MEMS, two basic motion forms, linear (perpendicular) and tilting motions of a rigid micro-beam, are fully studied with the forced and free oscillations. Firstly, based on the decoupled method, the damping coefficients at different Knudsen numbers are calculated. A consistent result is obtained compared with the compact damping model. In addition, the nonlinear phenomenon of damping force at a high moving velocity is reproduced. Next, the forced linear oscillations are studied. It can be found that the nonlinear damping is only generated at a low-frequency high-velocity oscillation, so the influence of oscillation frequency must be introduced to construct the damping model. Consequently, the advantage of the coupled framework is to study the large linear displacement problems of a micro-structure, such as shock problem of a high-$g$ MEMS accelerometer~\cite{Parkos2013Near}. And for the high-frequency small-displacement oscillation, the decoupled method still exhibits a good performance to predict the damping force or torque. Besides, the influence of oscillation frequency also must be considered for tilting oscillation, as the damping torques obtained by the decoupled method are higher than that by the coupled method at a high oscillation frequency, especially for flow at a high Knudsen number. Finally, the free oscillation in the resonance regime are studied. The maximum perpendicular displacements or tilting angles calculated by the numerical method agree well with the theoretical solutions. Further work such as the improvement of the squeeze-film damping model and the prediction of nonlinear damping for the complex micro-structure can be continued to enlarge the application range of the DUGKS in MEMS.

\section*{Acknowledgements}
This work is sponsored by the National Numerical Wind Tunnel Project, the National Natural Science Foundation of China (No. 11902266, 11902264, 12072283), the Innovation Foundation for Doctor Dissertation of Northwestern Polytechnical University (CX202015), the Natural Science Basic Research Plan in Shaanxi Province of China (Program No. 2019JQ-315), and the 111 Project of China (B17037).

\clearpage
\bibliographystyle{elsarticle-num}
\bibliography{ref}

\begin{thebibliography}{10}
\expandafter\ifx\csname url\endcsname\relax
  \def\url#1{\texttt{#1}}\fi
\expandafter\ifx\csname urlprefix\endcsname\relax\def\urlprefix{URL }\fi
\expandafter\ifx\csname href\endcsname\relax
  \def\href#1#2{#2} \def\path#1{#1}\fi

\bibitem{Senturia1997Simulating}
S.~D. Senturia, N.~Azuru, J.~White, Simulating the behavior of {MEMS} devices:
  {Computational} methods and needs, IEEE Computational Science \& Engineering
  4~(1) (1997) 30--43.

\bibitem{Rebeiz2004RFMEMS}
G.~M. Rebeiz, {RF MEMS: Theory}, Design, and Technology, John Wiley \& Sons,
  Inc., 2004.

\bibitem{Lee2002Nonlinear}
S.~I. Lee, S.~W. Howell, A.~Raman, R.~Reifenberger, Nonlinear dynamics of
  microcantilevers in tapping mode atomic force microscopy: {A} comparison
  between theory and experiment, Physical Review B 66~(11) (2002) 115409.

\bibitem{Bao2007squeeze}
M.~Bao, H.~Yang, Squeeze film air damping in {MEMS}, Sensors and Actuators A:
  Physical 136~(1) (2007) 3--27.

\bibitem{Veijola2004Compact}
T.~Veijola, Compact models for squeezed-film dampers with inertial effects,
  Journal of Micromechanics and Microengineering 14~(7) (2004) 1109.

\bibitem{Tsien1946superaerodynamics}
H.-S. Tsien, Superaerodynamics, mechanics of rarefied gases, Journal of the
  Aeronautical Sciences 13~(12) (1946) 653--664.

\bibitem{Guo2009Compact}
X.~Guo, A.~Alexeenko, Compact model of squeeze-film damping based on rarefied
  flow simulations, Journal of Micromechanics and Microengineering 19~(4)
  (2009) 237--244.

\bibitem{Beskok2005microflows}
G.~E. Karniadakis, A.~Beskok, N.~R. Aluru, Microflows and Nanoflows -
  {Fundamentals} and Simulation, Springer Science+Business Media, 2005.

\bibitem{Veijola1995Equivalent}
T.~Veijola, H.~Kuisma, J.~Lahdenper{\"a}, T.~Ryh{\"a}nen, Equivalent-circuit
  model of the squeezed gas film in a silicon accelerometer, Sensors \&
  Actuators A Physical 48~(3) (1995) 239--248.

\bibitem{Gallis2004improved}
M.~A. Gallis, J.~R. Torczynski, An improved {Reynolds-equation} model for gas
  damping of microbeam motion, Journal of Microelectromechanical Systems 13~(4)
  (2004) 653--659.

\bibitem{Pandey2007effect}
A.~K. Pandey, R.~Pratap, Effect of flexural modes on squeeze film damping in
  {MEMS} cantilever resonators, Journal of Micromechanics \& Microengineering
  17~(12) (2007) 2475--2484(10).

\bibitem{Lee2009Squeeze}
J.~W. Lee, R.~Tung, A.~Raman, H.~Sumali, J.~P. Sullivan, Squeeze-film damping
  of flexible microcantilevers at low ambient pressures: {Theory} and
  experiment, Journal of Micromechanics \& Microengineering 19~(10) (2009)
  105029.

\bibitem{Li2010molecular}
P.~Li, Y.~Fang, A molecular dynamics simulation approach for the squeeze-film
  damping of {MEMS} devices in the free molecular regime, Journal of
  Micromechanics \& Microengineering 20~(3) (2010) 035005.

\bibitem{Rader2010DSMC}
M.~A. Gallis, D.~J. Rader, J.~R. Torczynski, {DSMC} moving-boundary algorithms
  for simulating {MEMS} geometries with opening and closing gaps, AIP
  Conference Proceedings 1333~(1) (2010) 760--765.

\bibitem{Bahukudumbi2003unified}
P.~Bahukudumbi, A unified engineering model for steady and quasi-steady
  shear-driven gas microflows, Microscale Thermophysical Engineering 7~(4)
  (2003) 291--315.

\bibitem{Diab2014Model}
N.~A. Diab, I.~Lakkis, {Modeling squeeze films in the vicinity of high inertia
  oscillating microstructures}, Journal of Tribology 136~(2) (2014) 021705.

\bibitem{Fan2001statistical}
J.~Fan, C.~Shen, Statistical simulation of low-speed rarefied gas flows,
  Journal of Computational Physics 167~(2) (2001) 393--412.

\bibitem{Fei2013diffusive}
F.~Fei, J.~Fan, A diffusive information preservation method for small {Knudsen}
  number flows, Journal of Computational Physics 243 (2013) 179--193.

\bibitem{Yao2011method}
Z.~H. Yao, X.~W. Zhang, X.~B. Xue, {IP-DSMC} method for micro-scale flow with
  temperature variation, Applied Mathematical Modelling 35~(4) (2011)
  2016--2023.

\bibitem{Guo2013Discrete}
Z.~Guo, K.~Xu, R.~Wang, Discrete unified gas kinetic scheme for all {Knudsen}
  number flows: {Low-speed} isothermal case, Physical Review E 88~(3) (2013)
  033305.

\bibitem{Zhu2017Performance}
L.~Zhu, P.~Wang, Z.~Guo, Performance evaluation of the general characteristics
  based off-lattice {Boltzmann} scheme and {DUGKS} for low speed continuum
  flows, Journal of Computational Physics 333 (2016) 227--246.

\bibitem{Wang1}
Y.~Wang, C.~Zhong, J.~Cao, C.~Zhuo, A simplified finite volume lattice
  {Boltzmann} method for simulations of fluid flows from laminar to turbulent
  regime, {Part I: Numerical} framework and its application to laminar flow
  simulation, Computers \& Mathematics with Applications 79~(5) (2020)
  1590--1618.

\bibitem{Wang2}
Y.~Wang, C.~Zhong, J.~Cao, C.~Zhuo, S.~Liu, A simplified finite volume lattice
  {Boltzmann} method for simulations of fluid flows from laminar to turbulent
  regime, {Part II: Extension} towards turbulent flow simulation, Computers \&
  Mathematics with Applications 79~(8) (2020) 2133--2152.

\bibitem{Xu2010unified}
K.~Xu, J.~C. Huang, A unified gas-kinetic scheme for continuum and rarefied
  flows, Journal of Computational Physics 229~(20) (2010) 7747--7764.

\bibitem{Chen2019Conserved}
J.~Chen, S.~Liu, Y.~Wang, C.~Zhong, A conserved discrete unified gas-kinetic
  scheme with unstructured discrete velocity space, Physical Review E 100~(4)
  (2019) 043305.

\bibitem{Zhong2020simplified}
M.~Zhong, S.~Zou, D.~Pan, C.~Zhuo, C.~Zhong, A simplified discrete unified gas
  kinetic scheme for incompressible flow, Physics of Fluids 32 (2020) 093601.

\bibitem{Zhong2021simplified}
M.~Zhong, S.~Zou, D.~Pan, C.~Zhuo, C.~Zhong, A simplified discrete unified gas
  kinetic scheme for compressible flow, Physics of Fluids 33~(3) (2021) 036103.

\bibitem{Wang2019Arbitrary}
Y.~Wang, C.~Zhong, S.~Liu, Arbitrary {Lagrangian-Eulerian-type} discrete
  unified gas kinetic scheme for low-speed continuum and rarefied flow
  simulations with moving boundaries, Physical Review E 100~(6) (2019) 063310.

\bibitem{Zhu2017Unified}
Y.~Zhu, C.~Zhong, K.~Xu, Unified gas-kinetic scheme with multigrid convergence
  for rarefied flow study, Physics of Fluids 29~(9) (2017) 096102.

\bibitem{Zhu2019Unified}
Y.~Zhu, C.~Liu, C.~Zhong, K.~Xu, Unified gas-kinetic wave-particle methods.
  {II. Multiscale} simulation on unstructured mesh, Physics of Fluids 31~(6)
  (2019) 067105.

\bibitem{Su2020Can}
W.~Su, L.~Zhu, P.~Wang, Y.~Zhang, L.~Wu, Can we find steady-state solutions to
  multiscale rarefied gas flows within dozens of iterations?, Journal of
  Computational Physics 407 (2020) 109245.

\bibitem{Yuan202novel}
R.~Yuan, S.~Liu, C.~Zhong, A novel multiscale discrete velocity method for
  model kinetic equations, Communications in Nonlinear Science and Numerical
  Simulation 92 (2020) 105473.

\bibitem{Yang2021direct}
S.~Yang, S.~Liu, C.~Zhong, J.~Cao, C.~Zhuo, A direct relaxation process for
  particle methods in gas-kinetic theory, Physics of Fluids 33~(7) (2021)
  076109.

\bibitem{Liu2015unified}
S.~Liu, C.~Zhong, J.~Bai, Unified gas-kinetic scheme for microchannel and
  nanochannel flows, Computers \& Mathematics with Applications 69~(1) (2015)
  41--57.

\bibitem{Wang2019generalized}
Y.~Wang, C.~Shu, T.~Wang, P.~Alvarado, A generalized minimal residual
  method-based immersed boundary-lattice {Boltzmann} flux solver coupled with
  finite element method for non-linear fluid-structure interaction problems,
  Physics of Fluids 31~(10) (2019) 103603.

\bibitem{Zhang2020Competition}
J.~Zhang, S.~Yao, F.~Fei, M.~Ghalambaz, D.~Wen, Competition of natural
  convection and thermal creep in a square enclosure, Physics of Fluids
  32~(10).

\bibitem{Geradin2015Mechanical}
M.~Géradin, D.~Rixen, Mechanical Vibration: {Theory} and Application to
  Structural Dynamics, John Wiley \& Sons, 2015.

\bibitem{Chigullapalli2012Non}
S.~Chigullapalli, A.~Weaver, A.~Alexeenko, Nonlinear effects in squeeze-film
  gas damping on microbeams, Journal of Micromechanics \& Microengineering
  22~(6) (2012) 65010--65016(7).

\bibitem{Hou2012Numerical}
G.~Hou, J.~Wang, A.~Layton, Numerical methods for fluid-structure interaction
  -- {A} review, Communications in Computational Physics 12~(2) (2012)
  337--377.

\bibitem{Hartono2007Squeeze}
Sumali, Hartono, Squeeze-film damping in the free molecular regime: {Model}
  validation and measurement on a {MEMS}, Journal of Micromechanics \&
  Microengineering 17~(11) (2007) 2231--2240.

\bibitem{Iannacci2013RF}
J.~Iannacci, G.~Resta, P.~Farinelli, R.~Sorrentino, {RF}-{MEMS} components and
  networks for high-performance reconfigurable telecommunication and wireless
  systems, in: Next Generation Micro/Nano Systems, Vol.~81 of Advances in
  Science and Technology, Trans Tech Publications Ltd, 2013, pp. 65--74.

\bibitem{Chen2012a}
S.~Chen, K.~Xu, C.~Lee, Q.~Cai, A unified gas kinetic scheme with moving mesh
  and velocity space adaptation, Journal of Computational Physics 231~(20)
  (2012) 6643--6664.

\bibitem{qian1992lattice}
Y.~Qian, D.~d'Humi{\`e}res, P.~Lallemand, Lattice {BGK} models for
  {Navier-Stokes} equation, EPL (Europhysics Letters) 17~(6) (1992) 479--484.

\bibitem{thomas1978gcl}
P.~D. Thomas, C.~K. Lombard, Geometric conservation law and its application to
  flow computations on moving grids, AIAA Journal 17~(10) (1979) 1030--1037.

\bibitem{Pan1999Squeeze}
F.~Pan, J.~Kubby, E.~Peeters, A.~T. Tran, S.~Mukherjee, Squeeze film damping
  effect on the dynamic response of a {MEMS Torsion} mirror, Journal of
  Micromechanics \& Microengineering 8~(3) (1999) 200.

\bibitem{Farhat2006Provably}
C.~Farhat, K.~Zee, P.~Geuzaine, Provably second-order time-accurate
  loosely-coupled solution algorithms for transient nonlinear computational
  aeroelasticity, Computer Methods in Applied Mechanics \& Engineering
  195~(17/18) (2006) 1973--2001.

\bibitem{Newmark1959a}
N.~M. Newmark, A method of computation for structural dynamics, Journal of the
  Engineering Mechanics Division, ASCE 85~(3) (1959) 67―94.

\bibitem{Rainald1996Improved}
R.~L{\"o}hner, C.~Yang, Improved {ALE} mesh velocities for moving bodies,
  Communications in Numerical Methods in Engineering 12~(10) (1996) 599--608.

\bibitem{Xu2015Direct}
K.~Xu, Direct modeling for computational fluid dynamics: {Construction} and
  application of unified gas-kinetic schemes, World Scientific, 2015.

\bibitem{archambeau2004code}
F.~Archambeau, N.~Méchitoua, M.~Sakiz, Code {Saturne: A} finite volume code
  for the computation of turbulent incompressible flows - {Industrial}
  applications, International Journal on Finite Volumes 1~(1) (2004) 1--62.

\bibitem{Zhu2016implicit}
Y.~Zhu, C.~Zhong, K.~Xu, Implicit unified gas-kinetic scheme for steady state
  solutions in all flow regimes, Journal of Computational Physics 315 (2016)
  16--38.

\bibitem{Williamson1996}
C.~H.~K. Williamson, Vortex dynamics in the cylinder wake, Annual Review of
  Fluid Mechanics 28~(1) (1996) 477--539.

\bibitem{Singh2005Vortexs}
S.~P. Singh, S.~Mittal, Vortex-induced oscillations at low {Reynolds} numbers:
  {Hysteresis} and vortex-shedding modes, Journal of Fluids \& Structures
  20~(8) (2005) 1085--1104.

\bibitem{Li2019Mode}
X.~Li, Z.~Lyu, J.~Kou, W.~Zhang, Mode competition in galloping of a square
  cylinder at low {Reynolds} number, Journal of Fluid Mechanics 867 (2019)
  516--555.

\bibitem{Guo2015Discrete}
Z.~Guo, R.~Wang, K.~Xu, Discrete unified gas kinetic scheme for all {Knudsen}
  number flows: {II. Compressible} case, Physical Review E 91~(3) (2015)
  033313.

\bibitem{Parkos2013Near}
D.~Parkos, N.~Raghunathan, A.~Venkattraman, B.~Sanborn, Near-contact gas
  damping and dynamic response of high-g {MEMS} accelerometer beams, Journal of
  Microelectromechanical Systems 22~(5) (2013) 1089--1099.

\end{thebibliography}

\end{document}